# The State of AI Ethics

## Volume 5

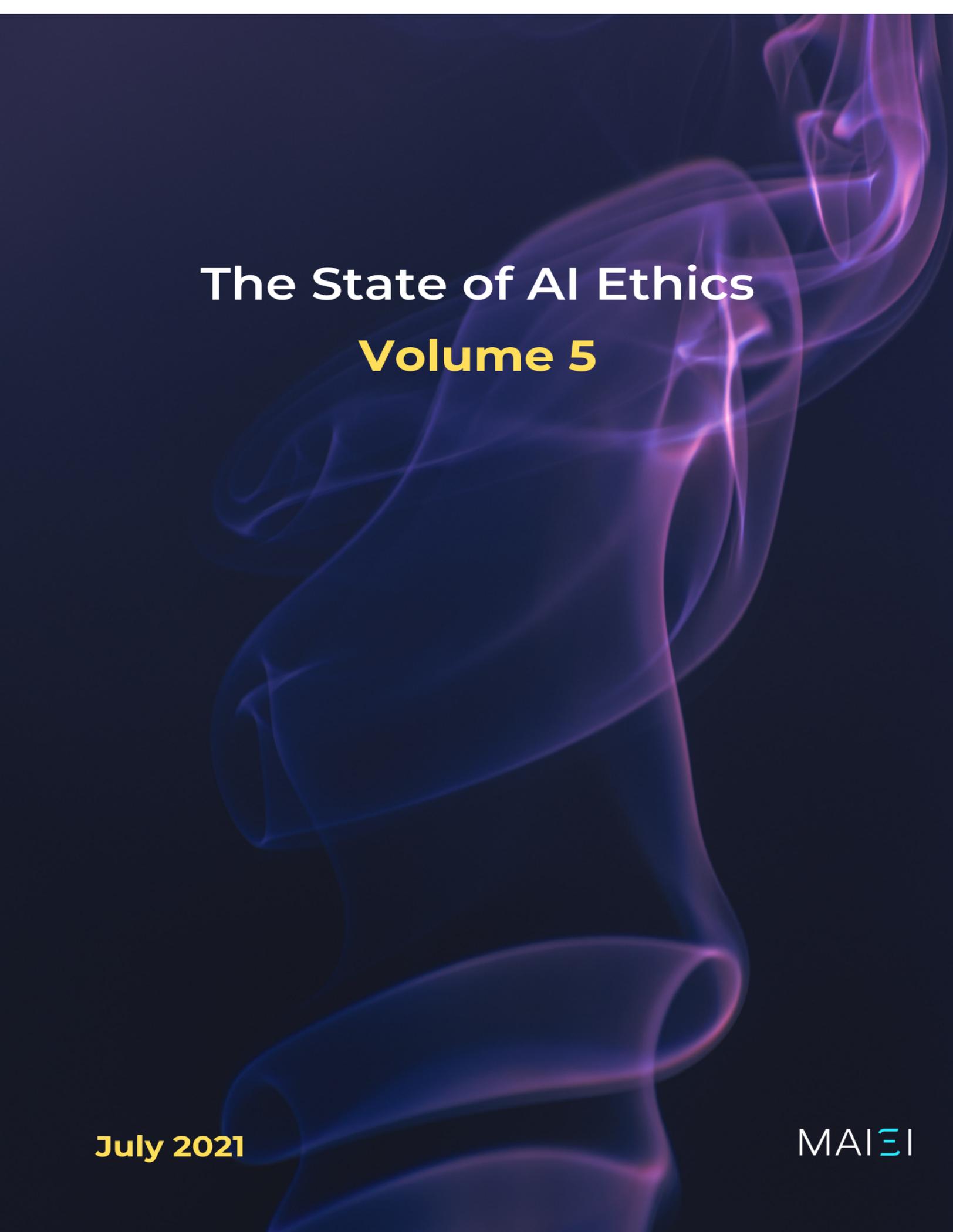

**July 2021**

MAIEI

This report was prepared by the **Montreal AI Ethics Institute (MAIEI)** — an international non-profit organization democratizing AI ethics literacy. **Learn more on our website** or subscribe to our weekly newsletter **The AI Ethics Brief**.

This work is licensed open-access under a **Creative Commons Attribution 4.0 International License**.

Primary contact for the report: **Abhishek Gupta (abhishek@montrealethics.ai)**

**Full team behind the report:**

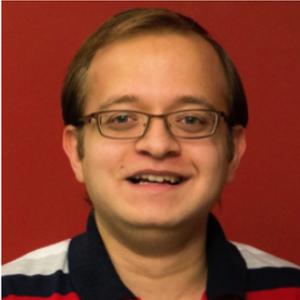

Abhishek Gupta
FOUNDER, PRINCIPAL RESEARCHER, AND DIRECTOR

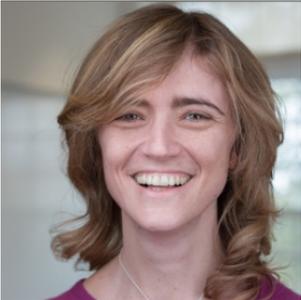

Marianna Ganapini, PhD
FACULTY DIRECTOR

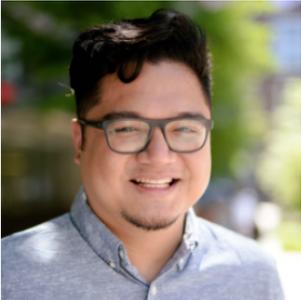

Renjie Butalid
CO-FOUNDER AND DIRECTOR

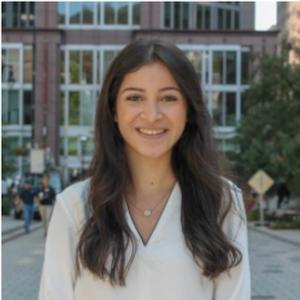

Masa Sweidan
BUSINESS DEVELOPMENT MANAGER

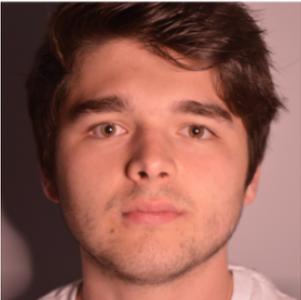

Connor Wright
PARTNERSHIPS MANAGER

**Special thanks to our contributors:**

Iga Kozlowska, Sarah Grant, Andrea Pedeferri, Alexandrine Royer, Victoria Heath, Muriam Fancy, Shannon Egan, Marcel Hedman, Natalie Klym, and Karim Nader.

# Table of Contents

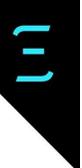

---

***Note:** The original sources are linked under the title of each piece. The work in the following pages combines summaries of the original material supplemented with insights from MAIEI research staff, unless otherwise indicated.

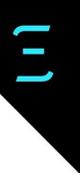

# Founder's Note

Welcome to another edition of the State of AI Ethics report from the team at the Montreal AI Ethics Institute! We've been thrilled with progress in the field this past quarter, especially as we've seen more calls for practical and actionable guidance for practitioners in the field, rather than just abstract, theoretical frameworks for evaluating the societal impacts emerging from the use of AI systems. In particular, we've seen a rise in interdisciplinary approaches to addressing these concerns as you'll observe from the pieces featured in this edition of the report.

We also got an overwhelming response from past readers of the report who wanted to learn more about our staff and community's perspectives on the developments in the field and heeding that call, we've added a new section in the report titled "What we're thinking" that will highlight some of our observations in more detail. We also have a special contribution from a stellar group of scholars who came together to discuss how we can reconceptualize critical race studies borrowing ideas from quantum computing titled "Critical Race Quantum Computer: A Tool for Liberation" that I strongly urge readers to peruse.

We also chose to engage in a more topical analysis of the state of our field by looking through the areas of "AI and Creativity", "AI and Environment", and "Geopolitics and AI" to offer readers an in-depth look at some emergent areas that we believe will be essential in understanding how the field is moving forward in some underexplored areas and what we can expect going forward. We also have our evergreen "Outside the boxes" section with an eclectic mix of topics for those who want to get a broad coverage of areas in the field. Finally, we have featured profiles in our "Community Spotlights" showcasing work from intrepid scholars and activists reshaping our understanding of society and technology in different parts of the world. I hope you enjoy this report and see you in the "Closing Remarks."

---

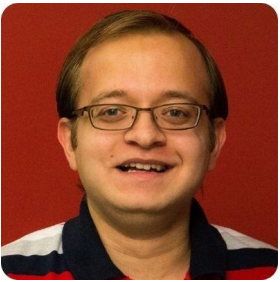

**Abhishek Gupta (@atg_abhishek)**
Founder, Director, & Principal Researcher
Montreal AI Ethics Institute

Abhishek Gupta is the Founder, Director, and Principal Researcher at the Montreal AI Ethics Institute. He is a Machine Learning Engineer at Microsoft, where he serves on the CSE Responsible AI Board. He also serves as the Chair of the Standards Working Group at the Green Software Foundation.

# 1. What we're thinking by MAIEI Staff and Community

## From the Founder's Desk

**Tradeoff determination for ethics, safety, and inclusivity in AI systems**

**(Original article by Abhishek Gupta)**

Design decisions for AI systems involve value judgements and optimization choices. Some relate to technical considerations like latency and accuracy, others relate to business metrics. But each require careful consideration as they have consequences in the final outcome from the system.

To be clear, not everything has to translate into a tradeoff. There are often smart reformulations of a problem so that you can meet the needs of your users and customers while also satisfying internal business considerations.

Take for example an early LinkedIn feature that encouraged job postings by asking connections to recommend specific job postings to target users based on how appropriate they thought them to be for the target user. It provided the recommending user a sense of purpose and goodwill by only sharing relevant jobs to their connections at the same time helping LinkedIn provide more relevant recommendations to users. This was a win-win scenario compared to having to continuously probe a user deeper and deeper to get more data to provide them with more targeted job recommendations.

This article will build on The importance of goal setting in product development to achieve Responsible AI adding another dimension of consideration in building AI systems that are ethical, safe, and inclusive.

**Why tradeoff determination?**

The primary purpose that tradeoff determination serves is to foreground inherent tensions between the various goals for a project.

Let's consider a scenario where the business metric is revenue through ads on a platform. A design decision that can help with that is implementing something like infinite scroll that keeps the user on the platform for as long as possible by continuously serving up more related content interspersed with ads. The ads themselves can be made more appealing or relevant to the user by utilizing their personal data which might increase the number of click-throughs that you get on those ads.

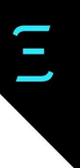

There are many concerns here. Perhaps the most obvious one is the invasion of privacy and contortion of the consent of the user as to how their personal data is used. Another is a reliance on a known dark design pattern that skews towards increasing time spent on the platform but doesn't necessarily talk about the quality of that time spent.

There are other choices like the bias vs. variance tradeoff that you might encounter as you inch towards utilizing more complex models, you run the risk of making things opaque from an explainability perspective. This might matter in case you want to justify that the ads on your platform are not discriminatory based on sensitive attributes. A more complex model might improve performance but at what cost?

When thinking about tradeoff determination here, highlighting this explicitly where there is a tension between what might be socially good for the user with what is good for business is the first step in helping address problems effectively.

**How to do tradeoff determination effectively?**

There are 3 initial steps that one can take to start off with tradeoff determination.

**1. Explicitly list the tradeoffs**

As highlighted in the example above, when there is clarity in the first-order effects of the techniques and design decisions being made, they should be listed out explicitly. Once that is done, adopt a systems thinking approach that takes into account second-order effects of these designs.

The second-order effects are subtle to track and can manifest in unexpected ways. Yet, they are often responsible for a large chunk of the harm because there aren't any explicit safeguards that are put in place to protect against those harms. An explicit listing also helps keep these tradeoffs centre of mind for everyone working on the system.

**2. Check these decisions against your goals**

As mentioned in The importance of goal setting in product development to achieve Responsible AI, goals can serve as the North Star in keeping us accountable to ourselves in terms of what we are trying to achieve with the system. As you go deeper into the specifics of a project, it is likely that details can derail the original vision and goals for the project. This is a candidate for introducing unwanted harm resulting from the use of the system.

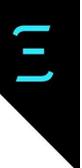

As you make technical and architecture decisions, checking frequently against the goals and thinking about what tradeoffs it results in can help orient you in the right direction. For those familiar with John Boyd's OODA (Observe, Orient, Decide, Act), this forms the crucial step of Orient to make sure that what follows is going in the right direction.

**3. Complementarities vs. win-lose scenarios**

Not all decisions that are in support of business goals need to be a losing scenario for the user in terms of their welfare. Going back to the example of LinkedIn job recommendations, we can see how a design that replaces extensive data collection with peer recommendations can continue to meet the business goals of wanting people to use the platform because of relevant job postings without needing to resort to invasion of privacy and misuse of personal data.

This is an opportunity to get creative and as more practitioners enter the field of AI, you can use this as a differentiating quality for yourself: achieving ethical alignment without sacrificing business needs.

**What are some actions you can do in the AI lifecycle for tradeoff determination?**

One of the key things you need to do in the AI lifecycle once you have done some of the steps above is to monitor for the tradeoffs. While it is great to think about them in the beginning and make choices that are aligned with ethical considerations, given the dynamism and complexity of AI systems, without constant monitoring, you run the risk of emergent harm that can diminish the impact of your work.

Setting thresholds for acceptable behaviour of the system is a concrete way to achieve the above. This might include things like the amount of time a user is spending on a platform at a stretch and if having a brief notification popping up asking them to take a stretch and walk outside can break a negative pattern. We already see things like this in fitness trackers and the call screens on Android phones.

Finally, it is not just sufficient to identify tradeoff determinations. Yes, acknowledging that there is a problem is always the first step but we need to move beyond. The way to do that is to associate actionable remediation measures with each of the tradeoffs that you list. This helps the stakeholders break inertia and meaningfully act on the recommendations to improve system outcomes.

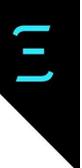

Hopefully this idea of tradeoff determination is something that you feel natural about and can already see where in your design and development phases of the AI lifecycle you can integrate them.

## Systems Design Thinking for Responsible AI

**(Original article by Abhishek Gupta)**

There is rarely a bare AI system. Typically, AI capabilities are included as a subcomponent of a larger product or service offering. Such a product or service offering exists in a social ecosystem where there are specificities of culture and context.

So, when thinking about building an AI system and the impact that it might have on society, it is important to take a systems design thinking approach to be as comprehensive as possible in assessing the impacts and proposing redressal mechanisms.

**What is systems design thinking?**

In essence, it is thinking about the components of the system in relation to the environmental elements which they will interact with when that system is deployed. This is in contrast to more closed-off approaches that look at a product/service in isolation from where it will be deployed. The purpose of using the systems design thinking approach is that it helps to unearth hidden feedback loops and externalities that might impact people in ways that can have ethical consequences like violations of privacy and facing biases.

Systems design thinking encourages the developers and designers to bring into the fold the larger sociotechnical context in which the system will be deployed to the table and map out how and where their product/service will impact those elements, both directly and indirectly.

**How to do systems design thinking for Responsible AI?**

A great place to start with acknowledging that you don't necessarily have all the pieces of information to apply this approach to the systems that you are building, both within your team and perhaps even your organization. But, that's ok!

The first step to address it is to bring in external expertise and stakeholders such as those with domain expertise, policy makers, etc. who can help you articulate better some of the challenges that will arise when you try to deploy an AI system in your target environment. Those with experience could also help to point out various system dynamics and cultural specificities that can impact the behaviour of your system.

This can entail doing some field work to gather intelligence on how people will actually use your product/service and the lab is not a good proxy for that. There are far too many variables in terms of what can happen out in the real world which can remain hidden in our blindspots.

Drawing system diagrams that include the external elements explicitly and then marking reinforcing and balancing feedback loops is another step that can help you articulate these impacts more holistically.

Also, pay attention to the learning component of the system. That is, pay attention to all the flows through which the system will learn as it interacts with the world which will be important for building resilience into the system so that it doesn't stray far outside operational parameters from a safety and reliability perspective.

**What are some things to keep in mind?**

In particular, it is critical that you think about the resilience of the system to being gamed. Such a system is going to be deployed in a complex, messy environment where there are adversaries who may try to game the system to eke out benefits that they are not entitled to or to harm the performance of the system so that it becomes less performant for those who need it the most. In such cases, building failsafes that can degrade gracefully is important.

Finally, keeping in mind the core idea of systems design thinking, emphasize the second-order effects and second-order inputs that will affect the system. These are often hard to tease out but with the help of domain experts and those who have practised such approaches in the past, you can improve not just the resiliency of the system but its overall performance because of a better understanding of the problem that you are solving and the implications that the solution that you are proposing will have.

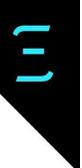

## The importance systems adaptability for meaningful Responsible AI deployment

**(Original article by Abhishek Gupta)**

So far we've covered a few ideas on how to deploy Responsible AI including "The importance of goal setting in product development to achieve Responsible AI", "Tradeoff determination for ethics, safety, and inclusivity in AI systems", and "Systems Design Thinking for Responsible AI" which have shown us that borrowing ideas from adjacent and related domains can aid in the process of making Responsible AI a reality rather than a pipe dream.

Let's use this post to explore the idea of Systems Adaptability as another tenet that can help us achieve Responsible AI in practice.

**Dynamism**

Humans are dynamic! (No surprises there!)

As is the environment within which we operate.

*"All models are wrong, but some are useful." - George Box*

Necessarily, when we have an AI system, we have to make some simplifying assumptions about the world within which it is going to be deployed. Such assumptions translate the high-fidelity real world into a lower-resolution digital twin where we perform our optimizations and we hope that they generalize to the real world.

But, hope is not a strategy! (I mean technically we don't hope of course; we test, evaluate, verify, and validate before releasing it into the world. If you're not doing that, please do so!)

Keeping this in mind, we need to make sure that we are able to adapt to this dynamism in a deliberate way. The reason we need to do this is because rigidity in an AI system in a dynamic world leads to ethical challenges.

Let's take an example.

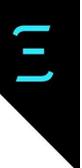

Think about a content moderation system that is updated every quarter because of the large training and deployment costs. This might happen because of large costs, immature infrastructure, or the belief of the team that they don't need to make those updates more frequently. Or any number of other reasons. So what can go wrong?

Language (if we take the case of unimodal content moderation) takes on new semantics quickly in a digital and interconnected world. Think about political movements that can adopt a word to become a call-sign or perhaps how words can be repurposed into hate speech online. This can happen in the span of a few weeks. A quarterly update to the system can lag the changes happening in the real world limiting their efficacy and letting hate speech pervade unchecked on the platform.

A non-adaptive system or a slowly adapting system has the potential to become misaligned with objectives and lead to more harm than what it is meant to mitigate in practice.

So, what can we do to address these challenges?

**Suggestions on how to do this**

This might seem trite given some of the articles that I linked at the top of this post but it bears repeating because it is that important: bring the right stakeholders to the table! In this instance, it particularly includes community stakeholders who will be the subjects of the system and those who are around them who might play a role in administering the system. The reason this is important is because we never know how a system will be used exactly in the field and how it might be abused (intentionally or unintentionally). The exercise of bringing together these stakeholders will help to elucidate those concerns and bring them to the centre.

Having internal red teams whose job is to break the system can also help with finding ways that the system can be misused. It is a common practice in security testing and it should be something that becomes more commonplace with AI systems as well. The field of machine learning security is one that focuses on that. Such red teams can also help to stress test the system that unearths new ways in which the system can start behaving erratically. This becomes particularly important in the case of mission-critical systems such as those used in transport and medicine.

Adaptability also comes from having the ability to rapidly swap out non-working or misbehaving models. This requires a mature MLOps infrastructure to succeed. I spoke more about this idea in "Adopting an MLOps mindset for AI Ethics".

## Building for resiliency in AI systems

**(Original article by Abhishek Gupta)**

We've come across ample failures of AI systems by now, as popularized again and again in reporting and research work that I hopefully don't need to make the case that we need to think about resiliency as an active component of building ethical, safe, and inclusive AI systems.

(If you are not convinced, take a look at The State of AI Ethics Report: Volume 4 that showcases some of that!)

But, the key question here is what should we be thinking about when we want to build such resilient systems? I would recommend the following as key considerations to build more resilient AI systems:
- Adversarial examples
- Criticality of system
- Failover mechanisms
- The "spaghetti" factor

The key takeaway that I would recommend for the folks reading this is: resiliency is a property of the entire system and hence approaches to build resiliency should take a systems-level approach.

### 1. Adversarial Examples

No discussion on the robustness and resiliency of AI systems is complete without a discussion of adversarial examples. These are carefully crafted inputs to an AI system that are meant to trigger misclassification of otherwise undesired behaviour from an AI system. There has been a lot of research in this space, in particular the NSCAI has also made it an important part of their recommendations for fielding AI systems in a responsible manner. While a full discussion of adversarial examples is beyond the scope of this article, I encourage you to keep these at the forefront in designing your AI systems. In particular, I encourage you to keep abreast of the latest literature in the domain, and to engage in a simple exercise: what is the worst result that can arise from my input to this AI system? The answer that you get from it is something that might just be possible and protecting against triggering those behaviours through more robust defences in the training lifecycle, better model choices, and applying some of the points as mentioned below will help in inching towards a more resilient AI system.

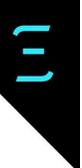

## 2. Criticality of system

Some of the things that we build end up being more critical than others. Think of managing customer payment history data vs. added emoji functionality in your web app. On a more serious note though, there are many systems that are critical in delivering service to customers and users which have very little room for downtime or leeway for degradation in performance.

Such systems deserve to be treated with special care and the advice presented in this article should be prioritized for immediate or worst-case near-term application rather than delaying it. While it may seem obvious, this happens all the time in the form of technical debt accumulation and technical debt also comes in different flavors. The kind that we're talking about in the critical systems is the worst in that it bites really hard when left unaddressed for long.

Working with other engineers and stakeholders on the team to get a sense for the criticality of the system will be essential in correctly prioritizing these measures on the product roadmap.

## 3. Failover mechanisms

Systems will fail!

Yes, read that again - systems will fail! And it's not an indictment against the quality of software developers, data scientists, and machine learning engineers that you have. It is just a function of the degree of complexity that is presented by these systems that presents challenges that will be unforeseen. Planning for systems failing is a better approach than hoping that they don't. Anyone that tells you otherwise either doesn't have real experience building systems or is trying to pull a fast one on you by selling you something that is going to overpromise and underdeliver. You've been warned!

*As one of my favourite quote says: "Hope is not a strategy." - Ben Sloss, SRE, Google.*

So, now that we got that out of the way, what is the best way to build resiliency through the use of failover mechanisms? First, a failover mechanism is an alternate pathway that will take over the performance of the primary function of a system in case the module or subsystem that usually performs that system fails for some reason. The goal of the failover mechanism is to provide a reprieve to the staff to figure out what went wrong and give them enough time to fix the original subsystem. In the meantime, the customers and users of the system hopefully see only minimal disruptions in the quality of service.

In the case of building ethical, safe, and inclusive AI systems, we have the opportunity to use these mechanisms as a way to guarantee that service, especially to those that really depend on the system, continues to exist when things go wrong. In addition, when an ill-behaving system needs to be pulled offline, a simpler, deterministic failover mechanism can take over to continue to provide service and not further deteriorate the experience for the customers and users because of the ill-behaving system.

**4. The "spaghetti" factor**

Oh yea! Well, there was going to be some talk of food here and given all the metaphors that we have in software development that touch on food, I figured why not use one of those to describe a problem that we face all too often in complex AI systems.

The "spaghetti" factor refers to the problem that arises when you have code that interleaves through a lot of different systems, both indirectly through API calls (loosely decoupled hopefully!) and directly through interacting with upstream data processing elements. Brittleness in these pipelines where the system's downstream functioning becomes highly contingent on proper functioning of upstream services makes smooth functioning a challenge. This subsequently severely impacts the resiliency of the system because minor updates to upstream code can wreak havoc in other code downstream. Following some good architecture design patterns and SOLID principles can actually get you quite far!

More so, when designing the system, make sure to list out as many dependencies, indirect and direct, as possible to get a full sense of the picture, both in terms of what can affect the "intelligent" component of the system but also how it can affect the rest of the system as well.

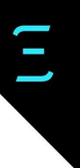

# Sociology of AI Ethics

**Algorithmic Impact Assessments – What Impact Do They Have?**

**(Original paper by Jacob Metcalf, Emanuel Moss, Elizabeth Anne Watkins, Ranjit Singh, and Madeleine Clare Elish)**

**(Research summary by Dr. Iga Kozlowska)**

**Overview**: Algorithmic Impact Assessments (AIAs) are a useful tool to help AI system designers, developers and procurers to analyze the benefits and potential pitfalls of algorithmic systems. To be effective in addressing issues of transparency, fairness, and accountability, the authors of this article argue that the impacts identified in AIAs need to as closely represent harms as possible. And secondly, that there are accountability forums that can compel algorithm developers to make appropriate changes to AI systems in accordance with AIA findings.

**Introduction**

Writing an Algorithmic Impact Assessment (AIA) is like painting an impressionist landscape of Van Gogh's swirling cypresses or Monet's floating water lilies. Well, maybe not exactly, but stay with me.

First, what are AIAs anyway? Metcalf et al. define AIAs as "emerging governance practices for delineating accountability, rendering visible the harms caused by algorithmic systems, and ensuring practical steps are taken to ameliorate those harms."

If you're an AI developer, your company may already have instituted this practice. You may have experienced this practice as a document your team has to fill out that answers questions about the machine learning model or algorithmic system you're building, maybe with the help of some ethical AI subject matter experts.

While praising these efforts as a good start, the authors focus on AIAs' existing shortcomings.

Specifically, they describe two key challenges with doing AIAs in such a way that they truly prevent harms to people:
- Impacts are only proxies for real harm that people can experience
- AIAs don't work without an accountability mechanism

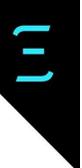

**Impacts as proxies for harms**

The authors argue that impacts don't necessarily measure harms and may, in worst-case scenarios, obscure them. Describing real harms is very difficult because it's a truth that, like many social and psychological aspects of humanity, is hard to evaluate and represent in words, let alone to quantify.

For example, in your AIA, you may measure how far your model deviates from your fairness benchmark, which may be based on a company policy (or just group consensus) that model outcomes shouldn't diverge more than 10% across some demographic characteristics (let's say age, gender, race) and their intersections. That metric measures the impact of your, say face recognition model, on your customer's ability to get equal quality of service. The impact is that there will be no more than a 10% difference in predictive quality between, for example, young Black women and older white men.

But the metric is not measuring the emotional or psychological harm done to individuals let alone entire communities when they get repeatedly misrecognized by the algorithm. It does not capture even the more easily quantifiable harms like an economic loss that can stem from such biased systems. The impact is only an indirect measure of underlying harm.

In other words, just like with an impressionistic painting, we may get a mediated sense of reality through the painting, but we don't come close to the underlying lived experiences of actually being in a field of Van Gogh's sunflowers. We don't even get a view that we might see in a pre-modernist painting where the name of the game was to convey a scene with as close to photographic precision as possible. The impressionistic view of impacts is still useful, but unlike with modernist painting, there is immense value in getting as close to a true representation of reality as possible, knowing that it will never be 100% perfect.

Moreover, when doing an AIA, it is difficult to evaluate its comprehensiveness because there is no objective standard against which to measure. When do you know you've adequately predicted all impacts that could potentially face your customers and other indirect stakeholders of your product?

Instead, the quality of an AIA is determined through the consensus of experts who happen to be at the table. And we know that not all voices, particularly those of marginalized communities, are going to be present at the table. Likewise, few product development teams hire people with psychology or social science backgrounds, so those perspectives are likely to be absent.

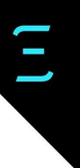

In other words, much like art, which is judged based on expert and/or public consensus and has no singular objective standard, the adequacy of AIAs is currently judged by what the authors call "epistemic communities" that are not necessarily representative of all voices needed to actually prevent harms.

**AIAs need accountability**

Just as there is no one who can legitimately tell an artist that they must change their work, with AIAs there is, as of yet, no authority that can mandate that an organization make changes to its AI systems based on what is found in an AIA. With no "forum" of accountability, as the authors call it, a company can write an AIA and yet make no mitigations to the AI system that would actually reduce harm.

Here is where the art metaphor really breaks down. Whereas we obviously don't want a regulatory agency enforcing certain artistic practices or styles—that is called censorship—in the case of AIAs, the authors argue, some accountability body is required. Such a mechanism is necessary to ensure that organizations do AIAs in the first place, do them well, and actually act on them. Doing an AIA just to check a box without it informing the design and development of the AI system does not reduce harm to users.

**Between the lines**

Completing an AIA may not be as profound or satisfying as painting an impressionist masterpiece. But it certainly is an art form that requires skill, knowledge, and the social construction of a world of algorithmic accountability. And, like a great piece of art, it can encourage us to reflect, motivate us to act, and hopefully create change for the better.

It's too early to tell just how common AIAs will become or how effective they will be in changing the shape of algorithm-based technology. Classification and prediction algorithms have already proven to cause real-world harm to those least advantaged, whether it's in the context of criminal justice or child abuse prevention. AIAs are a great immediate intervention, but without robust accountability measures, they might fall short of what most of us truly want: technology that amplifies—not erodes, human dignity.

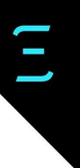

## Slow AI and The Culture of Speed

**([Original paper](#) by John Tomlinson)**

**(Research summary by Iga Kozlowska)**

**Overview**: The sociology of speed considers how people experience temporality in different contexts and how humans make meaning out of the social construct of time. John Tomlinson argues that we've moved from a culture of "mechanical speed" that dominated the 19th century as the Western world industrialized to "telemediated speed" at the turn of this century. This "immediacy" is an accelerated experience of time where information, goods, and people can be accessed immediately and effortlessly anywhere, anytime. While this is not categorically undesirable, Tomlinson considers how this imperative limits our imaginaries of the good life.

**Introduction**

*"The cult of productivity and our obsession with employing every minute wisely have become the great unquestioned virtues of our age."– Judy Wajcman*

We live in a one-click culture. Our digitally mediated world, in the last two decades, has all but obliterated the time between "now" and "then" and the space between "here" and "now." We are accustomed to interacting with things and people directly and immediately. The "middle term" is made "redundant," as Tomlinson puts it. The gap between desire and fulfillment has closed. With one click. Is this our version of the good life or can we imagine something better?

Our obsession with speed is driven by what Tomlinson calls "fast capitalism," which is a form of late capitalism where information and communication technologies, aka the Internet, accelerate the imperative to consume and therefore produce, thereby creating a vicious cycle. It is aided and abetted by a culture that still equates speed with positive values like efficiency and productivity whereas slowness it associates with laziness, idleness, waste, and even dimwittedness or backwardness. The cult of efficiency, underpinned by Frederick Taylor's scientific management of the early 20th century, still reigns supreme, particularly in the tech industry which is producing Tomlinson's "telemediated" world. In fact, efficiency and productivity are values that reign so supreme, that they sometimes obliterate other human values like dignity, pleasure, freedom, leisure, and yes, idleness (see Bertrand Russell's In Praise of Idleness).

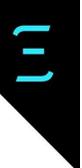

While Tomlinson doesn't address AI specifically, extending his concept of telemediated immediacy, I will argue that, in the context of AI, we need to take a step back and consider which social processes can or should be sped up through algorithmic intervention and which should not be. As we'll see, sometimes ethical AI means slow AI.

**Human dignity and work**

Obviously, not all digitally mediated experiences should be decelerated. It has been a long-standing design imperative, from the telegraph to Zoom, to make the user experience smooth and seamless. We want fast connectivity. We want our YouTube videos to buffer quickly and our Google documents to load rapidly. There is no reason why checking out an e-book from my local library should take five clicks. Amazon can do it in one, and it's the immediacy of the one-click to which we've become accustomed and now expect, nay demand!

However, for many human experiences, such as work, where consequential decisions are made about life opportunities, we need to think twice about whether we design for speed at all costs. In Alec MacGillis's recent book about Amazon, Fulfillment, we learn how automated surveillance systems "measure productivity" by calculating each employee's "time off task." These productivity scores are then used to make algorithmic suggestions on whom to retain and whom to fire. A quote from one of the company lawyers illustrates this:

"Amazon's system tracks the rates of each individual associate's productivity and automatically generates any warnings or terminations regarding quality or productivity without input from supervisors" (emphasis mine).

Hundreds of employees are fired this way. Others struggle to have enough time to use the restroom for fear of the algorithm catching them "off task." Such scoring has the potential to remove human bias in firing decision-making (though more research is needed to determine if that is actually true) and no doubt adds speed to the decision-making process thus generating "time-savings" for the supervisors who no longer have to review each case manually. But what are the consequences of this type of treatment for the people involved and their communities? It's unlikely that someone who is not given enough time to use the restroom can do their best work, to say the least. Social bonds and a sense of belonging and community at work are very important features of our social lives and could be impacted negatively knowing that, as a worker, I can be fired at any minute by an algorithm without even the decency of human input.

For information workers too, the immediacy demanded by digital technologies and the removal of in-person interaction due to the COVID-19 pandemic have led to "digital exhaustion." A recent "hybrid workplace" study by Microsoft found that employees feel the burden of accelerated, always-on digital exchange. While immediate contact with teammates through email, chat, and video calls, and sometimes all at once, seems efficient, effortless and time-saving (a walk down the hall no longer required!), there are undesirable sociopsychological consequences to this kind of accelerated communication: stress, anxiety, feeling harried, inability to focus, feelings of loss of control, and exhaustion from always being on and available. In the workplace, time is money, but sometimes it pays to slow down.

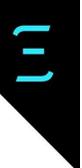

**Designing slow AI**

We've already seen how AI-based consequential decision-making in the workplace like firing and hiring is just one social context where the consequences can be harmful enough to warrant a second look at the cost of speed to human dignity. Other scenarios include healthcare diagnosing and treatment, bail and sentencing in the criminal justice system, policing and arrest-making, grading student exams and assignments, and the list goes on.

In addition to the more common concerns around fairness, accountability, and transparency, designers and developers should consider how accelerating a decision to digital speed impacts all the stakeholders in that process. Designing for slowness may not be popular, but it is not a new idea (see Hallnäs & Redström 2001) and is especially pertinent in the age of AI. The question each AI designer should ask themselves then is, how is the automation of this task speeding up the social rhythm of that activity? And what are the potential benefits and harms of that acceleration to all stakeholders involved?

For example, a teacher may benefit from automated exam graders by saving the time that it would have taken him to perform that activity manually, and maybe the students benefit too because now the teacher can invest that time in quality interaction with the students. But is there anything that is lost in that time gained? How might this rapidity eliminate the opportunity for the teacher to better get to know his students' writing style and learn more about them through their writing? How could that affect the teacher-student relationship? Maybe the student is grateful because the teacher has been biased toward her for one reason or another and always gave her a lower grade than she deserved. Or maybe the student feels like why bother trying hard when the only "person" that will read her paper is a machine that by definition cannot care about her work, learning, and development as a human being.

Through user research in the early design phase of the development lifecycle, these kinds of questions should come to the fore. Potential harms should be identified and mitigations considered. For example, for automated decision-making systems, you may require a "human-in-the-loop" so that the AI system doesn't trigger an action immediately but instead gives a human time to interpret the results, check with other experts, make sense of it, and then confirm the next step or override the system. Requiring human intervention slows down "the experience," but it can mitigate harms that would result from the system making the wrong decision.

Another example might be slowing down the user experience to elicit healthier, human-centred behaviour online. For example, in any personal data collection scenario, the user should be made aware of what is being collected, why, and what control they will have over their data. In situations with particularly important consequences to the user's privacy, we may want to slow the experience down by putting in a "roadblock" or a "speed bump" like a meaningful consent experience. This may require the user to read more information and click a few more buttons, but it also allows them to make a more informed decision. Equally in the context of social media, psychologists have documented that the online disinhibition effect sometimes makes us say or do things that we wouldn't otherwise do in person. Therefore, designers could consider helpful tips in the UI or pop-ups that help us stop and think before making a post. Meaningful human control over an AI system often requires slowing it down so that the user can pause, think, and then act. Rather than feeling out of control, the user is again made to feel like they are in the driver's seat instead of the algorithm.

**Between the lines**

Published in 2005, Tomlinson's book already feels a bit outdated and almost naively innocent given how much our world has "sped up" even in the last decade and a half. However, that is testament to the strength of his thesis, in that if applied to the latest digital technologies, like AI-based systems, it not only holds true but helps to illustrate just how effortlessly fast capitalism has obliterated the gaps in time and space. What were once physical, manual social processes are now ones that are quickly becoming digitized and automated. As Tomlinson argues, this is not necessarily bad, but nor should speed be taken at face value as a social good either. Automation does not always equal efficiency, and efficiency is not always the value we should be solving for. There are many roads to progress, and not all of them lead us through efficiency. In other words, we need a more nuanced approach to AI-based automation that examines the social contexts of each application and the range of values that people want to be expressed and enacted through AI. This is where sociology can help.

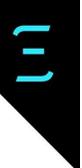

# The Sociology of Race and Digital Society

([Original paper](#) by Tressie McMillan Cottom)

(Research summary by Dr. Iga Kozlowska)

**Overview**: Tressie McMillan Cottom brings together the concepts of platform capitalism and racial capitalism to study how modern-day economic changes wrought by digital technology are reshaping ethnicity, race, and racism. She explores how ideas of race and racial relationships and inequalities are produced and reproduced as more and more of our social lives are mediated online. She argues that by privatizing these interactions the Internet obscures much of these racialized relationships between producers and consumers and that the most vulnerable in society are brought into the fold but usually on exploitative terms.

**Introduction**

Is the Internet racist? That's certainly not how Tressie McMillan Cottom would formulate her research question, but in short, the intersections of race/racism and digital society are her key research areas. In this paper, McMillan Cottom argues that the sociology of race has largely ignored the digital, and where the Internet is studied, it is often without a coherent theoretical underpinning of race, ethnicity, and racism. She proposes exploring this space through platform capitalism and racial capitalism and where the two meet. More specifically, she sees racial capitalism through two emergent phenomena: obfuscation as privatization and exclusion by inclusion. Let's explore these concepts first and then apply them to the design of AI.

Platform capitalism tends to obfuscate the relationships between producers and consumers behind the digital screen. It hides the large amounts of data that it collects and locks them within walled gardens, making it difficult for consumers, the public, and researchers to access. By privatizing more and more social interactions through digital technologies, opaque commercial interests increasingly structure our relationships. Trade secrets and security are often reasons given for a lack of transparency.

Platform capitalism excludes through "predatory inclusion" which is the "logic, organization, and technique of including marginalized consumer-citizens into ostensibly democratizing mobility schemes on extractive terms." For example, online degrees, in theory, expand access to higher education but they also prey on predominantly lower-income African-American women to take out predatory loans. This results in huge costs to the student, particularly if they default, and big profit for the for-profit educational institution and the private loan lenders. We see

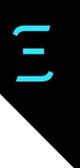

similar exploitation in the "gig economy" (more from McMillan Cottom on The Hustle Economy).

Thus given these recent phenomena, McMillan Cottom argues that "the study of race and racism in the digital society should theorize networked scale, the logics of obfuscation, and the mechanisms of predatory inclusion." Here the theories of racial capitalism – how networked capitalism reshapes global racial hierarchies and desires – come in handy to better understand how our online and offline lives are shaped and reshaped in racialized ways. So how can the concept of racial capitalism help inform the work of those who design and build platform services?

**Designing Racial Capitalism**

As McMillan Cottom describes it, the availability of Internet communications in the last couple of decades has reshaped the economy, producing an informal economy of part-time, gig workers, consultants, freelancers and entrepreneurs who find and get paid for work online rather than through a traditional full-time state/firm employer-employee relationship. This is enabled through platforms that bring together service providers and buyers like TaskRabbit, Upwork, Instacart, Uber, Lyft, and Amazon. This ecosystem of digital employment and services provides those who are unemployed or underemployed or who simply can't make ends meet with a regular full-time job with an opportunity to make extra cash on a one-off basis without benefits and usually in extractive conditions (little control over scheduling, limited recourse to abuse on the job, digital surveillance etc.). This informal economy relies on the most precariously situated workers in the formal economy, often women, people of colour, and immigrants. This racialized capitalist structure, rather than providing economic opportunity, serves to exacerbate racial and economic inequalities and shift the burden and risks of work from employers onto workers furthering the divide between capital and labour.

Knowing this, how can technology designers avoid contributing to these processes? Particularly in the space of AI? While many of the solutions will be on a macro-structural scale requiring public policy interventions, there are some things that the technology itself and those that build it can change. Let's consider some AI design examples at all points of the machine learning development lifecycle.

**Model Training**: When designing facial recognition technologies for ride-sharing apps, for example, the algorithm needs to be assessed on racial impact to ensure it does not bias against people of colour, since misidentification can lead to job loss or lost pay and aggravate racial economic inequality. Preventing such harms may require retraining the model on better data, which may mean collecting a new dataset.

**Data Collection**: When collecting data to improve AI algorithmic accuracy, care must be taken to ensure that the data is racially representative of the problem being solved by the technology. The data collection must match the purpose for which the algorithm trained on that data will be used. The process of data collection must also be culturally sensitive and non-exploitative. This means issues like transparency, meaningful consent, data subject rights, and appropriate remuneration given the cultural and economic context must be considered. While the inclusion of people of colour into training datasets is important so that models can be trained to avoid racial bias, this inclusion must not be predatory, for example taking someone's image without their consent.

**Model Deployment**: Finally, any time algorithms that are to be used for performance evaluation or hiring/firing decisions, at a minimum, must not be racially biased. Because of the sensitivity and impactful consequences of this algorithmically-based decision-making, a human in the loop approach must be considered to avoid automated actions without human review. Additionally, workplace conditions should not be deteriorated through the use of technology (e.g. surveillance mechanisms) that diminishes workers' freedoms, privacy, and dignity. For example, driver monitoring systems or warehouse worker tracking systems should consider issues around notice and consent, minimization of data collection, time and place of personal data storage, right to object to automated processing, and the right to contest automated decision-making etc. Technology designers and builders should speak up when there is no way to design a system that is not racially and/or economically exploitative given the socioeconomic context in which that technology will be deployed.

**Between the lines**

Just as sociologists of digital society must consider race and racism so race scholars must no longer relegate the Internet to the theoretical periphery. The same goes for practitioners. AI/ML researchers, data scientists and engineers, and UX designers can no longer put questions of race/racism and economic inequality to the side. Questions of racial and economic inequality in the age of digital transformation and platform capitalism cannot be peripheral as these very social institutions are shaped and reshaped by the technology we build. The story doesn't end at "build it and they will come." Tech builders must ask the inevitable next question: "and then what?"

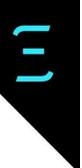

# Office Hours

**We interviewed 3 experts who teach Tech Ethics. Here's what we learned.**

**(Original interviews by Marianna Ganapini)**

Last time we asked you about best practices in teaching Tech Ethics. Now we bring to you the ideas, experiences and suggestions of 3 thought leaders with a long track record in developing Tech Ethics curricula:

*Karina Alexanyan (15 years of experience at the intersection of social science/tech/media/education)*

*Philip Walsh (Teaches philosophy at Fordham University)*

*Daniel Castaño (Professor of Law & Founding Director at the Center for Digital Ethics at Universidad Externado de Colombia)*

They will tell us about their teaching philosophies, their course content, and why they think teaching Tech Ethics is so important!

**What is your background? What courses do (or did) you teach connected to Tech Ethics and who's your audience?**

**Karina**: I have over 15 years of experience researching and working at the intersection of technology and society. My academic research explores how our values manifest in the tools we create, and how those tools shape us in turn. My professional work helps academic and social impact organizations apply these insights for social benefit. My background includes PhD research in global media and society, work with Harvard's Berkman Klein Center on issues pertaining to New Media and Democracy, work with Stanford University on Industry & Academic research partnerships addressing key issues in human sciences and technology, and work with the Partnership on AI on responsible AI development. I have taught courses in Communication and New Media at NYU and with Stanford Continuing Studies, and advise two education start-ups.

**Philip**: I have a PhD in philosophy. My published work focuses on phenomenology and philosophy of mind. I teach two classes related to Tech Ethics: Philosophy of Technology, and AI,

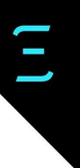

Sci-Fi, and Human Values. These are both undergraduate philosophy elective courses at Fordham University in New York City.

**Daniel**: I'm a Colombian lawyer. I got my LL.B at Universidad Externado de Colombia, an LL.M. and JSD at the University of California – Berkeley. My scholarship focuses on the regulation of complex issues like technology, environmental protection, and public health under uncertainty and in light of different legal, political, and institutional arrangements. To that end, it maps the architecture of decision-making to postulate a methodology to transform the enlightened principles, rule of law values, ethics, and morality into legal norms, private compliance protocols and tech products.

I've been a law professor at Universidad Externado de Colombia Law School since August, 2010. I've focused mostly on teaching courses about the legal and ethical challenges of radical technologies like AI, Blockchain, IoT, and AR/VR. It's been quite a journey because many of these topics and discussions may appear foreign to many people in Colombia. I also started the Center for Digital Ethics at Universidad Externado where I have been advising the Provost and leading an initiative that includes the creation of a new tech & data science department, undergraduate, and graduate degrees. We will launch the new data science undergraduate degree next fall.

**Why do you think it is important to teach Tech Ethics courses? And to whom?**

**Karina**: My current professional focus is on the nascent realm of ethical, responsible, and respectful technology. I believe that a diverse workforce is an essential aspect of ensuring that technical innovations are aligned with public interest. The people building our tools should mirror those that will be impacted by these tools. All of society will benefit if a wide range of experiences, perspectives, and expertise is brought to technology development. That is why I believe it is important to educate young people about the social and technical implications of AI – so that the widest range of people sees in themselves the possibility of contributing to building the tools that will shape our future. To help advance this vision, I am working with The AI Education Project and other industry and education partners on bringing their content to high schools and community colleges in California.

**Philip**: I began answering this question by saying that it is especially important for business and computer science students to study Tech Ethics, but actually I think that's too narrow. I think everyone should study it. I know that's a bland answer, but I believe it. Of course, business and computer science students are the likely future decision-makers at tech companies, but Tech Ethics is bigger than that. Other students from other majors should also study Tech Ethics because they should understand that there is an important role for non-business and non-technical people in shaping the future of technology. For example, a recent student of mine was a biology major. Taking my philosophy of technology course opened her eyes to the intersection of Tech Ethics and the healthcare industry. Another student was an English major and is now pursuing journalism focused on the tech industry. The list goes on. Everyone should study Tech Ethics because it ends up intersecting with everything.

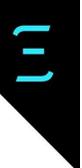

**Daniel**: I think that we are living in what Luciano Floridi calls the "infosphere", that analog-digital world where our lives unfold. I believe it is critical for everyone to understand the promises and perils of living in the "infosphere", regardless of their career or background. It is also very important to teach how to address those promises and perils from a cross-functional perspective

**What kind of content do you teach? What topics do you cover? What kinds of readings do you usually assign?**

**Karina**: I am collaborating with the AI Education Project which is an education startup that has developed an engaging and accessible curriculum to help all students thrive in the age of artificial intelligence. We believe that AI literacy should be a basic component of every student's education. Our online lessons and evidence-based activities were developed with students, teachers, and parents in mind, designed to help students understand the social, political and economic implications of 'the new electricity.' Our multidisciplinary curriculum addresses the skills and knowledge necessary to thrive in and positively contribute to a society where AI and automation are increasingly a part of every industry. The five key modules touch on AI Foundations; Data & Models, Societal Impacts, AI in Action and AI in Your Life.

**Philip**: How much time do you have? For the comprehensive answer, here is the link to my course syllabi: https://philipjwalsh.com/teaching

For Philosophy of Technology I start with some classic work in the field focusing on whether we can think of technology as having a single, unifying "essence" (Heidegger and Borgmann). I then move through some contemporary work on the possibility of artificial general intelligence, human enhancement technology, algorithms and big data, and then privacy and surveillance. For AI, Sci-Fi, and Human Value I used Brian Cantwell Smith's recent The Promise of Artificial Intelligence as our primary text, paired with classic and contemporary science fiction to frame our discussion of the issues. This was a seriously fun class to teach. We read/watched: Frankenstein, R.U.R. (Rossum's Universal Robots), Westworld, Black Mirror, Prometheus, Ex Machina, Her, and of course: 2001: A Space Odyssey.

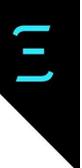

These classes aren't framed as "Tech Ethics" per se. Rather, you might say we cover a lot of meta-ethics of technology. We delve into the epistemic, metaphysical, and existential issues brought on by technology. The human condition is now a technological condition, and that's what I want my students to understand. For example, in addition to ethical value I also emphasize cognitive and epistemic value. I think most debates about AI neglect this topic. One "reading" that I use for both classes and have found to be very successful with students is an episode of Barry Lam's podcast Hi Phi Nation called the "Pre-Crime Unit," which is about algorithmic policing, along with a corresponding paper by Renee Bolinger.

**Daniel**: I teach product-oriented workshops, undergraduate, and graduate courses aimed at discussing the regulatory and ethical questions raised by AI/ML algorithms, Blockchain, IoT, and AR/VR. I usually cover the history and theoretical foundations of tech ethics since the works of Norbert Weiner until the modern approaches that we can find nowadays. I also cover questions about the nature and scope of tech ethics, the difference between ethics and law, ethical principles, ethics by design, and "enforcement" methods.

**What are some teaching techniques you have employed that have worked particularly well? For Tech Ethics, what kind of approach to teaching do you recommend?**

**Karina**: The AIedu curriculum is designed to be accessible to young people of all ages – from middle school through undergrad – although I think adults could benefit as well. Most importantly, it's designed to be easily integrated into coursework by teachers who have no experience with AI. It's a relatively self explanatory curriculum that takes students on a guided journey from "what is AI" through examples of how AI will affect future work/careers, and potential ethical concerns. The curriculum combines hands-on activities with videos, self assessments, and small exercises. At the end, there's a final project that challenges students to come up with an AI solution to a problem in a future career.

The curriculum is designed to give learners the tools they need to make their own decisions about tech ethics. Rather than direct students to "right" or "wrong" answers, we teach them how to think critically about difficult subjects, and how to relate questions about tech ethics to their everyday lives. Learners engage with topics like AI in mental health, data and privacy, and their own social media usage. The course encourages robust debates between students so they feel they have a rightful place in the larger conversation. If we want a diverse set of perspectives in tech ethics, we have to give students from all backgrounds the confidence to amplify their voices in a field that can initially seem intimidating.

The content is also intentionally diverse – the explanatory videos feature women and narrators of color, and the content includes examples of AI in unexpected places – like creative industries.

The content is also energetic, delivered with a playful and friendly tone that makes the technical material accessible to students from all backgrounds.

**Philip**: My strength as an instructor is lecturing. I like lecturing and (dare I say) I'm pretty good at it. I'm naturally extroverted and get very excited about things I'm interested in, so the students seem to respond well to that and we always have very engaging discussions in class. One general piece of advice about teaching that I have is to not underestimate your students. Assign difficult material. Even if you think it will be over their heads, this gives you a challenge: break it down for them. Explain how the puzzle pieces fit together. It will force you to get really clear on the material for yourself and lead to very engaged discussion in class. If the students see that you are working through the material for yourself, it feels like a collaborative enterprise.

Relatedly, I've always had more success assigning material that actually makes a claim. A lot of stuff on AI ethics and Tech Ethics doesn't claim anything. It just lays out a "landscape" of issues or summarizes a bunch of ethical principles that are commonly found in Tech Ethics "frameworks." That's all well and good, but honestly gets pretty boring.

Finally, I recommend letting students develop their own multi-stage research projects. This has been one of the most rewarding aspects of teaching these courses. I basically end up managing 20 research projects every semester, on all kinds of issues in technology. I learn so much from my students. Once again, this gives the class a very collaborative feel and the students respond very positively to that.

**Daniel**: I lecture on the basic theoretical framework and then assign case studies or product review workshops where students analyze the legal and ethical challenges raised by a tech product. For me, product-review and case studies have proven to be an effective teaching method to promote a cross-functional dialogue and bring students as close as possible to real-world scenarios.

**What challenges have you encountered, what are the things that have not worked for you and why?**

**Karina**: I reached out to the staff of AIEDU to see what kind of feedback they've gotten from the many instructors who have taught their course in high schools across the US. Here's what they said:

"The biggest challenge with teaching tech ethics, especially as it relates to AI, is that many students think learning about anything technology-related is only meant for computer scientists, mathematicians, and so on. We know that AI will touch everyone's lives regardless of their interests or career goals but we have to find a way to convey that to students earlier in their school careers, otherwise they self-select out of courses like ours. As educators, we should constantly interrogate our own lesson planning, teaching strategies, and messaging about tech and tech ethics if we want to attract a broad, diverse student audience to the subject. We all have to do a better job of integrating technology, humanities, and the arts so there is an entry point for every student. Another challenge that we've come across is finding ways for students to continue talking about tech ethics outside of the course. We know from AIEDU's student responses that learners who go through the material report high levels of engagement and interest in learning more about topics like AI ethics, but they often don't know where to turn. We tried implementing a project where students explained some of what they learned to family members or friends and hoped it would help facilitate an ongoing conversation about AI. Unfortunately, the students found it difficult to find loved ones that they could engage with on these topics. Now AIEDU is building a detailed resource of free programming that students can complete after the course if they are interested. We know we can spark students' interest in AI ethics but we also have to take responsibility for fanning that spark by finding creative ways for students to apply their learning outside of the classroom."

**Philip**: Sorry but I can't think of much to say here. I've been teaching these courses for a couple years and they have been uniformly great. As I mentioned above, I think it is best to assign difficult material that makes interesting claims. I've assigned overviews of ethical frameworks before and they just aren't that interesting. That's not to say they aren't valuable, but I find they are better suited as a supplement that students can consult and incorporate into their independent research projects.

**Daniel**: It is very hard to assign readings since most of literature on digital or tech ethics is in English. Maybe it is time to publish a comprehensive textbook on digital ethics in Spanish? I'm convinced that ethics and technology need to speak more Spanish. If anyone is interested in making this happen, please feel free to reach out!

**Full bios of interviewees:**

**Karina Alexanyan** has over 15 years of experience directing research & program initiatives at the intersection of social science, information technology, media, and education. She has worked with Stanford, Harvard, and Columbia University, as well as organizations such as Partnership on AI, All Tech is Human, and the m2B Alliance (me2BA.org), and currently advises two educational start-ups.

**Philip Walsh** received his Ph.D. in philosophy from UC Irvine in 2014. He currently teaches at Fordham University. His research and teaching focus on phenomenology, philosophy of mind, philosophy of technology, and Chinese philosophy. He blogs about philosophy of technology at thinkingtech.co.

**Daniel Castaño** is a Professor of Law & Founding Director at the Center for Digital Ethics at Universidad Externado de Colombia. LL.B – Universidad Externado, LLM & JSD – University of California at Berkeley. Consultant in new technologies, regulation, and digital ethics.

**The Future of Teaching Tech Ethics**

**(Original interview by Marianna Ganapini)**

How do you see the Tech Ethics Curriculum landscape evolve in the next 5 years? In this column, 3 experts tackle this and other important questions about the development of an effective, inclusive and comprehensive Tech Ethics Curriculum for the future.

They tell us about their teaching experience and about the gaps in this field. First, meet the experts:

*Merve Hickok — Founder of AIethicist.org, a global repository of reference & research material for anyone interested in the current discussions on AI ethics and impact of AI on individuals and society.*

*Dorothea Baur — Principal and Owner at Baur Consulting AG, which advises companies, non-profit organizations and foundations on matters related to ethics, responsibility and sustainability with a distinctive but not exclusive focus on finance and technology.*

*Ryan Carrier — Executive Director at ForHumanity, a non-profit founded to examine the downside specific and existential risks associated with AI.*

**Here are some highlights of the interviews below:**

- While teaching Tech Ethics, we need to acknowledge that a discussion around 'power' is key to explain and understand how AI and disruptive Tech is changing our social-economical landscape.

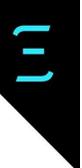

- For the next 5 years, there will be more empowerment and integration of Ethics committees and Ethics Officers, and an increased value in philosophical/moral thinking with regards to sustainable profitability in corporations.
- Our Tech Ethics Curriculum needs to address these changes. In the near future, there will be more universities and colleges developing multidisciplinary programs geared towards Tech Ethics.
- Tech Ethics should become a standard part of each degree in Tech.

**What is your background? What courses do (or did) you teach connected to Tech Ethics and who's your audience (e.g. undergrads, professionals)?**

**Merve**: I have BAs in International Relations and Political Science, and am a certified privacy professional. I provide both consulting and tailored training to organizations and individuals. I am a lecturer at the University of Michigan, School of Information for Data Ethics course under the Master's of Applied Data Ethics program. I also have an online self-paced course that I created (provided under RMDS Lab) for professionals with any background, who are interested in the fundamentals of AI ethics, bias and ethical decision-making.

**Dorothea**: My background is a Ph.D. in business ethics, several years of postdoctoral research, and 5 years of independent ethics consulting with a focus on tech and finance. I am teaching classes on "AI and ethics" at various universities of applied sciences. My class is part of degrees on "AI Management", "Digital Ethics", "Disruptive Technologies", "Digital Finance" etc. The audience always consists of heterogeneous groups of professionals across different industries. They all have previous degrees and several years of work experience.

**Ryan**: I founded ForHumanity after a 25-year career in finance, and I now focus on Independent Audit of AI Systems as one means to mitigate the risk associated with artificial intelligence. As for my teaching experience, I have had the opportunity to teach general ethics as a part of a course introducing the 'Independent Audit of AI Systems'.

**What kind of content do you teach? What topics do you cover? What kinds of readings do you usually assign?**

**Merve**: I provide details of multiple approaches that are currently highlighted in AI ethics discussions. I cover AI & tech ethics from a fundamental rights approach, principles-based approach and human-centric values-based approach and discuss the pros & cons of each. It is important for decision-makers, developers, implementers and policy-makers to understand what each of these means, what it means for business/agency/society and the harms and burdens that can manifest themselves.

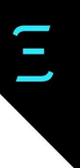

I also provide context on bias, where bias can creep into the system during its lifecycle and what are some of the practices to mitigate. A lot of my focus is around demonstrating the real-world applications (recruitment, predictive policing, healthcare, workplace) of these high-level approaches – in other words bringing theory into action. What does it look like in practice, in what situations they might work, how to ask the right questions and decide. Every organization and every person involved are at different maturity levels with regards to their understanding of impact and consequences. They are also coming from very different industries and backgrounds. So it is important to provide the fundamentals and tools to be able to dissect the AI systems and rhetoric yourself. As for topics, I cover the following: AI bias (especially in recognition and prediction systems), AI governance, policy, regulation, ethical frameworks, power, harms/burdens, inclusive design, exploitation/extraction, surveillance, manipulation, disinformation, social justice, techno-solutionism, diversity, data colonialism.

**Dorothea**: I mostly only teach one block of 4 or 8 hours in each programme. I usually start with a very brief introduction to ethics because the students usually do not have any prior knowledge about ethics. I then move on to show where you can 'anchor' ethics – i.e. at a state level (through legislation), at industry or corporate levels (through corporate AI ethics initiatives), at the level of individuals (through awareness training etc.). I spend the bulk of my time on highlighting ethical challenges based on use cases – e.g. algorithmic credit lending, facial recognition, emotion recognition, algorithms in education or in court etc. In these contexts, we discuss issues like bias, discrimination, privacy, accountability etc. In this type of continuing education, you don't usually assign readings, but I always recommend some literature and websites to people in case they want to know more.

**Ryan**: At ForHumanity we are deeply committed to establishing best practices in curriculum design for AI Ethics. At the moment we are working on building clear learning objectives for new Algorithm Ethics classes. As for topics, our main focus is better developing the idea of 'Ethical Choice' and building a curriculum that could empower the Ethics Committee by using our system of accountability, governance and oversight. This is because we see the Ethics Committee as a key tool for managing AI and Autonomous System risk. For more on this, please see "the rise of the Ethics Committee".

**What are some teaching techniques you have employed that have worked particularly well? For Tech Ethics, what kind of approach to teaching do you recommend?**

**Merve**: I use lots of case studies that range from corporate applications to those used by public entities to the ones we use in our private daily lives. I am trying to raise awareness of the fact that AI systems are ubiquitous in all facets of our lives. I want the participants in my courses to know that the focus should not only be on the systems they are asked to develop at work, but it goes beyond that: the impact of discriminatory systems or scoring applications is on both an individual and societal level.  I assign the readings curated in AIethicist.org. I also show impactful videos, movies and documentaries. If I am conducting training for a specific company/organization, I tailor it to the needs of the team and help them incorporate it into their processes and work. For the Master's level program or my online self-paced training, the students choose to take the class to advance their understanding of the issues and their impact. Regardless of the situation though, it is often hard to work with the trainees/students to help them really question these systems and deep-seated ideas.

**Dorothea**: The class size is usually quite small – not more than twenty students. Given that these are all professional adults, they are usually intrinsically motivated and keen to engage in discussion without me using a lot of specific teaching techniques. However, I also usually let them do group work, e.g. for case studies, where I assign different cases to the groups and let them explain what they have learned to their peers in plenary. In my context, the best recommendation is to keep tech ethics as applied as possible and to inspire people to reflect on what it means for their own work experience.

**Ryan**: In teaching, I have used a number of different techniques, such as crowdsourcing, lecturing, in both synchronous and asynchronous settings.

**In your opinion, what are some of the things missing in the way Tech Ethics is currently taught? For instance, are there topics that are not covered enough (or at all)? What could be done to improve this field?**

**Merve**: These are some of the things that are often not highlighted enough in this space:
- Focus on critical thinking skills and the ability to break apart concepts and resist false dichotomies that benefit certain groups more than others. Critical thinking is also key to the ability to separate prediction vs causality, and to the ability to separate pseudoscience from real science & research
- A discussion about power is crucial to understand how AI and disruptive tech is changing the landscape. This also entails understanding the history of for example civil rights, segregation, colonialism, public vs private entities, international organizations etc…
- A serious reflection on the negative impacts of AI across all domains of our lives (not just professional), and an understanding of corporate power and emergence of legal persons and corporate liability

- The willingness to develop and share the tools to resist unethical practices (importance of collective action here)
- A need for governance and accountability mechanisms in general

**Dorothea**: I don't have a good overview of the state of Tech Ethics across universities because I am focused on this very specific continuing education setting. I cannot say anything about Tech Ethics at BA or MA level, or engineering degrees, etc.

**Ryan**: What is missing is a real focus on an Independent Audit and the subsequent infrastructure of trust, specifically in the Governance, Oversight and Accountability of 3rd party-audits. I see dramatic demand for these skills due in part to their requirement in the process of Independent Audit of AI Systems. That means that we need courses able to train people who can fill that gap.

**How do you see the Tech Ethics Curriculum landscape evolve in the next 5 years? What are the changes you see happening?**

**Merve**: In the near future, I see more universities developing multidisciplinary programs and curricula geared towards Tech Ethics. At the moment, interdisciplinary education and professional work are still woefully lacking. And there is indeed a need for a shared language and respect and understanding of each other from both humanities and CS/Engineering sides. In the Master's course I'm involved in, 90%+ of the students are working in different fields and have different backgrounds. As a result, the conversations among students are very rich even though their understanding of Tech ethics tends to be very different. I think we need more of that interdisciplinary work and education.

**Dorothea**: I hope that Tech Ethics becomes a standard part of each degree in Tech – be it at undergraduate, graduate, or continuing education level. Everyone studying tech should be exposed to ethical questions in the classroom.

**Ryan**: For the next 5 years, I see more empowerment and integration of Ethics committees and ethics officers, and I see increased value in philosophical/moral thinking with regards to sustainable profitability in corporations. I would argue that with the rise of soft law and duty of care legal terminology in laws like GDPR, Children's Code and now the EU High-Risk regs proposals, the demand for skilled practitioners (ethics officers) trained in instances of Ethical Choice all throughout the design and development of algorithmic systems, will rise at the corporate level. The idea is that private entities will see these changes as a risk to the sustainability of their profits unless they learn how to properly risk-manage these issues. My prediction is that this will also transform the landscape of the Tech Ethics curriculum design and new courses will be taught to fill these gaps and address these needs.

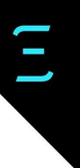

**Full bios of interviewees:**

Merve Hickok is the founder of www.AIethicist.org. She is an independent consultant, lecturer and speaker on AI ethics and bias and its implications. She aims to create awareness, build capacity, and advocate for ethical and responsible AI. She collaborates with several national and international organizations building AI governance methods, and has been recognized by a number of organizations for her work – most recently as one of the 100 Brilliant Women in AI Ethics™ – 2021. Merve has over 15 years of global level senior experience in Fortune 100 with a particular focus on HR technologies, recruitment and diversity. She is a Senior Researcher at the Center for AI & Digital Policy, a Data Ethics lecturer at the University of Michigan, School of Information, an instructor at RMDS Lab providing training on AI & Ethics and responsible AI development and implementation. Merve is a Fellow at ForHumanity Center working to draft a framework for independent audit of AI systems; a founding editorial board member of Springer Nature AI & Ethics journal; and a regional lead and mentor at Women in AI Ethics Collective, she works to empower women in the field.

Dorothea Baur has over 15 years of experience in applied ethics, starting with research, moving on to teaching and consulting. She began with a PhD on business ethics, and more recently focused her consulting and teaching on AI ethics. She has taught at leading European business schools like ESADE University Barcelona, University of St. Gallen, and Nottingham University Business School. Aside from running her own consulting company, she is currently very active as a lecturer on AI ethics in continuing education at various universities in Switzerland.

Ryan Carrier founded ForHumanity after a 25 year career in finance. His global business experience, risk management expertise and unique perspective on how to manage the risk led him to launch the non-profit entity, ForHumanity, personally. Ryan focused on Independent Audit of AI Systems as one means to mitigate the risk associated with artificial intelligence and began to build the business model associated a first-of-its-kind process for auditing corporate AIs, using a globally, open-source, crowd-sourced process to determine "best-practices". Ryan serves as ForHumanity's Executive Director and Chairman of the Board of Directors, in these roles he is responsible for the day-to-day function of ForHumanity and the overall process of Independent Audit. Prior to founding ForHumanity, Ryan owned and operated Nautical Capital, a quantitative hedge fund which employed artificial intelligence algorithms. He also was responsible for Macquarie's Investor Products business in the late 2000's. He worked at Standard & Poor's in the Index business and for the International Finance Corporation's Emerging Markets Database. Ryan has conducted business in over 55 countries and was a frequent speaker at industry conferences around the world. He is a graduate from the University of Michigan.

## What's missing in the way Tech Ethics is taught currently?

**(Original interview by Marianna Ganapini)**

*What's missing in the way Tech Ethics is taught currently? Two experts in this field, Heather von Stackelberg and Mathew Mytka, shared their ideas and experiences on these and other vital issues. See their full bios at the end of the article.*

**What is your background? What courses do (or did) you teach connected to Tech Ethics, and who's your audience (e.g., undergrads, professionals)?**

**Heather**: My Masters's is a cross-disciplinary social sciences degree with a focus on Adult Education. I also have undergraduate degrees in communication studies and biological sciences. I usually teach with colleagues who have CS/engineering graduate degrees. I don't, at the moment, teach a course that is solely on Tech Ethics, but I teach classes on AI literacy and AI management for business leaders, which have vital ethics components. I teach mainly to business professionals, though it's not unusual to have a substantial minority of post-secondary students in any given class.

**Mat**: I've got a relatively diverse 27-year career—everything from construction and journalism to product management and UX design. I studied social sciences and spent several years working at the intersection of information sharing, trust, and ethics. I've designed and delivered entrepreneurship programs to co-design modules for postgrads and data ethics to professionals and practitioners. For the past three years, I've been leading programs with Greater Than X, helping cross-functional teams to operationalize data ethics. We have also worked on national data sharing ecosystems like Consumer Data Right in Australia and Open Banking in the UK, and universities like Northwestern Kellogg School of Management's Trust Project. Late last year, we launched a social learning platform to help people design more trustworthy organizations and technology. One of the recently published courses on the platform is a primer to operationalizing data ethics. It builds on the work we'd been doing in our services business with global clients and years of working in these areas. The framework we cover in that course is more like an operating system to interface with modern product development workflows, factoring in proactive, reactive, and retroactive decision making and diverse stakeholder engagement.

**What kind of content do you teach? What topics do you cover? What types of readings do you usually assign?**

**Heather**: What I teach is very focused on application and real-world examples. It's usually not so much the technical aspects of AI but how the technical relates to the business and larger social and ethical impacts. So we talk a lot about how to choose a project that provides business value and is technically feasible. On the ethics side, we talk about ethical questions across the entire lifecycle – from selecting an appropriate project (that is, one that isn't inherently meant to harm people) to data collection, data labeling, choosing proxies and metrics, operationalizing concepts like fairness, to working with users, training them, evaluating, and watching for unexpected consequences.

Assigned readings aren't relevant for industry training, but I have a list of resources I recommend if people ask. We also have been working to develop reference sheets, summaries, and checklists that are more likely to be used by people in the industry than textbooks or academic papers.

**Mat**: The community at Greater Than Learning. as a broad thematic area, we've focused on helping people design more trustworthy products and services. We cover topics such as operationalizing data ethics, behavior design in organizational change, and how the collection and use of people's data influence trust. There are also areas on how to design privacy notices or terms of use and establish the workflows to do this in a modern business. Content is a small part of it. So you might learn about these topics via watching a video or exploring a scenario. Then reflective activities to connect it to your lived experience. There is a range of learning experiences, but most of the focus is on social learning amongst the small but diverse community of practice. While readings are not prescribed, we provide reference reading in courses, from journal articles to books. These might be articles from someone like Luciano Floridi to books like Social Physics from Alex Pentland. But because it's a social learning platform, content comes via community members sharing resources with each other.

**What are some teaching techniques you have employed that have worked particularly well? For Tech Ethics, what kind of approach to teaching do you recommend?**

**Heather**: We've had people work through a full proposal for an ML project, which requires them to make decisions about it and state the reasons for their choices. That seems to work well for getting people to think about the realities and application of the ideas and principles. Of course, this only works in more extensive, long-term courses. In sessions that are only a few hours, this isn't practical. The other method that seems to work with Ethics is the discussion of case studies. Provide an example of an organization with a notable ethics failure, and discuss why they did it and both what and how they should have done differently. Again, you need to have enough time for that discussion, which is difficult when you're teaching busy industry professionals.

**Mat**: Any extensive curriculum needs to cover a comprehensive set of fields, from philosophy and history to science fiction and co-design. But most of all, it needs to be practical. It needs to simulate the pressures that come with making ethical decisions when designing and building technology. You have to do it to learn it! Assigning applied learning projects that involve interdisciplinary focus is one technique we're exploring with the community at Greater Than Learning. When it comes to tech ethics, we need to focus on experiential knowledge and practical skills, helping with forming ethical "muscle memory." If ethics is about deciding what we should do, learning environments need to mimic the decision-making context. There is little point in learning about consequentialism or deontological approaches if the learner can't relate. And this approach has worked when I've been helping people in organizations where people are dealing with real trade-offs and commercial constraints, and they have a "skin in the game." Indeed, a perception of loss, risk, and consequence are essential motivators in learning even, more so when it comes to learning to navigate the grey area of tech ethics.

**In your opinion, what are some of the things missing in the way Tech Ethics is currently taught? For instance, are there topics that are not covered enough (or at all)? What could be done to improve this field?**

**Heather**: The thing that often frustrates me is how often deep research and best practices from the social sciences are ignored in STEM fields, especially Data science and ML. For example, all of the research and best practices on getting accurate and unbiased data annotation (which in ML is data labeling) or on validating a proxy. These are very relevant in ethics and preventing harm, but when I talk about them to CS people, they've often never heard about it before.

**Mat**: I don't have a clear view of the ways tech ethics is being taught everywhere. So what I express here is based on my limited reach into all the different formal and informal learning contexts. There is a wide range of topics being covered across this space. Universities and various organizations have courseware popping up. There's loads of content and reports to read, history to draw upon, or frameworks and principles to learn. So it's not for lack of material that there are gaps. What's missing is the practical focus. How on Earth do I raise an ethical concern in the workplace when I don't feel safe to do so? Or how might I work through a moral decision on a new product feature within a cross-functional team? What metrics matter, and how do we measure them? How might we balance our commercial KPIs with our responsibility to the communities in which we impact? How do we bring these communities into the design process upfront? The skills required for where the rubber meets the road are missing in most of what I've seen in the market. It's hard to "teach" tech ethics without the realities of actually doing it.

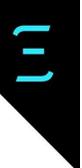

In general, the field is emergent, and there are intersecting movements. Be that responsible innovation and humane tech, data for good and tech ethics or many other permutations. I think we're all trying to work through the ambiguity, sense-making and find pathways to learn better. So from that perspective, there needs to be more coordinated collaboration across these diverse communities of practice. Be that to co-create curricula or to support ongoing and experiential learning. I believe we need an open-source social network for learning in this area that leads by example. It shows people how it's done and involves people in the process by also providing a case study for the ecosystem to learn from. There is a plethora of providers and platforms talking big on tech ethics and using technologies that misalign to the values they espouse. For example, if I come to your website and you're communicating all these things about ethical tech, it sets some expectations. If, at first, I get a cookie consent notice that gives me no real choice…well, that's misalignment! This 'ethical intent to action gap' eventually diminishes trust.

**How do you see the Tech Ethics Curriculum landscape evolve in the next five years? What are the changes you see happening?**

**Heather**: Both in post-secondary and in the industry, the big thing that is missing – and is slowly changing – is the operationalization of AI ethics principles. Globally, we're slowly converging on a generally agreed-upon set of principles. Still, we're only at the beginning stages of defining what they mean in terms of the day-to-day work of data scientists and ML scientists. Over time, I think we're going to see the development of norms, standards, and best practices for the development of ML that integrate and operationalize those ethical principles. Still, it's going to take a while.

**Mat**: I think there will be a plethora of curriculums evolving both via formal and informal methods. I've seen an increase in the number of universities offering courses on these topics. Then certifications are also starting to pop up. Workplaces are increasingly advertising for these new roles. And curriculum design in this space will be more collaborative and shifting to a practical focus: I see the various communities of practice coalescing and co-creating curriculums. It's already been happening, and there is some fantastic thinking and direction starting to take shape. And the demand for these types of courses is there.

The signal-to-noise ratio will be a problem, though in the sense that the commercial opportunity in the space brings with it organizations that are mainly in it for the money. Shiny courseware and big talk attract the audience. This means that upskilling in these areas is potentially being done in an inferior way. People might get certification as an ethical technologist, and they get a promotion or are hired based on this credential. Still, it is unclear that all certificates should have the same value.

It will be messy, and lots of misconceptions will develop. So we'll see a mix: inadequate approaches to curriculum design and more innovative approaches to co-creating and crowdsourcing curriculums. There's so much great work happening in this space, so I believe that what is happening is positive on balance.

**Is there anything else you'd like to add?**

**Mat**: Yes. If anyone reading this wants to collaborate on crowdsourcing curriculums, reach out. We can create better together.

**Full bios of interviewees:**

**Heather von Stackelberg**: Heather is a Machine Learning Educator at Alberta Machine Intelligence Institute (Amii). She has been teaching and developing educational material on the interactions between society and machine intelligence systems. Before this, she taught math, chemistry, and biology at MacEwan University in Edmonton and at two First Nations colleges.

**Mathew Mytka**:  Mat is a humanist, generalist, father, gardener, wannabe extreme sports enthusiast, and Co-founder of Greater Than Learning. He's worked across digital identity, e-learning systems, behavioral design, decentralized technologies, and spent the past few years helping organizations to design more trustworthy products and services.

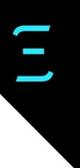

# Anthropology of AI Ethics

## The AI Junkyard: Thinking Through the Lifecycle of AI Systems

**(Original article by Alexandrine Royer)**

In mid-March, Netflix revealed the details behind our streaming consumption patterns that tally into the company's carbon footprint. DIMPACT, a tool that calculates digital companies' emissions, determined that one hour of streaming was equivalent to one ceiling fan running for four hours in North America or six hours in Europe. While such statistics do not seem shocking nor particularly alarming, they omit what we might be doing instead of spending an additional hour on our computer.

Streaming services also do not operate in the same way in a given environment as, say, a ceiling fan or a washing machine. For a more precise understanding of the environmental and social impacts of streaming, calculations ought to include the energy costs of charging our laptops to keep streaming, securing access to high-speed Internet, upgrading devices and discarding old hardware, and so on.

The widespread usage of streaming platforms shows how a single technological change produces ripple effects that modify our habits, our energy needs and our carbon footprint, even if they appear minute. As many anthropologists have argued, AI systems should be viewed as socio-technical systems instead of single bounded entities. The term further invites a holistic approach in understanding the social and environmental impact of having these algorithms run in our lives and what happens once they enter disuse.

Our inability to see and comprehend the lines of code behind the design of our favourite apps and platforms has helped foster the view of AI systems as operating in a virtual realm of their own. The invisibility of algorithmic systems contributes to the lack of transparency regarding the biases integrated within and integral to these systems. Kept away from peering eyes are the massive data centers required to keep these systems running and their polluting effects, leading us to quickly forget that tech is a double-edge sword capable of preventing and generating damage to the environment.

The implications of our long-term reliance on tech are often tinged with techno-dystopian discourses of artificial intelligence taking over and dispelling the end of the human race. Such alarmist views encourage a distorted view of AI systems' current capacities and what is probable and possible. Instead, I argue for a more critical inquiry of the social and environmental effects of AI systems that follows each step of the systems' life cycle and how they interact with previously existing structures along the way.

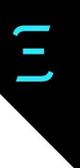

As highlighted by Eric Broda, ML systems' lifecycle from tech companies such as Microsoft and Google is often presented through the same stages of data understanding/project objectives, acquiring and engineering data, model development and training and model deployment and monitoring. For Broda, the requirements of model reproducibility, traceability and verifiability tend to be omitted or underplayed in these AI/ML lifecycles. To this criteria, I would add the sustainability of AI systems, the long-term consequences of keeping these systems running and their likely expiration date within the tech space.

To better understand the shortcomings of AI systems, Broda suggests introducing a life cycle catalogue, a type of "book of record", that "provides a viewport into the data science lifecycle" by allowing "data scientists to visualize, categorize, group, manage, and govern all of an enterprise's data assets". The lifecycle catalogue can be a valuable tool to estimate an AI system's impacts from within its code to its outside connections to other existing systems and deployment into our lives. It can serve to provide visible ethical 'checkpoints' for legislators and citizens alike to understand the implications of each stage of the AI/ML process.

The catalogue also pushes us to reflect on what happens to the end of life of a system, knowing that systems cannot exist forever and that they must dispose of the masses of accumulated data. Scholars such as Edina Harbinja have already pushed for legislating and regulating post-mortem privacy to protect the trails of data that individuals have left behind through a lifetime of online activities. But beyond the individual level, little is known as to how companies dispose of their data (and perhaps for good security reasons). Just as importantly, few reports have addressed the consequences of the dismantling systems that people have come to rely on.

With the lifecycle catalogue mindset, we can return to our initial example of Netflix. Netflix streaming services will have accumulated precise personal information over viewers' series and movie selection preferences throughout their deployment stage. If Netflix is suddenly unable to compete with other streaming service providers and must be discontinued, it will have to dispose of the masses of data on their users. Even if the data is anonymized, individuals can be easily traced back through cross-checking with user profiles on other public sites, such as IMDB.

Alongside these ethical privacy concerns are the environmental costs of keeping these systems running. In their continuous learning and monitoring phase, Netflix's algorithms will be tweaked to incentivize viewers to keep watching for hours on end, increasing both the company's computational power demands and the individual's energy consumption. Individuals looking to improve their streaming experiences will be encouraged to acquire high-speed internet and devices with better image quality. Tracking the sustainability of AI systems throughout their lifecycle will require monitoring all these factors.

Thinking holistically about AI systems also involves thinking about their endpoint. AI as socio-technical systems can create a series of connections but also ruptures and breaks. How much do we know about AI systems that have been discarded? To frame this question otherwise, what might be resting in the AI junkyard? And what can we learn from it?

# Beyond the usual

## Artificial Intelligence and Healthcare: From Sci-Fi to Reality

**(Guest column by Marcel Hedman)**

This past year has been filled with huge disruption due to the onset of COVID-19. However, this has not detracted from major leaps in the world of artificial intelligence. These leaps include algorithms to determine the 3D structure of proteins like DeepMind's AlphaFold 2 algorithm or huge systems that can generate original text or even code! Quickly, we are entering into an age where the concepts once reserved for Sci-fi movies are now seeming slightly more possible. While it is very clear that we are far from realising all-knowing AI systems (AKA General AI), this should not detract from the scale of the technology being developed. The advancements within systems that can perform a single task extremely well have been great to see.

Similarly, AI has been hugely disruptive in the realm of healthcare. Diagnostics, remote patient monitoring and hospital logistics are just three areas where AI has made an impact by leveraging the wealth of available data. Healthcare has proved to be a prime recipient for AI intervention due to the abundance of data that is collected, an almost a 1500% increase over the past seven years. Therefore, as developments such as the Apple watch makes healthcare data even more abundant, we are beginning to see even more opportunity for health AI growth. To distil these two complex fields, we will explore two questions:
- What is happening?
- What are the difficult questions that need to be tackled?

**What is happening?**

Below are three prominent spheres where AI is causing disruption within healthcare. AI has the potential to do the same tasks as humans but efficiently, faster and at a lower cost. However, the real value is not found in replacing doctors and medical personnel but in using these systems to augment their capabilities and free up resources to focus on only the most critical tasks.

**Diagnosis**

The first area that readily comes to mind at this health-AI intersection is its capabilities in assisting with the diagnosis of health conditions. Machine learning, the branch of AI which is responsible for lots of recent developments, finds patterns in data extremely well. This pattern recognition aptitude then directly translates into an ability to synthesise numeric and even graphical health data and make predictions.

According to PWC research, IBM's Watson for Health and Google DeepMind health are major drivers in this area. For example, in cancer diagnosis, studies have found AI systems that achieved a breast cancer detection accuracy comparable to an average breast radiologist. Furthermore, machine learning and systems neuroscience are also being combined to build algorithms into neural networks that mimic the human brain.

Excitingly, plenty of startups are finding success in applying these novel techniques towards diagnosis. One such is named, Lucida Medical, which utilises machine learning on MRI scans to accurately support cancer screenings.

**Personalised Medicine and Healthcare**

Next, we turn our attention to personalised healthcare. The ability to provide tailored healthcare at the individual level is something largely unseen. At most, populations are broken into subgroups based on gender and ethnicity but as is the case with all groups, the mean is not always representative – there are always outliers.

With the onset of cheaper genome sequencing alongside wearable IoT devices (e.g. smartwatches), personalised healthcare can really become a reality. However, there is a huge amount of data required to bring this about and this is where AI comes to play. For example, clustering algorithms can create drug recommendations based on similar patients who responded well to the treatment plan. However, these recommendations could be made based on the unique genomic profile of individuals as opposed to macro identifiers such as ethnicity.

**Patient monitoring and care**

Finally, we turn to the field of patient monitoring which can be both real-time or post-data collection. The use of wearable IoTs and other remote monitoring tools allows for a constant stream of data that can be input into an AI system to predict conditions before they manifest. This is especially valuable given the number of complications that arise once a patient is discharged from hospital care. By incorporating hardware and cloud computing advancements with these novel algorithms, vital signs can be measured and used to predict health deterioration ahead of time.

iRhythm is a company that does just that. The company uses data from its remote monitoring devices to predict heart conditions before they occur, and its service has already helped over two million patients.

**What are the difficult questions that need to be tackled?**

With all this advancement, it's easy to only see the bright side of this technology. However, companies such as OpenAI have developed systems so advanced that they have been unwilling to share them out of fear of what would be made possible by bad actors. It is essential for certain key questions to be answered before we see mass adoption of this technology.

How to overcome the implementation gap? The implementation gap refers to methodologies and techniques that have been proven effective in controlled environments but are not widely adopted due to a failure to navigate the broader macroeconomic landscape or the nuances of a certain industry. This gap is significant in healthcare as fewer than 50% of clinical innovations make it into general usage. Therefore, the need to overcome this gap is clear. A big driver to overcome this gap in healthcare is the concept of interpretability, how did the system arrive at the conclusion it did? Increasing interpretability and moving away from the so-called "black box" approach is vital for mass adoption.

How do we combat bias within datasets? The field of machine learning is exactly as the title suggests, it is a machine that is learning based on the data that it receives. This means that if the input data is biased, e.g. too few samples of a minority group, it will negatively affect its performance on that group at deployment. While medical data is abundant within healthcare, gaining access to that data is slightly more difficult due to data privacy laws. These laws have meant that creative approaches have been made to ensure that algorithms are trained on non-biased datasets.

One such example is from Microsoft, where they've developed federated Learning with a Centralized Adversary (FELICIA) that allows hospitals to share medical image datasets without violating patient privacy. It achieves this by generating a synthetic dataset utilising FELICIA and the contributions to FELICIA will come from hospitals globally. This approach combats local population bias as data can be shared across many regions and in theory, this will lead to better health diagnostics for all.

The two above questions are just two of many that will be tackled in the quest to mass adoption of AI within the healthcare space. As the level of education on the benefits and importantly the limitations of AI rises, the further deployment of these systems within healthcare seems inevitable. I for one am excited to see the benefits it can have for both individual and public health.

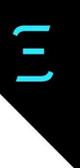

## The Chief AI Ethics Officer: A Champion or a PR Stunt?

**(Original article by Masa Sweidan)**

We have reached a point where the far-reaching impacts of AI's ability to identify, prioritize and predict can be felt in virtually every industry. Over the last couple of years, both researchers and practitioners have established that the power relations embedded in these systems can deepen existing biases, affect access to reliable information and shape free speech. Many organizations have attempted to stay relevant and keep up with the developments in AI Ethics by introducing the role of a Chief AI Ethics Officer (CAIEO), which has other titles including AI Ethicist, Ethical AI Lead and Trust and Safety Policy Advisor, to name a few.

Although the creation of this role seems to be a step in the right direction, many are questioning whether the presence of a CAIEO will truly boost the adoption of AI Ethics within organizations and effectively shift the focus from creating guiding principles to implementing effective practices. Before examining the challenges and responsibilities of the profession, it would be helpful to frame this discussion with some context.

The role of an Ethicist has been around for some time, especially in the field of healthcare. Also known as Clinical Ethicists or Bioethicists, these healthcare professionals are typically employed by large, academic medical centers to help patients, families, and medical teams solve health-related dilemmas. They often deal with questions pertaining to autonomy, beneficence, non-maleficence and justice in an attempt to make the "right" choice or decision.

Within the context of AI, it has proven to be quite difficult to define the responsibilities that fall under this role, because ethical issues around AI are uniquely complex and nuanced, meaning that social, legal and financial consequences must be considered. It is important to highlight the distinction here between the field and the profession, since "the AI Ethicist is only one piece to the puzzle of AI Ethics." As Olivia Gambelin explains, part of the confusion stems from the fact that the position itself is named so closely after the field, leading to the assumption that the individual with the title is the only one with enough power to create change in the area of AI Ethics.

This contributes to what I consider to be the root of most concerns regarding the CAIEO role. We should be extremely wary of having AI Ethics as a separate mandate that a single officer executes on their own, as opposed to the entire organization. If the other employees do not get involved in the process of ensuring that ethical AI standards are met, then all the responsibility falls on the shoulders of one person. Rather than being a siloed effort limited to risk management practices, it should act as a way to consolidate the organization-wide ethics-related activities. However, this is much easier said than done.

Due to the interdisciplinary and philosophical nature of the role, Natasha Crampton explains that "it is impossible to reduce all the complex sociotechnical considerations into an exhaustive set of pre-defined rules." To overcome this challenge, companies like Microsoft and Salesforce are developing processes, tools, training and other resources to ensure that their AI solutions reflect the original principles that were adopted.

The progress being made is certainly exciting, but many still question the incentives behind creating this role and wonder if it is just an add-on feeding into the bottom line. Arguments against the initial adoption of this role often include terms such as "ethics-washing" or "PR stunt" to convey the idea that this is an attempt for companies to avoid regulation. Ben Wagner further elaborates by stating that "the role of 'ethics' devolves to pre-empting and preventing legislation," so financial and political goals can be masked by the language of ethics.

This may seem extreme, but it does touch on an important point that should be mentioned but deserves its own analysis. With many global companies, there is a disconnect between scale and governance. Skeptics may view 'ethics' as the new 'self-regulation' for private companies that are unwilling to provide real regulatory solutions. As the dangers become increasingly clear, it seems that the answer should be better governance, not self-governance.

Moreover, there are several individuals calling for a more general Ethics Officer, who would be responsible for dealing with the ethics behind practices within the organization and could seek training on the ethical aspects of AI if they lack the necessary knowledge. The problem I see with this suggestion is that it removes the specificity of AI-related challenges from the role, which is crucial to detect if organizations plan to overcome them.

In this scenario, it is easy to anchor a CAIEO role with the general field of Business Ethics. This well-researched area can take shape in different forms, ranging from the study of professional practices to the academic discipline, and it tends to touch on various aspects of a firm's relationship with its consumers, employees and society. However, the added context of AI creates new issues that can impact millions of people, meaning that "AI Ethicists are no longer dealing with person-to-person ethical issues, but rather machine to person." Therefore, a specific position needs to be carved out within an organization to examine the increasingly complex implications of this modern technology.

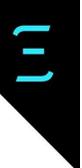

After looking at the ambiguity surrounding this role, there is no doubt that the efficacy of this position will ultimately boil down to the individual that is hired. This job involves analytical thinking, proper communication and relationship management, but most importantly, it requires trust. After all, these champions will be responsible for leading the efforts to integrate their operational processes into the DNA of the organization.

I foresee that the ethical approaches pertaining to AI design and development, such as external participation and transparent decision-making procedures, will continue to improve. However, there is one rule that must remain the same: AI Ethics can not substitute for fundamental human rights. Moving forward, it seems that the organizations employing a key driver at the executive level, such as a CAIEO, to build wider competence and adherence at the firm level will be the leaders in this space.

## The Paradox of AI Ethics in Warfare

**(Original interview by Natalie Klym)**

The interviews in this series explore how today's AI practitioners, entrepreneurs, policy makers, and industry leaders are thinking about the ethical implications of their work, as individuals and as professionals. My goal is to reveal the paradoxes, contradictions, ironies, and uncertainties in the ethics and responsibility debates in the growing field of AI.

I believe that validating the lack of clarity and coherence may, at this stage, be more valuable than prescribing solutions rife with contradictions and blind spots. This initiative instead grants permission to be uncertain if not confused, and provides a forum for open and honest discussion that can help inform tech policy, research agendas, academic curricula, business strategy, and citizen action.

*Interview with Michael Conlin, inaugural Chief Data Officer and Chief Business Analytics Officer (2018-2020), US Department of Defense, May 2021*

Perhaps the most fundamental paradox when discussing AI ethics emerges when exploring AI within a domain that is itself regarded by many as unethical. Warfare is arguably the most extreme case. Such domains represent harsh realities that are nonetheless better confronted than avoided. For this interview, I was largely inspired by Abhishek Gupta's writings on the use of AI in war and his summary of the paper, "Cool Projects" or "Expanding the Efficiency of the Murderous American War Machine?" that investigated AI practitioners' views on working with the U.S. Department of Defense. (See Gupta's summary here, full paper here.)

**Michael, you were the DoD's first Chief Data Officer (CDO), hired in July 2018. What did the creation of this position signify?**

The DoD, like the rest of the public sector, was about 35 years behind in terms of a data strategy. There was actually an initial wave of CDOs hired in government about 10 years earlier. But the goal back then was to protect or safeguard data. Ten years later, the goals were the complete opposite. There was a shift from secrecy to sharing–making data available to the right people and ensuring quality of the data, for purposes of better decision making.

**What motivated you to join the US Department of Defense, and why were they interested in you specifically?**

They were interested in me because I brought commercial sector know-how to public service. That is a generalized trend in the public sector — to learn from private-sector best practices.

In terms of my personal motivation to join the DoD, I knew they had a big problem with data fragmentation and I wanted to solve it. I enjoy leading digital transformation. They are the largest organization in the world. There are 13,000 IT systems within the department. So there was no department-wide view or decision-making possible. Furthermore, the authoritative structure of the organization added to the balkanization of data. Every system had its own authorized access, i.e., no single person had authority over all the data. The opportunity to offset both data fragmentation and an authoritative organizational culture was interesting to me. And I was going to be senior enough to accomplish something.

The fact that it was an inaugural role was exciting. I never thought I was good enough at anything to be the first one to do it, but it intrigued me as a distinction. They had initially talked about a "data czar," a title I found entertaining, and then they changed it to Chief Data Officer.

There was also an element of patriotism. I wanted to serve my country and contribute to the safeguarding of our nation.

Within my capacity as CDO, I saw a more specific opportunity to make a positive difference in the way AI is being adopted and implemented in this country. I was particularly concerned with some of the irresponsible practices I had seen coming out of Silicon Valley with regards to AI.

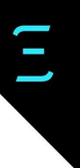

The general attitude is captured in Facebook's motto, "Move fast and break things," but in some cases, these people were breaking people's lives.

I studied psychology as an undergraduate so I understand the basics of statistics, testing, and measurement. I understand that data has to be a valid representation of the real world throughout its life cycle. But many of the people I had encountered in Silicon Valley were not careful with regards to these basics, and this offended me as a professional and as a human being. I wanted to make a difference.

**So ethics and responsibility were not exactly part of the private sector "best practices."**

Correct. There's a lot of talk about principles, but not a lot about how to actually apply these in practice. As part of my professional development, I participate in a series of study tours that take place twice a year. During my time at the DoD, these tours took me to Silicon Valley, New York, and London. I got direct exposure to how people were integrating principles into their techniques and methods, or not, as the case may be.

I would add that it's not just Silicon Valley that needs to be more careful. The Covid crisis exposed just how complicated things can get even in the most well-intentioned, i.e., "AI for good," contexts. In the early days, data-driven solutions for containing the spread of the virus proposed by AI researchers were typically framed as a choice between death versus privacy. That kind of framing certainly encouraged privileged members of society to consider giving up their privacy, but for many individuals and communities, this dichotomy doesn't apply, especially when taken in historical context. In other words, the risk associated with collecting and sharing their demographic data has, historically, been systemic death or violence, including by state actors. The Indian residential school system in Canada during the late 1800s, and the ongoing reservation system, is a case in point and explains much of the resistance by some members of the indigenous community here and elsewhere to collecting and sharing data.

The Ada Lovelace Institute in the UK describes the emergence of a third stage of AI ethics that focuses on the actual application of the principles developed in earlier stages. Does that ring true for you?

I actually spent time at the Ada Lovelace Institute as part of my study. They had a very practical, clear-eyed, and down-to-earth way of looking at things. I loved their focus on actual outcomes. They represented the antithesis of Silicon Valley's "move fast and break things" attitude. They encouraged a more thoughtful and responsible approach of considering possible positive and negative outcomes, as opposed to just going ahead and seeing what happens. It was about carefully considering the consequences of your actions.

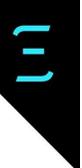

In terms of your earlier comment regarding the goal of making data available to the right people so they can make better decisions, can you elaborate on what kinds of decisions were being made at the DoD?

The DoD is involved in several types of activities including military and humanitarian missions, and a lot of business administration. Back office operations represent about 75% of the department's budget. My function as Chief Data Officer and subsequently Chief Business Analytics Officer was primarily focused on the business mission, e.g., things like financial management, logistics and supply chain, human resources, medical, real estate acquisition. We were trying to answer critical business questions from an enterprise-wide perspective. What's the total number of truck tires of dimension X in all of our warehouses? Where are we renting office space that is within 20 miles of other Federal government-owned office space that is less than 50% occupied? What's the total amount of money we spend with a given electrical utility? Is it better to continue running our own medical school or should we pay for aspiring doctors in uniform to attend the medical school of their choice?

Very little of my work was related to battlefield applications. In fact, only a very small percentage of what the DoD does is about targeted, precision fire, i.e., killing people and destroying things, and assessing the damage after the fact. Ideally, the DoD "holds the adversary at risk" realistically enough that shooting never starts. That said, a lot of the data supports the back office, humanitarian, and military missions simultaneously. You can't always easily separate them.

**What are the military equivalents of the "business" questions you outline above?**

Battlefield/warfare activities were outside my purview, and there's not much I can say here without violating the terms of my security clearance. But the most obvious example I can give you would be Battle Damage Assessment (BDA). For example, let's say pilots execute a sortie, which means a combat mission, against a designated target. The "After Action" task is for analysts to review imagery to assess the performance of the sortie based on a BDA. The fundamental question is, did we take out the target we intended to take out? This is a complex data problem that requires increasingly sophisticated levels of detail and accuracy that match the increased accuracy of weapon systems.

**So how did your learnings about ethics come into play?**

I focused on building ethical principles into protocols for curating and analyzing data, across the life cycle. My main concern was whether the data provided a valid representation of the world.

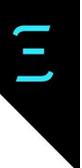

People like to talk about Artificial Intelligence because there's a certain sizzle to the term. As a data practitioner, I know that AI, ML (Machine Learning), BA (Business Analytics), and BI (Business Intelligence) are all variations on a theme. You take some code and feed data into it. The code identifies patterns, relationships and probabilities. Then the code uses those to make decisions, predictions, or both. So the validity and accuracy (I'm using statistical terms here) of the data is critical.

For a long time, one of the fundamental guidelines of IT has been GIGO – garbage in, garbage out. It's still true. And it's more important than ever because of two things. First, there's system bias and the tendency to believe something is correct because the system says so. Second is the sheer number of decisions we're permitting code to make on behalf of organizations in both the commercial sector and public sector. Only we try to dress it up by referring to the code as algorithms. And when we've fed the data into the code we call the result trained algorithms. But what that obscures is that all code is flawed. I say that as someone who made a living by writing code. Even when the code is relatively clean, you have to select the right code for the purpose. Then you have to feed it the right amount of good-enough data (ProTip – all data are dirty and incomplete). Finally, you have to know how to interpret the results. So there are many ways the entire value chain can go wrong and deliver unintended consequences.

Now step away from the role of the data practitioner. As a responsible adult and citizen I have a responsibility to do the decent thing… to behave responsibly. We all do. The DoD role as CDO gave me a platform from which to have a direct positive impact on the way the DoD approached this responsibility.

Look, the DoD, and Federal government in general, is filled with dedicated, smart, highly professional people committed to public service. I understood that superficially before I joined them. Once I was on the inside I got a much greater appreciation of how committed they are to ethical behavior, by which I mean they were well aware of the potential for both good results and bad results in every decision. "Friendly fire" and "collateral damage" are not new concepts to the military.

That is a very important point. We tend to attribute pre-existing issues that intersect with AI as being specific to AI, as if AI itself is the issue.

**Can you talk about the DoD AI ethics principles released in February 2020?**

The principles had been in the works for over a year before release. They had been developed by a group of political appointees, which in my opinion was full of people attempting to create cover for a corrupt agenda. They slapped labels onto projects that made them sound good and interesting, but they had no clue how the project deliverables could be misused. And they promoted projects that would have compromised personal data by exposing it to commercial firms. The principles themselves are useful, as long as they are put into practice through meaningful disciplines.

There were 5 principles: Responsible AI, Equitable AI, Traceable AI, Reliable AI, and Governable AI. For me, the details regarding principles themselves were less interesting than the question of how to put them into practice. There are minor differences between one principle and another, what's important is the outcome.

AI ethics is not a binary, either/or, thing. It's a matter of degree and probability. There's a ladder of creepiness when it comes to things like surveillance. The goal is to figure out where everyone is on that ladder, and stop the upward movement.

On a more general level, I believe that AI is an ideology. I reject the idea of an AI technology race, i.e., that we in the USA should mimic the Chinese approach. Number 1, we don't share their values with respect to the balance of power between the individual and the collective. Number 2, we don't share their enthusiasm for central authority. Number 3, "copy cat" is a losing strategy in business. We have to remain true to our ideals.

Some of the early criticisms regarding the establishment of AI principles across all sectors include the usual "lack of teeth" and mere "ethics whitewashing." But, in the case of the DoD there's an obvious contradiction or paradox in the notion of applying ethical principles to activities related to, as you put it, "killing people and breaking things," regardless of what percentage of overall activities they represent. Did this come up for you during your time at the DoD and how so? Do you personally see a contradiction?

Many people view the DoD as a war machine, and they therefore label me as a warmonger. When I speak in public, I know those people are going to reject my messages, but I expect that and see it as the sign of a diverse audience–and that's a good thing. I have a habit of starting speeches by saying, you're not going to agree with a lot of things I say. I embrace that spirit of disagreement.

I was in high school when the Vietnam war was going on, and by my sophomore year, I was within 2 years of being eligible for the draft. I was taught that if you're drafted and you didn't go to war, you were a coward and a traitor. And if you weren't drafted but you volunteered, you were an idiot.

The way I saw serving the DoD as CDO, it wasn't about enabling people to kill, rather, it was to help defend the country against enemies. That said, we, meaning the US, have created most of those enemies in central Asia and the Middle East. And by 'we' I specifically mean the US government, not the US DoD. The DoD is an instrument of the Federal government. Our elected officials determine where and how the DoD operates.

In my role as a citizen, I work through the political process to shape the decisions our elected leaders make. In my role as CDO & CBAO I worked to make the DoD effective in carrying out the will of the elected leaders. There are elements of both roles that I'd say were problematic, but being an adult and a leader means deciding what problems you want to live with.

*Michael Conlin is Chief Technology Officer at Definitive Logic.*

# 2. The Critical Race Quantum Computer: A Tool for Liberation by Michael Lipset, Jessica Brown, Michael Crawford, Kwaku Aning, & Katlyn Turner

**Illustrations created and adapted by Jessica Brown from "Open Peeps" by Pablo Stanley, CC0**

**Grappling with Uncertainty and Inequity Amidst a Society in Crisis**

As collaborators, we met over zoom during the COVID-19 pandemic. 2020 had a sobering effect in that there were so many facets of inequality, inequity, and injustice which became too blatantly invisible to ignore -- from the murders of George Floyd and Breonna Taylor, to the white-supremacist demonstrations in Summer 2020, to the blatantly racist rhetoric around voter suppression and "legitimate votes" in the November 2020 election. The COVID-19 pandemic itself made inequity even more visible. The differences between zoom workers, essential workers, the unemployed, and the elite class were laid bare, and the country debated whether it was biology or culture (rather than inequity) underlying the reason certain populations got sicker from COVID-19. Behind every public health discussion -- whether to open restaurants indoors, whether to require masking, or whether to travel -- were assumptions regarding what responsibilities individuals had to protect others from the spread of such an inequitable disease. Friends and family erupted into arguments over the morality of eating inside a restaurant, visiting Mexico, attending a Black Lives Matter protest, or going for an unmasked walk. It's fair to say that in 2020, inequity was forced into our national consciousness in a way that it hadn't ever been before -- and the responses each of us as individuals, communities, states, and counties had -- have changed the world in a way that we likely won't fully understand for years.

In our zoom discussions, what began as connections made based on shared interests became regular meetings--with the topics of inequity, divisive rhetoric, and current events often taking a forefront in these conversations. One day, one of the cisgender white men among us asked, "How do you **make certain** you aren't replicating oppressive structures in your interactions?"

Another of us offered, "You can't **make certain**. Every context is different, every scenario unique. Instead, you have to develop the ability to read each unique moment at every level. Our tendency to ask questions like this one, which presumes an answer where no definitive answer should be found, is of limited utility to that work. It sets up a false binary."

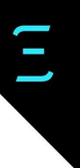

Another noted: "This question in general is particularly important work for people who hold privileged identities -- *e.g.* white, cisgender -- in America. When one's social location is that of 'the oppressor', almost every interaction holds the potential for the reification of systemic forms of violence. **The suggestion that one should be able to 'make certain' not to replicate oppressive structures is fraught."** Tema Okun's "White Supremacy Culture" carries a reminder that the presumption and expectation of perfectionism, certainty, and either-or thinking (what we refer to here as binary thinking) are functions of white supremacy culture (Okun, 1999). Equity does not function on certainty, perfectionism, and binary thinking, but instead on a plurality of states existing at once.

In response to the question of certainty and reproducing oppressive structures, we likened the work of anti-racism to a quantum computer. While normal computers run on binary code where bits operate as either 1s or 0s, in a quantum computer, code gets written for qubits, which exist as 1, 0, or simultaneously as both.[1] The metaphor seemed to work on a number of levels. Rather than thinking of issues relevant to the work of anti-racism in binary ways like the way our question was posed we agreed that complexity, nuance, and the willingness to push each other to think more deeply about this question were necessary. Instead of *making certain*, we have to *do our best to make certain* and acknowledge that certainty is an imperfect metric. One must be constantly vigilant in their abilities to 'read the room,' assess dynamics of power and privilege, be conscious and open to how gender, class, ability, language, etc play out differently in each situation, and how the inputs and outputs of each situation have ramifications far beyond the situation itself. Additionally, as individuals, we're more likely to act in a way such that our actions may support the liberation of some, but not all, and thus exist between the two ends of a liberation-oppression spectrum. **In other words, a single human being is far more like a qubit than a classic bit.**

From what little we knew about quantum computers, the comparison between these concepts, the work of critical race theory, and the practice of anti-racism seemed well-aligned and helpful to supporting the understanding of these complex topics[2]. We wondered what other parallels might be drawn between these spaces. We considered concepts such as amplitude, quantum superposition, and probability strings (Hidary, 2019; Trabesinger, 2017, and identified analogous pairs in critical race theory (e.g., the amplitude of a qubit (Hidary, 2019) mapped onto the critical framework people must develop to do anti-racist work (Freire et. al, 2018; Kendi, 2019);

---

[1] Additionally, qutrits have also been identified, i.e. bits that can exist as 0, 1, 2 or all three simultaneously. Quantum computer scientists use the term 'qudit' to refer to any bit that has three or more possible states. For the sake of uptake and simplicity, we recognize the existence of these more evolved qudits but will use qubit throughout this paper. As quantum computer technology evolves, this heuristic can and should evolve with it to consider humans as evermore complex (qudits).

[2] We recommend "Quantum Computers" on the Data Skeptic podcast with Prof. Scott Aaronson from the University of Texas at Austin.

quantum superposition mapped onto the recognition of multiple, valid lived realities (Delgado et. al, 2017; Okun, 1999); the concept of fractals put forth by adrienne marie brown (2017) in *Emergent Strategy* shared important and powerful similarities to probability strings.

We recognized the potential of bringing two seemingly unrelated concepts into conversation with one another. *The Medici Effect,* by Franz Johansson, identifies the potential for innovation when connecting two seemingly disparate topics or ideas. Johansson would call concepts like quantum anti-racism an *intersectional* idea.[3] Intersectional ideas, he claims, change the world by giant strides in new directions whereas *directional* ideas evolve in generally predictable, linear patterns. (Importantly: this is not to be confused with the concept of intersectionality, a foundational paradigm in Black feminism first framed by Kimberle Crenshaw (Crenshaw, 1989) that offers a framework to understand how aspects of one's identity--race, class, and gender, for example--overlap to frame one's experience of the world.)

A critical and timely similarity is how these technologies are perceived. Those in the quantum computing space working to communicate the technology to the general public run up against the challenge of people avoiding the topic altogether for its seeming complexity (Aaronson, 2021). Critical race theory has become particularly shrouded in propagandist obfuscation, especially in recent months. Attempts by Republican politicians in states like Tennessee, Idaho, Iowa, and Oklahoma have sought--somewhat successfully--to outlaw the inclusion of critical race theory in public schools at the state level, largely through the promotion of gross misunderstandings of the concept (Sawchuck, 2021). Another example--in a high-profile story, the scholar Nikole Hannah-Jones was originally denied tenure at her alma mater, the University of North Carolina, for the controversy ignited by her Pulitzer-Prize winning project, "1619"--which draws from critical race theory among other disciplines to tell the story of how the construct of race developed and was made canon in US history.

Arguably, those who dismiss critical race theory also wish to avoid the topic for its complexity: the complexity of one understanding of "history" juxtaposed with another valid and less-well-known one--one that acknowledges oppression, genocide, and instituionalized segregation.

In the same way that quantum computing may serve to help delineate anti-racism and critical race theory, perhaps this metaphor might also help clarify some of the concepts in quantum computing for a broader audience.[4]

---

[3] Not to be confused with the concept of intersectional feminism as developed by Kimberlé Crenshaw (1991).
[4] While we advocate for a greater uptake of critical race theory, we do not yet advocate for a greater uptake of quantum computers. Such technology has not yet been tested and, with most technologies, is likely to be both a bane and a boon.

To achieve a more just and liberated world, people must operate more like the qubits we actually are, we surmised — complex beings operating between many spectra simultaneously — rather than residing within a simplistic binary like bits, as pictured in Figure 1. The goal of a given scenario may be to avoid the replication of oppressive contexts, but it may also be to eradicate those contexts altogether. Humanity must evolve beyond binary understandings of concepts that require more complexity than a "0s-and-1s" understanding if we want to build a world that operates with life, humanity, and love at the foundation. The metaphor of a quantum computer felt important to understanding how we might maintain a complex, non-prescriptive attitude towards the everyday work of liberation. The outcome of this work is a heuristic we're calling the Critical Race Quantum Computer (CRQC) and a concept we're calling Quantum Anti-Racism.

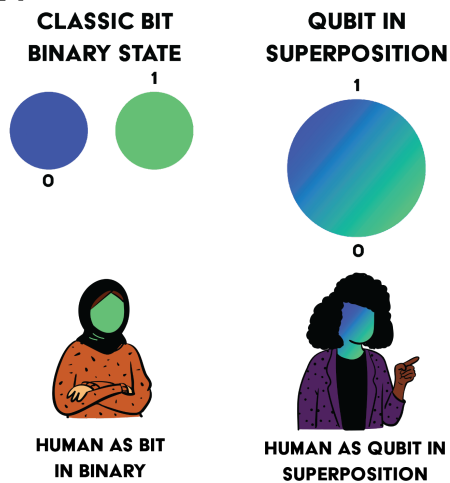

**FIGURE 1**

**Tenets of Critical Race Theory**

The critical race quantum computer draws from the literature on critical race theory and the work of anti-racism. The critical race theory (CRT) movement is "a collection of activists and scholars engaged in studying and transforming the relationship among race, racism, and power" (Collins 2002, Crenshaw 1989, Davis 2011, Delgado *et. al* 2017). CRT draws from critical legal theory, radical feminism and was originally taught as an approach to legal practice to support future lawyers in their abilities to challenge the neutrality of constitutional and other forms of law (Delgado *et. al*, 2017).

Although more tenets of critical race theory have been added to the CRT canon, the following six features are considered fundamental to CRT (Crenshaw, 1989; Delgado *et. al*, 2017):

1) **ordinariness**, which refers to the everyday, everywhere nature of racism that makes the work of anti-racism more challenging
2) **interest convergence**, or the notion that change away from racist policies and practices only occurs when it favors the dominant group
3) **race as a social construction**, or the recognition that race and racism are not rooted in biology but are created by society as a way to benefit and uphold white supremacy

4) **the different ways in which different racial groups are racialized by the dominant group**, which can be seen in the ways Jewish and Italian people, once considered non-white, have since been moved into the racial category of white to bolster the white populace
5) **intersectionality**, or the notion that no one can be only categorized by their race, we are also gendered, sexualized, ableized, and more
6) **the "unique voice of color,"** or the idea that people of color are capable of communicating the uniqueness of their experiences to white people and can share stories of what it means to exist in their own unique ways to broader society

**Precedence: Quantum Mechanics and Social Theory**

There is precedent for the combination of quantum mechanics and social sciences. In 1983, Flora Lewis wrote "The Quantum Mechanics of Politics" for *The New York Times* in which she connected quantum mechanics to democracy. In her article, she writes:

> *The best possible society, which can never be anywhere near ideal, has to be the one best able to correct its mistakes, to develop new ideas, adjust to new circumstances…*
>
> *I'm trying to urge an attitude that admits we really don't know anything for sure, and still considers it worthwhile to keep trying to find out… Prudence, patience, a willingness to see that the best intentions don't guarantee the best results, and above all, a capacity to see an infinite need for corrections - to my mind, these are what we need…*
>
> *The quantum mechanics of politics, then, demands from us an understanding that flux is neither good nor bad but inevitable.*

We agree with Lewis, that the best society is not described by some ideal, but by an ongoing ability to correct its own mistakes. Additional past criticisms of quantum mechanics and the social sciences that unfold along the following lines: 1) that quantum theory doesn't apply at macro scales but only on the microscopic level and, 2) that the application of quantum science as a metaphor does not bring scientific reasoning to the physical nature of social systems.

While we know quantum theory is also used to examine and understand physical reality, what we're talking about here should be understood explicitly in the metaphorical sense. We make no claim that anti-racism and quantum mechanics bear relationships to one another beyond the use of quantum computing as a metaphor for anti-racism to further the development of individual and collective critical consciousness. We put forth our idea for what this is (*i.e.*, a tool for learning), while also opening it up to critique, and the idea itself evolving.

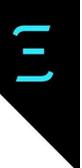

**The *Why* Behind a Critical Race Quantum Computer**

Our "why" is relatively straightforward. **We believe that the more people that exist with a developing sense of critical consciousness the closer our world will get to being an equitable, socially just, liberated space for us all by enabling us to better self-correct as a human community.** The #BlackLivesMatter, #MeToo, and modern social justice movements are the contemporary banners under which age-old calls for a less exploitative and more equitable society have historically been cast. Today, as we bear witness to the ongoing oppression of Black, Brown, and Indigenous people, as well as the backlash to naming and amplifying the ideas that posit this oppression exists -- we embrace the imperative of doing everything in our power to contribute to a society where everyone has the right to breathe, live, and thrive.

The space wherein people doing work to develop the critical consciousness of others has been a hotly contested and fraught space for quite some time. On the one side, current right-wing ideology posits that discussion of racial inequity is dangerous and should be silenced. However, on the left of the political spectrum, the work around inequity, awareness, and education has been critiqued for a myriad of reasons. Developers of important concepts in the field of Diversity, Equity and Inclusion (DEI) — the banner under which much of the work to develop critical consciousness and anti-racism among individuals and organizations occurs — are often critiqued for deepening the fragility of white audiences and pushing them further towards prejudice (DiAngelo, 2018; Schick 2000; Schick 2014). Literature dedicated to developing anti-racist positionalities among the public at-large saw tremendous growth in the wake of George Floyd's murder in the summer of 2020 (Flood, 2020). In the US, the top two-selling books were Ibram X. Kendi's *How to Be an Anti-Racist* and Robin DiAngelo's *White Fragility* (Flood, 2020; McWhorter, 2020). Race2Dinner, an organization charging white women upwards of $2500/head to attend a dinner where two women of color engage participants in deconstructing racism within themselves, also saw a rise in patronage (Noor, 2020). Critics point out that unlearning racism is important but only the beginning step of living a life that fosters liberatory outcomes.

In the wake of the ongoing attack on Black, Brown and Indigenous lives across the US and throughout the world, attempts by the privileged elite to educate themselves through reading or hosting dinner parties have been critiqued as insufficient and contributing to the pacification of white peoples' own white guilt. The work of developing more anti-racist individuals is difficult, complex, risky, and in need of approaches that emphasize collective liberation at a systemic level. DEI, for the left, doesn't go far enough, and for the right, makes matters worse.

The technology sector also has not been immune to calls for greater accountability around anti-racism, but has arguably seen less uptake therein. Given the recent proliferation of literature at the intersections of critical race theory and Science and Technology Studies (STS), or what Ruha Benjamin (2019) calls race critical code studies, as well as increasing calls for critically conscious approaches to the current use of technology that include concepts like design justice, coded equity, and tools from the tech industry that contribute to, rather than detract from, the liberation of all people, we offer the concept of the CRQC in support of this emerging body of work (Benjamin 2019, Buolamwini *et. al* 2018, Costanza-Chock, 2021, D'Ignazio & Klein 2020, Eubanks 2018, Hampton 2021, Noble 2018, O'Neil 2016, Turner *et. al* 2021)

We also present this framework to support the development of better technology. Technologies made in absence of deep recognitions toward our pluralistic human community and what makes it so vibrant do not in fact serve everyone. Insofar as we have arrived at the technological epoch, we must recognize digital technology as a permanent part of our existence and thus include it in whatever model of liberation we collectively build towards . (Bailey 2019, Benjamin 2019, Galloway 2020, O'Neill 2016)

As we stand on the precipice of companies like Google, IBM, and Microsoft achieving quantum supremacy, we must recognize the unknown potential of a world where computers operating on binary code may soon be supported by computers operating in quantum. If binary computers have been shown to replicate the same biases as the people coding them, producing a technological whitewashing of algorithms that reproduce societal injustices in housing, education, credit scoring, healthcare, surveillance and more, imagine the pitfalls of a society operating with quantum? Companies running algorithms under the veil of AI have reproduced racist social practices and admitted along the way that they aren't sure how the algorithms themselves operate (Bishop, 2018, O'Neill 2016). If the uncertainty of these processes exists within the world of binary largely as a result of embedded human bias as scholars like Kathy O'Neill, Ruha Benjamin, Joy Buolamwini, Safiya Umoja Noble and others have shown, imagine the potential for such programs to run amok in the increasingly complicated world of quantum.

**A Critical Race Quantum Computer For *Whom*?**

We imagine the Critical Race Quantum Computer being used by people with no or developing critical frameworks as a tool to support the understanding of a non-binary, complex existence that actively works to see the ways in which power and privilege work to the benefit of some and the detriment of others. For those working in tech, this may be a valuable entrypoint into conversations around how tech and anti-racism must intersect. For those inside tech making things, this is a tool meant to support them in their making and life more generally so as to maintain a level of criticality necessary to live and advance liberatory-- as opposed to oppressive-- contributions to our planet.

Scholarship on the decolonization and development of socially just technology points towards the important need for those creating our technology to develop anti-racist critical frameworks. Cathy O'Neil's 2016 *Weapons of Math Destruction* claims that "Big Data processes codify the past. They do not invent the future. Doing that requires moral imagination, and that's something only humans can provide. **We have to explicitly embed better values into our algorithms**, creating Big Data models that follow our ethical lead. Sometimes that will mean putting fairness ahead of profit" (p. 204). Some have called this digital justice or coded equity (Benjamin, 2019; Noble, 2018). Regardless of the name, a clear need exists for tools designed to support the development of technologies that contribute to a more equitable, liberated planet for all people. This tool helps fill that need by supporting the development of the mindsets necessary to code for equity and justice.

To further unpack how the CRQC functions, we understand that a computer needs code. One thing we can provide is the identification of that code; i.e. critical race theory and its accompanying body of work. The code to the CRQC is the knowledge established to-date that promotes and supports the end goal of collective liberation--written by people like Sojourner Truth, James Baldwin, Martin Luther King Jr., Kimberle Crenshaw, and others.

While we as authors each hold nuanced perspectives on the specifics of the shared, collective end-goal we have stated, we hope this concept gets taken up as a form of technology, equitably coded, as one contribution to the movement towards digital justice. We now turn to our own positionality and its possible influence upon this tool to better understand how our own approach to justice and liberation might influence this contribution.

**By Whom?**

We contribute to this conversation as people concerned, first and foremost, with the learning and development of an anti-racist, intersectional race critical framework for all. Benjamin (2019) identifies the need for our contribution as educators in the technology space when she says, "Justice… is not a static value but an ongoing methodology that can and should be incorporated into tech design. For this reason, too, it is vital that people engaged in tech development partner with those who do important sociocultural work honing narrative tools through the arts, humanities, and social justice organizing" (pp. 192-193). Our personal politics as authors are important as our beliefs compose the many sides of our personal abilities to function like qubits.

We employ an intersectional feminist definition of anti-racism as that which should be understood when working with the CRQC because we espouse such work ourselves. We believe in a definition of justice that recognizes the baseline work of liberation for all people as being most well-represented by the liberation of trans Black women. Scholars of CRT argue that the most oppressed person in the world is not just Black women, but transgender Black women. Transgender Black women face overt, legalized, and de facto sexism, racism, misogynoir, transphobia, and transmisogynoir. For example, transgender women of color face increased homelessness, medical discrimination, hiring discrimination, and domestic violence--as compared to white transgender women and cisgender people of all races (Graham 2014, Jackson *et. al*, 2018, Jackson *et al*. 2020, Mulholland 2020, Williams 2016). The liberation of transgender Black women would represent progress to the degree of liberation for all. Therefore, anti-racism recognizes the intersecting mechanisms of power and oppression as they act not just along lines of race, but also gender, sexuality, class, ability, language, time, space, and scale as well. "Intersectional" has already been defined by scholars as including considerations of the ways in which power and privilege operate in conversation with each other across age, gender, ability of all kinds, socio-economic status, sexuality, religion, and geographic location (Crenshaw, 1989; Piepzna-Samarasinha, 2011), as well as across the micro/macro scale of societal interconnectedness (Kaba and Murakawa, 2021; maree brown, 2020; Mingus, 2016). This definition holds particular importance since it's precisely this definition we seek to unpack and better understand through the framework of the CRQC.

But to the extent that our perspectives are simply those of ~8 billion qubit-like humans on planet earth, the shared tool holds far more importance than our own perspectives as authors. We also recognize that in the same ways that a coder's biases impact the codes they write, so too are our own biases likely to show up throughout the establishment of this tool and concept. We acknowledge the CRQC as an open concept tool for continuous co-creation among those who take it up and those who write the code.

**The Critical Race Quantum Computer (CRQC)**

The concept of quantum anti-racism draws the following parallels between critical race theory as a practice that impacts and is impacted by microcosmic and macrocosmic forces and the ways in which a quantum computer depends upon the alignment of many qubits to produce a given outcome. Let's say the stated goal of anti-racist work is a world where everyone has access to quality food, clothes, shelter, housing, mental health support, medical care, physical safety, and emotional wellbeing. Scholars of critical race theory, from which anti-racism work draws heavily, have identified this goal as one that requires the education of individuals who, when working together, compose our established institutions and systems in support of a liberated world. In order to understand how the development of an anti-racist, critical race framework — or an anti-racist worldview — might be supported by this metaphor, let's begin with a broad description of a quantum computer and some of the more important concepts to quantum computing.

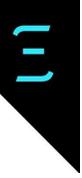

Quantum computing has become one of the next areas of high interest from such major technology companies as Google and IBM. Quantum computers are claimed to operate at 100 million times the speed of today's binary supercomputers (Singh, 2015). Quantum computers execute algorithmic computations more quickly because they operate on **qubits**, or bits that can exist as 1s, 0s, simultaneously as 1s and 0s, or even more states depending on the computer's design. This state of simultaneity is known as **quantum superposition**.

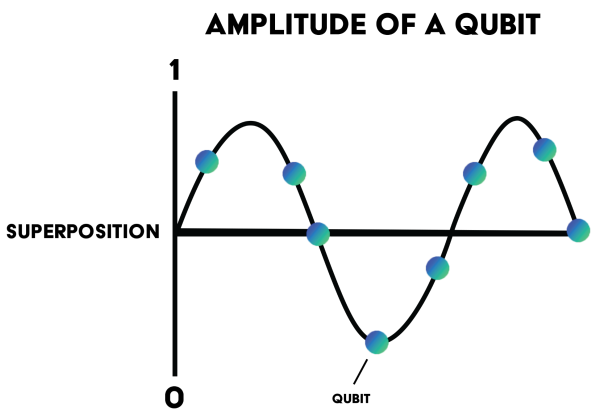

FIGURE 2

AMPLITUDE OF A QUBIT

Quantum computers operate on the basis of multiple qubits working together, so it takes more than one qubit to form a quantum computer. For every possible state a qubit can hold, quantum computer scientists assign it a measurable **amplitude**, which can be roughly translated into the likelihood of a qubit to exist in one particular state or another (see Figure 2). Quantum computers work by coordinating these waves of amplitudes among a series of qubits in order to produce a certain outcome. This coordinated wave is known as a **probability string** because it strings along a series of qubits to produce a probable outcome. A probability string is simply a series of bits or qubits in a coded sequence. Figure 3 illustrates the difference between a classic 3-bit probability string and a quantum 3-qubit probability string. In the space of quantum computing, classic probability strings are referred to as local, whereas quantum probability strings are referred to as global (Dumon, 2019).

FIGURE 3

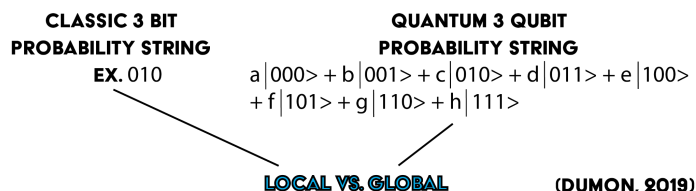

CLASSIC 3 BIT PROBABILITY STRING
EX. 010

QUANTUM 3 QUBIT PROBABILITY STRING
$a|000> + b|001> + c|010> + d|011> + e|100> + f|101> + g|110> + h|111>$

LOCAL VS. GLOBAL          (DUMON, 2019)

In quantum computers, the amplitudes of different qubits can **interfere** with each other. So, if something can happen positively in one qubit and negatively in another, those qubits can cancel each other out. For quantum computers to work properly, they have to exploit the interference between qubits in such a way as to get the paths corresponding to a given outcome to interfere with each other, as seen in Figure 4 ("Quantum Computing", 2017). At the same time, the computer needs to do something to boost the amplitudes of the qubits that align with the desired outcome (Zickert, 2021).

Importantly, interference, or the prevention of a given outcome, can occur in a quantum computer as a result of **decoherence**, or unwanted interaction between a quantum computer and its external environment (Hidary, 2019). Quantum computer scientists have shown that you don't actually need to eliminate decoherence in order for a quantum computer to function properly, you simply have to reduce decoherence enough such that the loss of certain qubits to decoherence would not destabilize the rest of the qubits. These decohered qubits are called **noisy qubits** ("Quantum Computing", 2017). In order to minimize the impact of noisy qubits, quantum computers require **constant measurement and analysis** - also known as error correction - of the computer to see if an error has occurred and to do so in parallel across all of the qubits (see Figure 5) (Trabesinger, 2017). The point is that even with noisy qubits, it's still possible to do an incredibly complex quantum computing problem.

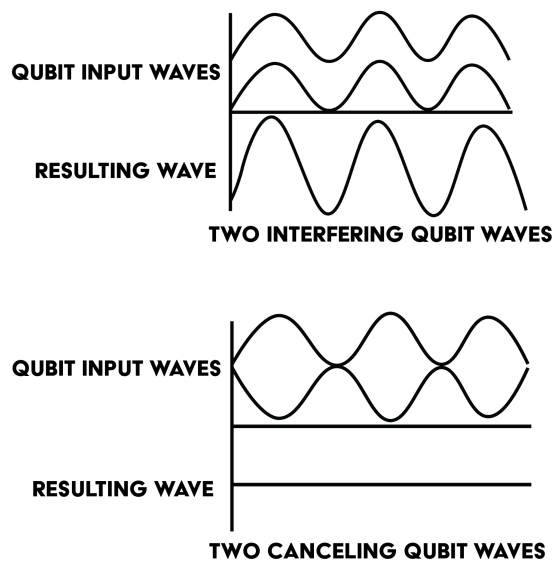

**FIGURE 4**

So, what does this have to do with the work of anti-racism and critical race theory? Well, nothing until the parallels between the ways we imagine building a more just world are made explicit in connection to the functioning of a quantum computer. Let's analyze this metaphor using the goal of anti-racism, stated above, as the quantum computing problem at hand, i.e., how can we build a more just and liberated world for all, including this planet we call home? **In order to collectively compute this problem we must each operate with the understanding that we are more like qubits than bits. We are human beings capable of existing between the ends of many spectra, capable of doing harm and contributing positively to our surroundings at any given time. In the case of binary computers, this spectrum is between 1 and 0, but in the case of humans this spectrum may be between racist or anti-racist.**

Those with less developed and less critical understandings of racism think racism primarily occurs at the person-to-person level, that people either *are* racist (e.g., Proud Boys) or *aren't* racist (e.g. everyone else). The truth of the matter is that racism can either unfold as a result of purposeful action (such as in the case of a Proud Boy or Klan member) or through inaction (such as through claims made by individuals that racism "isn't my problem"). Anti-racism, in contrast, is something someone *actively* does in opposition to racism in policy and practice (Kendi, 2019).

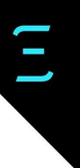

**Therefore, one can exist as racist (1), anti-racist (0), or, more importantly, someone conscious of their existence as a dynamic human being (qubit) aware of the possibility that they may exist between 0 and 1 at any given time (see Figure 6). As people, at any moment, our quantum superposition then becomes the state in which we exist knowing that we are capable of perpetuating systemic forms of violence, disrupting and dismantling the systems that perpetuate systemic forms of violence, or, more likely, often doing both at the same time.**

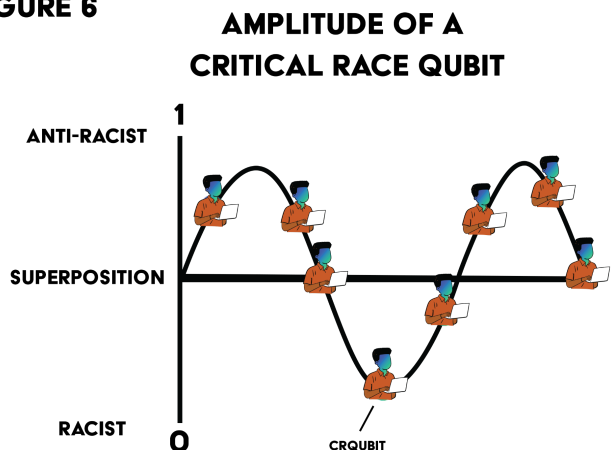

FIGURE 6

Quantum superposition also refers to the important recognition in critical race theory that we each experience reality differently, yet in valid ways (Delgado et. al, 2017). For example--at the time of its founding, the United States espoused the underlying claim of "*liberty and justice for all*" if you were an able-bodied, cisgender, heterosexual man racialized as white. This was seen as radical by many because it was a more open and less oppressive ideology than the colonial powers--where simply being an able-bodied, cisgender, heterosexual man of any ethnicity was not enough to merit "liberty and justice"--one also needed titles, for example. At the same time it espoused this paradigm, the United States allowed and encouraged oppressive norms for people who were not considered part of "all"--for example women, people of color, the disabled, and religious minorities (Dunbar-Ortiz 2014, Kendi 2016, Ortiz 2018, Zinn 2015). **These realities are true at the same time, and the notion that two opposing truths can exist at the same time fits into this function of the CRQC.** The amplitude of our qubit is then the degree to which we are either more capable of contributing to the unequal and systemically violent status quo or disrupting and dismantling that status quo. Our individual amplitude can be anticipated by the strength of our individual critical frameworks, which themselves are also ever-evolving, changing, and in flux. In a dynamic world where things are constantly changing, we have to continue working to correct ourselves.

When we as anti-racist human qubits work in alignment with each other, organized toward a shared goal, we set up the conditions for a probability string social justice to unfold. As mentioned earlier, a classic computer's probability string can be referred to as local whereas a quantum probability string is referred to as global. The notion of fractals (maree brown, 2017), which suggests the work of liberation must begin with ourselves and our immediate communities, spreading out through our proximal social networks and throughout society from there, could be considered the social justice analog to a quantum probability string. Without each other, we do not make a movement, and without other qubits there is no CRQC. At the moment, the factory-like nature of much of society's systems serves to dehumanize and separate people from one another. Though we have the potential to recognize our qubit-like, multitudinous nature, we often operate like factory workers, thinking linearly or in binary and simplistic ways. One might call this state of myopia 'local.'[5] The image of a series of people who recognize our qubit-like nature working within the framework of critical race theory in community and collaboration with one another, in full recognition of each of our social locations and imperfections, aware of the ripple effect our actions can have, might be referred to as 'global' (see Figure 7).

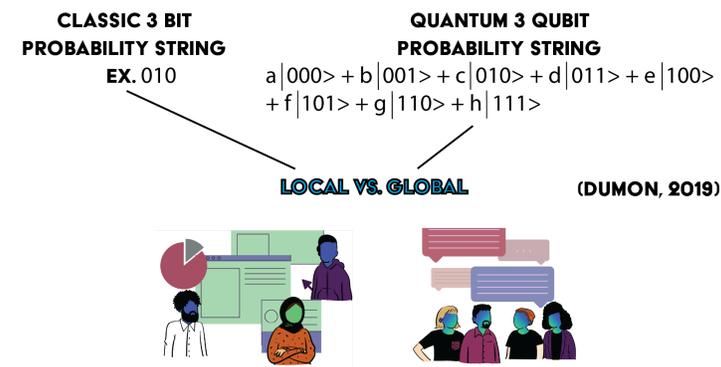

**FIGURE 7**

CLASSIC 3 BIT PROBABILITY STRING
EX. 010

QUANTUM 3 QUBIT PROBABILITY STRING
$a|000> + b|001> + c|010> + d|011> + e|100> + f|101> + g|110> + h|111>$

LOCAL VS. GLOBAL (DUMON, 2019)

WHEN PEOPLE WORK WITH BINARY FRAMEWORKS, WE END UP POLARIZED, IN SILOES, UNABLE TO COLLABORATE AND SPEAK WITH ONE ANOTHER (LEFT). WHEN WE'RE ABLE TO BREAK OUT OF THIS BINARY AND ROOT OURSELVES IN NUANCE, WE SEE, HEAR, HONOR AND WORK WITH EACH OTHER (RIGHT).

Of course, those without anti-racist critical frameworks produce an amplitude that interferes with the broader goal of anti-racist work. The work of the conservative right and the moderate middle often run in opposition to the work of progressive, activist communities. They interfere or, in the case of the moderate middle, actively interfere sometimes and passively interfere at others. In 1963, none other than Martin Luther King, Jr. (2017), in his "Letter from a Birmingham Jail" named the moderate white middle as perhaps the most dangerous contingent in the US for its surreptitious ability to appear progressive while simultaneously advocating for change to be made "at a better time" when there's greater political alignment. If the human anti-racist quantum computer consists of all humans as qubits, then the metaphorical critical race quantum computer itself won't work until a critical mass of anti-racist qubit-like humans exists.

---

[5] We recognize the importance of local initiatives to establish sustainable consumption and building towards justice. That is the foundation of fractals, start with those around you. There may be better language to use here.

More people need to develop their anti-racist critical frameworks in order for the collective probability string to produce an anti-racist — and generally anti-oppressive — society.

Until that critical mass has been established, decoherence of anti-racist qubits as a result of "outside factors" — let's call these factors the presiding status quo of kyriarchy, a term describing the ways in which domination and submission exist simultaneously across varying levels of society and wherein a given individual may be viewed as having privilege in some situations but as oppressed in others (Fiorenza 2001, Osborne 2015, Pui-Lan 2009), or the matrix of domination (Collins 2002) — will continue. The work of building and maintaining the CRQC, therefore, requires the same constant measurement and analysis as a real quantum computer. In the same ways that we as individual qubits must constantly retool our personal critical frameworks, so must those of us working to build a critical, liberatory mass movement keep our eyes on the state of the world in order to determine at what point we have reached critical mass. It's that moment, the point at which a critical mass of anti-racist human qubits has been established, that our human quantum computer will begin to function as a computer, rather than a bundle of individual qubits.

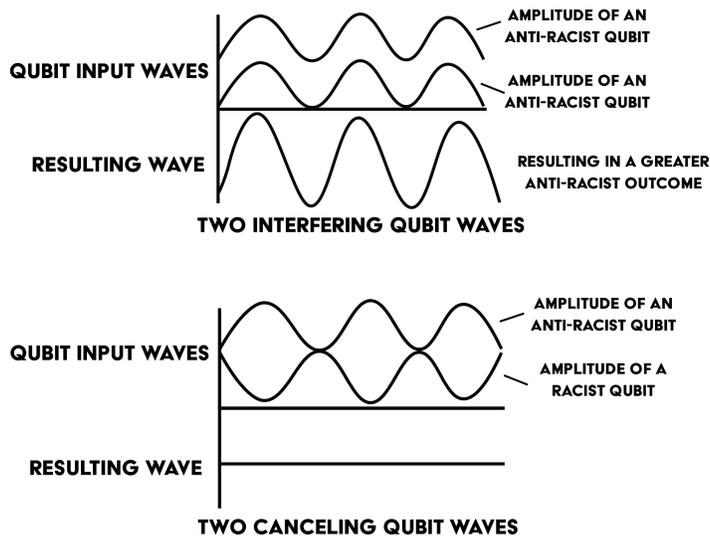

FIGURE 8

Decoherence, in a quantum computing sense, would mean the loss of a qubit's superposition and a reversion to the binary state of either 1 or 0. In the critical race theory understanding of quantum computing, decoherence would mean a loss of one's critical framework and a reversion to a state of either-or, binary thinking. To understand decoherence in this way, one might consider ideological polarization the result of a loss of critical race superposition. Perhaps, however, this is where some might feel the metaphor falls flat. Indeed, there are instances when radical action that appears to some as ideological extremism or polarization may, to others, be read as a necessary act of revolution. For example, the burning of the Minneapolis Third Police Precinct in the wake of George Floyd's murder was, in the immediate aftermath, called violent and extreme by right wing commentators. Civil rights activists, Black Lives Matter organizers, and some more progressive talking heads referred to it as an understandable and necessary response to living under oppressive circumstances that were literally suffocating people. On the other end of the trial of former police officer and convicted murderer Derek Chauvin, one might quite convincingly argue that his conviction may not have happened had the Third Precinct and other local institutions not been destroyed in response to his actions and the circumstances that allowed him to kill George Floyd in the first place.

We are not working to extrapolate mathematical concepts from quantum theory and claim what works at the microcosmic quantum level works the same way in macrocosmic sociological circumstances. The CRQC is a heuristic and only a heuristic, a tool to support people in developing their abilities to act in liberatory ways. However, if we take the metaphor of the CRQC to its foregone conclusion as part of our thought experiment, we find some interesting potentials for impact. For example, quantum computer science claims that the telltale sign of a 'working' quantum computer is its ability to decode all existing cyber-security measures ("Quantum Computing", 2017). If the metaphor we have set up holds, then a human anti-racist quantum computer, once functional, could mean that we as a society have broken down the barriers between each other that require our current forms of security. It is possible that once we achieve the goal of **quantum supremacy** — the name quantum computer scientists have given to the first, fully functional quantum computer — we will have moved past white supremacy and entered into a world where the only security we have as a planet is the maintenance and protection of the computer itself. Noisy qubits — or rightwing extremists, conservative distractors, and generally bigoted people — will be so few and far between as to exist without causing the decoherence of anti-racist qubits.

Our goal in presenting this metaphor is to support the development of more anti-racist qubits capable of maintaining an anti-racist framework so as to eventually reach the tipping point necessary for the achievement of quantum supremacy. We recognize the problematic nature of the word 'supremacy' given its historic use in relation to whiteness. Here, however, quantum supremacy is the supremacy of the people working in community to establish and maintain each other's liberation. In this sense, we use this language purposely to reclaim the word 'supremacy' and identify the importance of transferring, as has been the rallying cry at civil rights protests for generations, "All power to the people."

**Conclusion**

The critical race quantum computer is an aspiration at the moment. As a planet, we have a long way to go until we realize an anti-oppressive shared existence. It would be antithetical to the logic of our metaphor to suggest that any one of us is ever going to become a perfect critical race qubit-like person. In that way, we do not claim to have conceived of this metaphor perfectly, either. We imagine this contribution as one that will grow as others take it up, interpret it, and include/critique it through their own work.

As with everything contained within the CRQC, each "coder's" script must be read critically. James Baldwin wrote the code. Malcolm X, Martin Luther King, Jr., Angela Y. Davis, Assata

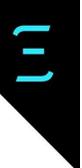

Shakur, Huey P. Newton, Bunchy Carter, Mary Wolstonecraft, adrienne marie brown, Patricia Hill Collins, bell hooks, Ibram X. Kendi, Ta Nehisi Coates, Mikki Kendall, Moya Bailey, Jane Elliot, the authors cited in this article, and so many more have been and are currently writing the code. These critical race "programmers" haven't written "code" that could be considered gospel or scripture outright. Each has been, can, and should be critiqued.

adrienne maree brown (2020), for example, recently published *We Will Not Cancel Us: And Other Dreams of Transformative Justice* wherein she delineates the important nuances of cancel culture. In this contribution, brown discusses the benefits and pitfalls of cancel culture, providing guiding questions to those seeking accountability for harms wrought as well as resources for further learning to be done around this topic. The CRQC would categorize brown's contribution as code supporting the development of us as qubits more capable of contributing to an equitable, liberated society by developing our individual and collective abilities to hold each other accountable for our actions without doing further harm in the process. Her code does so by developing our abilities to "cancel" people, but to do so transformatively and only when and where appropriate. It requires critical thought in order to know when, where, why and how to implement it.

With the rapid pace of technological evolution and the inevitable incorporation of quantum computers into our ways of life, recognizing the importance of a critical race approach to quantum computing and technology may allow us to get out ahead of potential pitfalls that may occur as a result of their use. In his recent book, *Post-Corona*, Scott Galloway argues that better technology is technology that reaches everyone, is inclusive of everyone and, thus, acknowledges that the user comes first. "Better," according to Galloway, means scalable to everyone in an equitable way and in a way that gets technology in everyone's hands without doing harm. While notions of scale have become fraught in their own right, the concept of liberation must be scalable or freedom itself, realized as a collective experience rather than an individual one, cannot exist. The concept of a critical race quantum computer is available to all. We hope people think about it, use it, share it, and develop their critical race theory (code).

In the time since we began writing this paper up to its publication, the discourse surrounding critical race theory has only become more hostile and divisive. More states and communities within states have sought to ban it, and indeed it has become somewhat of a testing of waters for factions of right-wing politicians--*e.g*. *how much of a potential for culture war is there, here*? How salient, triggering, and galvanizing is the concept of CRT: is it enough, for example, to shift the country towards the right again politically in 2022 and 2024? The polarization regarding CRT has played out on a national stage: from banning it in schools to the attempted silencing of those like Nikole Hannah-Jones who use it in their work. A timely question we consider now is in this vein: what happens if the "noisy qubits" gain an amplitude such that they are able to establish policy and enforce systems that further oppress, harm and marginalize? In other words: what happens when the influence of policymakers opposed to CRT and DEI grows to the point where it prevents others from developing their own critical consciousness and frameworks?

We are in the process of designing workshops based on the concept of the CRQC. These are meant to be rigorous resources to support interpersonal, institutional and systemic change towards more conscious contributions to the movement. We are also considering the development of a podcast series to support the understanding of this concept. But, more importantly, we believe the identification of the CRQC and the dissemination of this article contributes to the development and, hopefully, ultimate realisation of something akin to the CRQC. Our goal is not for everyone to think, act and behave the same way, but to recognize that we are all imperfect and, thus, in need of support, collaboration and error correction to be the best we can be individually and collectively. We don't each know everything, nor does any one of us have all the answers. From this space of humility we can dialogue, co-create, and build a better world together.

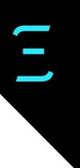

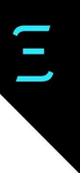

Trabesinger, Andreas. Quantum leaps, bit by bit. *Nature* 543, S2–S3 (2017).

https://doi.org/10.1038/543S2a

Williams, S. (2016). #SayHerName: Using digital activism to document violence against black women. *Feminist media studies*, *16*(5), 922-925.

Zinn, H. (2015). *A people's history of the United States: 1492-present*. Routledge.

---

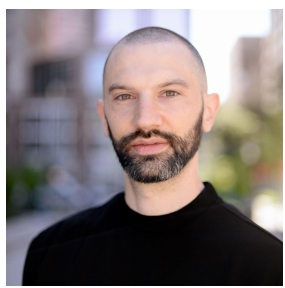

### Michael Lipset

Lipset is a scholar, author, educator, artist, creative producer and Hip-Hop head. His work sits at the intersections of critical arts pedagogies, liberatory UX design, education change, pushout re-engagement and social justice. He holds a Ph.D. in culturally sustaining school change from McGill University and an Ed.M. in the Arts in Education from the Harvard Graduate School of Education. He is a Course Lecturer in the Department of Integrated Studies in Education where he teaches on the topics of digital media and technology in education. He most proudly serves as the Director of Social Impact for the High School for Recording Arts and as a Researcher at the Center for Policy Design.

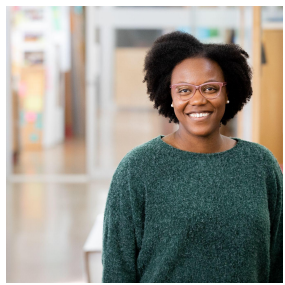

### Jessica Brown

Jess is an organizer and educator dedicated to creating education models where Black and brown learners can thrive [from cradle to grave]. She is a Designer and Lecturer at Stanford University's Hasso Plattner Institute of Design (commonly known as the d.school). Her work focuses on the intersections of K12 learning, neuroscience, culture, and design. Throughout her career, Jess has worked in coalition with communities to dismantle barriers and bend the future towards liberation. She has taught youth and adults in out-of-school programs, led community-centered design processes, and co-created organizations that hold institutions accountable to equity, students' wellbeing, and systemic change.

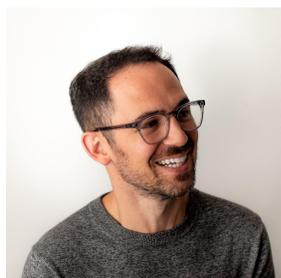

### Michael Crawford

A relentless learner with 14+ years of experience in academia, nonprofits, and startups, Michael J. Crawford is constantly connecting and creating to help people activate their best selves. He's the creator and host of the micropodcast consideranew, and he's been an invited mentor, presenter, facilitator, author, and podcast guest, exploring wide-ranging topics, such as education, technology, entrepreneurship, and the future of learning. Michael earned a BA in Psychology from the University of Michigan, an MS in Sport Psychology from Michigan State University, and a PhD in Educational Psychology from the University of Kansas.

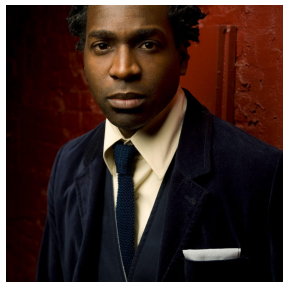

### Kwaku Aning

Currently I am the Director of the Center of Innovation and Entrepreneurial Thinking at the San Diego Jewish Academy but I have had the privilege to work in various roles within public charter and private schools over the past 19 years within education. In my current and previous roles I have had the privilege to work with students on various STEAM and Project Based Learning projects utilizing robotics, artificial intelligence, augmented reality, virtual reality, and projection mapping. One of the highlights of this work occurred in the fall of 2017 when I produced four short 360 video documentaries for the United Nations on Fijian youth and the effects of climate change.

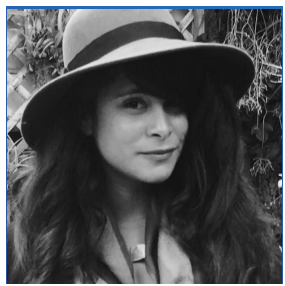

### Katlyn Turner

Dr. Katlyn Turner is a Research Scientist within the Space Enabled research group. In that role, her primary research includes work on inclusive innovation practices, and on principles of anti-racist technology design. She additionally mentors students, works on proposal writing efforts, and helps to communicate the team's work. Dr. Turner earned her PhD in Geological Sciences from Stanford University, where she researched novel nuclear waste forms. From 2017-2019, Katlyn was a postdoctoral fellow at the Project on Managing the Atom & the International Security Program at Harvard Kennedy School's Belfer Center for Science & International Affairs, where she researched environmental and socio-political impacts of nuclear energy. Dr. Turner additionally holds an M.S. in Earth & Environmental Sciences from the University of Michigan, and a B.S. in Chemical & Biomolecular Engineering from the University of Notre Dame. Dr. Turner is passionate about issues of diversity, justice, inclusion, and accessibility within society-- particularly in higher education and within STEM employment sectors.

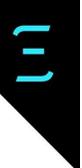

# 3. Creativity and AI

**Opening remarks** by Ramya Srinivasan (AI Researcher at Fujitsu Research of America Inc.)

"Art is unquestionably one of the purest and highest elements in human happiness. It trains the mind through the eye, and the eye through the mind. As the sun colors flowers, so does art color life." – John Lubbock

In the book "The Art of Civilization", Dr. Didier Maleuvre, a professor of French and comparative literature, states that creativity not only reflects culture, but also influences and shapes society. As philosopher Sean Dorrance Kelly argues, creativity is not just novelty---creativity should trigger innovation, and eventually find a place in the society. This reasoning also applies to AI art as echoed in a recent panel discussion concerning the "Future of creative AI", an event co-hosted by Art-AI festival, EVA 2021 (the international computer arts society's annual conference) and the Leicester Computer Arts Archive. Panelist Ernest Edmonds, a pioneering computational artist, argued that creative AI should improve the quality of life at a large scale. Elaborating on this point, the speaker envisioned an interactive process for future creative AI systems---involving "action, response and influence". All of these three features are applicable to both humans and AI involved in the creative process, suggesting thereby a collaborative co-creative ecosystem. While terms like "action", "response", "collaboration" and "co-creation" may not be new in the context of human-AI systems, the notion of "influence" is perhaps beginning to gain importance in the field of computational creativity. Here, influence not only refers to how the created work can shape society in terms of changing culture, people, policies, and relations, but also in terms of the value they contribute to the society.

Speaking of influence, perhaps, one of the strongest ways in which computational creativity can bring about societal change is by the creation of ethically grounded art. Indeed, artists are exploring this space in a variety of ways. Be it the work of transmedia artist Stephanie Dinkins who explores if AI art can help spark new dialogues about social equities, or artist Rashaad Newsome whose works asks the audience to examine their role in a culture shaped by oppression and discrimination, engaging new ways are being investigated in pursuit of enhancing social awareness and fostering empathy.

When we talk about society, empathy, and awareness, it doesn't just concern people, but also relates to our environment. Thus, when examining the influence of computational creativity, one cannot discount environmental impacts. Of late, there have been rising concerns with respect to the [ecological cost of Crypto Art](). As digital arts curator Melanie Lenz (another panelist at the Future of Creative AI discussion) succinctly summarized, ethics, empathy, and sustainability are likely to be some of the important aspects governing the development of future creative AI systems.

The future of computational creativity looks both exciting and challenging. As creative tools open the doors for non-domain experts to create artworks, it also increases the chances of creating artworks that may be biased. For example, as new media artist, critic, curator, and writer [Ellen Pearlman]() explained in a [recent CVPR workshop]() panel discussion, many cultures do not have digitized versions of their artworks, leading to selection bias in the datasets. Furthermore, as many creative AI systems do not involve domain experts (e.g. art historians, anthropologists, etc.) in the pipeline, issues of bias only get compounded, remarked Ellen. With creative AI applications being employed across diverse use cases, the long term influences of these technologies on society remains to be seen.

"Art doesn't have to be pretty; it has to be meaningful" - Duane Hanson

---

**Dr. Ramya Srinivasan**

Dr. Ramya Srinivasan is an AI researcher in Fujitsu Research of America, Inc. Ramya's background is in the areas of computer vision, machine learning, explainable AI, and AI ethics; with some of her research spanning diverse application areas such as creativity, finance, and healthcare.

# Go Deep: Research Summaries

**Creativity in the Era of Artificial Intelligence**

(Original article by Philippe Esling, Ninon Devis)

(Research summary by Abhishek Gupta)

**Overview:** The current focus on mimicking human capabilities at the intersection of creativity and AI is counterproductive and an underutilization of the potential that AI has to offer. This paper details the various aspects of creativity from a value-creation and process perspective, highlighting where AI might be a useful mechanism to augment rather than replace our abilities.

**Introduction**

When we think of AI-based creativity today, most of our attention is drawn to the GAN-style works that are a mash-up of existing creative works be those visual, literary, or auditory. But, through an examination of the epistemology of creativity, the paper illustrates that we are over-indexing on only certain aspects of creativity while ignoring others. The paper talks about how we can better harness the potential of AI to expand the meaning of creativity and create experiences and art that is beyond the limitations that humans impose on them. This also includes the limits inherent to the current crop of AI technologies in how they are designed. An analysis of these limitations reveals that perhaps co-creativity is what we need to be aiming for and move from the notion of artificial intelligence in creativity to artificial creativity.

**Epistemology of creativity**

Creative work has significance not just because of its content but also because of its positioning in the social ecosystem that surrounds it, often highlighting societal tensions. This requires the accumulation of sufficient knowledge around that subject and then a critical mass of movement on that subject for the artwork to become relevant. There are three primary drivers for any piece of creative work: novelty, quality, and relevance.

Creativity plays several roles in society such as that of improvement, pushing us to imagine what the future can look like. It also plays the crucial role in self-expression for individuals that are a part of society. And finally, that of transformation of knowledge that helps to create innovations by combining existing tools and techniques in novel ways. The paper positions the creativity process as a selection-variation algorithm that helps us explore the space of possibilities, choosing to pursue some and ignore others which shapes societal structures.

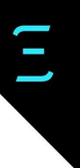

Since intelligence and creativity tend to be quite intertwined, the authors provide a crude distinction between them classifying intelligence as a convergence methodology while creativity being a divergent methodology. Taking this on as a lens, we can now think of how we might want to evaluate the creative outputs from a machine. There are many criteria such as ideational fluency (amount of answers), originality (unusualness of answers), flexibility (variation in the concepts elaborated), and elaboration (precision and details of the answers) that are used to evaluate divergent thinking tasks and which can be ported here to evaluate creativity emerging from AI.

Given that AI systems are set up to optimize against certain targets, this poses inherent limits to the space of possibilities that it can explore. The solutions within that space can have infinite variations though. There are of course similar constraints on human creative output as well: through the social structures within they are present.

**Intrinsic limits of AI for a self-contained creative agent**

In the Wallas model of creativity that spans the stages of preparation, definition, incubation, illumination, and verification, we can see the role that AI plays in the preparation phase that includes information gathering. Since AI is able to process large amounts of information and surface potential connections between distant but related items, it can aid in the creative aspect of generating novelty. Another aspect is the ability of AI systems to transform data into different representations (the idea of representation learning) which can help human creators reimagine their own knowledge base in a different light sparking creativity along different axes using the same material. However, AI ends up being quite limited in the problem finding and illumination phases of the creative process, and that might just be limitations of current techniques; nonetheless, it provides us with a roadmap for where AI can be used effectively in the creative process.

Going back to the idea of novelty, looking at traditional machine learning approaches, outliers tend to be discarded as the system is hunting for dominant patterns that can be generalized. Given this inherent limitation, approaches such as those from reinforcement learning where there is an exploration-exploitation tradeoff might be more appropriate to deploy AI. But, this requires the specifications of rewards and success functions which are still limited by the ability and perspective of the human that can place artificial constraints on the AI system.

Coming up to the concept of relevance, since that is something that can be done only in the context of a social ecosystem, this becomes challenging for an AI system to do so in a self-contained manner.

**Co-creativity as a way forward**

So, perhaps rather than emphasizing and evaluating the outputs from such systems, the focus should be on the creative process and the interaction between the human and machine artists. Thinking back to how AI can help organize existing material in a different fashion, it can serve as a creativity facilitator. This comes with the caution that human limits transfer onto the limits of AI in the way that we define the problem and evaluate the outputs from the system, hence the focus on process rather than the output. Thus, we can view artificial creativity as an emergent phenomenon where the interaction between machine and human is what matters the most. Finally, a benefit of AI being able to process large, multivariate data is that it can expand our perception of creativity beyond the dimensions that we're used to thinking in and move towards wholly novel forms of creative art.

**Between the lines**

The paper provides a well-rooted framework to reason about how we should think about AI systems in the context of creativity and how we might go about incorporating them into our creative workflows. For the most part, if we continue to push machines to emulate human creativity, we will end up with cheap imitations rather than expansions of the creative horizon. The benefits that machines bring are unique and beyond the abilities of humans; at the same time, humans will continue to be ones who can situate the work emerging from such processes to impart meaning to them. Hence, the focus in this work on the interaction and process much more so than the outputs itself is what is interesting. Certainly something for those in the intersection of AI and art to think more about.

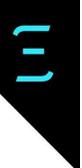

## AI vs. Maya Angelou: Experimental Evidence That People Cannot Differentiate AI-Generated From Human-Written Poetry

([Original paper](#) by Nils Kobis, Luca D. Mossink)

(Research summary by Victoria Heath)

**Overview**: Can we tell the difference between a machine-generated poem and a human-written one? Do we prefer one over the other? Researchers Nils Kobis and Luca D. Mossink examine these questions through two studies observing human behavioural reactions to a natural language generation algorithm, OpenAI's Generative Pre-Training (GPT-2).

**Introduction**

*"Science tells us that the essence of nature is empathy."*

Who wrote that? An algorithm or a human being? *Spoiler alert* That philosophical sentence was generated by an open-source algorithm called the New Age Bullshit Generator, a rudimentary example of a text or natural language generation algorithm (NLG). From autocompleting our emails to creating news stories, NGLs are increasingly present. Not only can these systems be used to create new forms of plagiarism but they can also be used to create and disseminate mis/disinformation. Therefore, our ability to decipher what is created by an algorithm and what isn't is important.

While previous studies have tested the capabilities of algorithms to generate creative outputs, there is a gap in research observing people's reactions to these outputs. Researchers Nils Kobis and Luca D. Mossink sought to fill that gap by using poetry generated by OpenAI's advanced NLG, the Generative Pre-Training (GPT-2) model, to measure and analyze:
1. People's ability to distinguish between artificial and human text
2. People's confidence levels in regards to algorithmic detection
3. People's aversion to or appreciation for artificial creativity
4. How keeping humans "in-the-loop" affects points 1-3

**Studies 1 and 2: General methodology**

Before diving into the results of Kobis and Mossink's research, let's take a quick look at their methodology. They created two different studies, Study 1 and Study 2, that each had four parts:

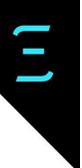

Part 1 entailed "creating pairs of human-AI poems." In Study 1, these were written by participants in an incentivized creative-writing task. In Study 2, they used professionally written poems. In Study 1, the researchers selected which creative outputs would be used for judging, something they refer to as keeping a human in the loop (HITL) while in Study 2, they tested HITL as well as what would happen if the poetry was randomly sampled (HOTL). In both studies, they used GPT-2 to generate poetry from an algorithm.

Part 2 entailed "a judgement task" modelled after the Turing test, in which a judge tries to decipher between two participants which is the machine and which is the human. In Kobis and Mossink's studies, participants acted as "third party judges" tasked with indicating which creative text they preferred. The researchers also told some participants which poems were written by a human (i.e. transparency) but kept that information hidden from others (i.e. opacity).

Part 3 entailed a financial incentive to "assess people's accuracy in identifying algorithm-generated creative text," an aspect to this study that makes it unique.

Part 4 entailed the judges indicating "their confidence in identifying the correct poem." In Study 1, there was no incentive. In Study 2, however, judges received financial incentives for "correctly estimating their performance."

Studies 1 and 2: Results

In Study 1, judges showed a preference (57%) for human-written poems over the GPT2-generated poems. Contrary to the researchers' hypothesis, "judges did not reveal a stronger preference for human-written poetry when they were informed about the origin of the poems." Results also showed that judges were able to accurately distinguish the poems about 50% of the time—indicating that people are "not reliably able to identify human versus algorithmic creative content." On average, however, the judges were overconfident in their ability to identify the origin of the poems. "These results are the first to indicate that detecting artificial text is not a matter of incentives but ability," they conclude.

While their findings from Study 1 were generally replicated in Study 2, they observed that when the machine-generated poems were randomly sampled (HOTL) vs selected by humans (HITL) there was a stronger preference for the human-written poems. There was also a notable increase in preference for algorithm-generated poems in the HITL group. Further, they found that people were more accurate in identifying the artificial poetry when the pieces were randomly chosen (HOTL).

**Discussion: Should we fear the robot poet?**

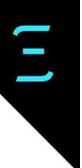

Nils Kobis and Luca D. Mossink's research generally affirms what other studies have shown: people generally have an "aversion" to algorithms, especially algorithmically generated content that people perceive as "emotional" rather than "mechanical." This could indicate that certain creative professions, like journalism, are more likely to be disrupted by AI in comparison to others, like poetry or music. Another significant finding of this research is the influence humans can have on perceptions of artificial content. "We provide some of the first behavioural insights into people's reactions to different HITL systems," they explain. This should inform discussions around algorithmic accountability. While keeping humans in the loop can help "monitor and adjust the system and its outcomes," Kobis and Mossink write, it also allows us to "crucially shape the conclusions drawn about the algorithm's performance."

Further research is required into humans' behavioural reactions to NLG algorithms. By using an incentivized version of the Turing Test, they argue, we could learn more about the use of NGL algorithms in other creative domains, such as news or social media. Kobis and Mossink also argue that creating studies comparing HITL and HOTL is necessary to produce "reliable and reproducible findings on the nexus of human and machine behavior." They conclude the article by pointing out that although algorithms' ability to mimic human creative text is increasing, "the results do not indicate machines are 'creative.'" Creativity requires emotion, something machines don't possess (yet).

**Between the lines**

When it comes to AI and creativity, there are significant issues we must contend with and questions we must ask. Like, how do AI-generated creative outputs fit within the realm of copyright and intellectual property law? Should the developer of an algorithm own the copyright of its creative outputs? Creative Commons (CC), the nonprofit organization behind the open CC licenses, says there isn't a straightforward answer. "It brings together technical, legal, and philosophical questions regarding "creativity," and whether machines can be considered "authors" that produce "original" works," wrote CC's Director of Policy Brigitte Vezina in 2020. Pending further research and debate, they argue, any outputs by an algorithm should be in the public domain.

While this research affirms that humans aren't fond of artificial text, it also indicates that we can't really tell the difference—something that will only increase as these systems become more sophisticated. Therefore, it's important that we prepare ourselves for a future in which these systems impact creative industries, both positively and negatively. Personally, I dread the day I'll be replaced by an algorithm that can more efficiently (and affordably) spit out witty Tweets and attention-grabbing headlines. Thankfully, that day is not today.

# On Human-AI Collaboration in Artistic Performance

([Original paper](#) by Alessandro Saffiotti, Peter Fogel, Peter Knudsen, Luise de Mirando, Oscar Thörn)

(Research summary by Victoria Heath)

**Overview**: How can AI systems enhance the interactions between a human and an artificial performer to create a seamless joint performance? In this paper, researchers create a new model for human-AI collaboration using an AI system to mediate between the agents and test the model in three different performance scenarios.

**Introduction**

Artificial intelligence (AI) is here to stay—the question is, how will these systems be integrated into our daily lives? Further, what will be the impact? These questions have led many to the conclusion that we must develop effective collaborative models which encourage collaboration between humans and AI-based artificial agents rather than augmentation or replacement models. This aligns with what is outlined in the Ethics Guidelines recently adopted by the European Commission which "insists that humans should maintain agency and oversight with respect to AI systems."

To further research on collaborative models, researchers Alessandro Saffiotti, Peter Fogel, Peter Knudsen, Luise de Mirando, and Oscar Thörn turn towards the creative arts and create a new collaboration model that uses an AI system as a mediator between a human and artificial performer. They test this model in three different case studies: 1) between a pianist and a virtual drummer, 2) between dancers and a virtual drummer, and 3) between a pianist and a robot dancer. Following these tests, the authors affirm their hypothesis that the AI system is capable of effectively creating a "harmonious joint performance" between the human and artificial performer.

**A model for human-AI collaborative artist performance**

The collaboration model created by Saffiotti et al is rather simple: an AI system is "used as a mediator to coordinate the performance of two autonomous agents," the human performer and the artificial performer. Artificial performer, in this case, can include any agent that "generates physical outcomes." In the following case studies, the artificial performers are a

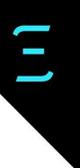

virtual drummer and a humanoid robot. The model has three features: 1) supervisory (the AI system does not generate a creative output directly), 2) reactive (the AI system analyzes the human's artistic expression and adapts the parameters of the artificial performer to match it), and 3) proactive (the AI system can set the performance parameters proactively based on the human performance). These features, explain Saffiotti et al, implement "the two directions in human-AI co-creativity," with the ultimate goal of aligning the "artistic expressions of the two performers, each one realized through its specific expressive means." The AI system used in this collaborative model contains two distinct modules:

**Features extraction**: this module uses inputs from the human performer (past and present) to generate estimates of the value of a set of variables related to the "musical expression" of the performer. In the case studies outlined below, this can take the form of keypress velocity, rhythmic density, etc.

**Parameter generation**: this module takes the variables from the features extraction to "decide the values of the execution parameters of the artificial agent." By doing so, it is providing the artificial performer with information to adjust its performance based on the musical expression of the human performer. For example, these parameters can include the intensity and complexity of drumming.

It's important to note that the AI system itself does not generate performance outputs (e.g. music) but performance parameters instead. The architecture of their AI system, explain Saffiotti et al, relies on a knowledge-based approach in which knowledge from the human performers is encoded into the system as "fuzzy logic" using multiple-input multiple-output Fuzzy Inference Systems. For example, the knowledge provided by the human performers was expressed in terms like, "If rhythmic complexity on the lower register is high, then rhythmic complexity of drums should increase strongly." All of the case studies below utilize this system with some minor tweaks.

**The model in action: three case studies**

The first test conducted using this collaborative model was between a human pianist and a virtual drummer at two public concerts in Sweden. Before these improvised jazz performances, the tempo and style were agreed upon and used as a setting for the virtual drummer, while the other parameters were set in real-time using the model. Feedback from the performer following the concerts was positive, describing the AI-controlled drummer as "human-like" and "proactive" and "surprising" in its performance. The researchers note that this feeling of "proactiveness" or anticipatory behaviour may be the result of the use of expectations in parameter generation where the system is using patterns to anticipate the artist's behaviour and adjusting the parameters accordingly.

The second test was between two human dancers and a virtual drummer at Music Tech Fest in 2019. In collaboration with the dancers, they decided which "features" to analyze and how the drummer should "react." For example, the dancers' distance from each other would change the virtual drummer's drumming pattern. The third test was between a human pianist, a robot dancer (Softbank Robotics' commercial robot Pepper), and a virtual drummer during a live performance. The robot and the virtual drummer received parameters set by the AI system (similar to the first case study).

**How should collaborative performances be evaluated?**

While the authors received unstructured feedback following each test (primarily positive), there remained questions as to how to measure these performances in order to understand the model's effectiveness. Is the "subjective experiences from humans" enough to evaluate the model?

In order to gather more structured feedback on the effectiveness of their model, Saffiotti et al created an online user study with over 90 subjects. They created two versions of a case study with the human pianist, virtual drummer and robot: a test version which used the AI model to decide the parameters for the robot's performance and a control version which selected those parameters randomly. By doing so, they aimed to test the collaboration aspect of the performance. The results from this study indicate that their model successfully aligns the artificial agent's performance with the human's, leads to a sense of "collaborative performance," and may lead to a higher perception of "artistic value."

**Between the lines**

While the case studies outlined above affirmed the researcher's hypothesis—an AI system would effectively create a "harmonious joint performance" between the human and artificial performer—they acknowledge that more research needs to be done. In particular, Saffiotti et al aim to replace their knowledge-based approach with a more data-driven approach in order to "complete the hard-written rules, or to adapt them to the artistic taste of a given musician." By making the feature extraction and parameter generation modules more precise and robust, they can further understand the role an AI system can play in mediating between a human and a machine.

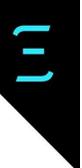

# From the Gut? Questions on Artificial Intelligence and Music

(Original paper by Florian Koempel)

(Research summary by Connor Wright)

**Overview**: Can AI actually create music, or can it solely produce such a phenomenon? Should their creations be covered by copyright law? Whether a musical purist a technological advocate, the very soul of music comes into question in this topic, especially surrounding AI creativity.

**Introduction**

Would you be able to tell the difference between an AI-produced and human-produced musical piece? Should this matter? Creatives have been investing in AI for a long time, but considerations over AI's ability to now be able to produce music has sparked a debate over whether AI can be creative, whether what it produces is actually music, and whether it has a copyright claim to what it outputs. With the author examining a brief history of AI in music and the concept of AI as an author through copyright laws in the EU and the US, AI deriving copyright protection is still not quite there yet. However, the very essence of what music is could change should this claim eventually be ratified, especially owing to whether AI can be creative.

**Creativity is key**

When mentioning AI and music, what immediately comes to mind is the topic of creativity. Can an AI algorithm charged with composing a piece of music actually be creative? Or, is being human essential to the creative process?

What is for certain, is that AI is able to produce music. Companies such as AIVA have harnessed the power of AI to programme algorithms to produce music from all sorts of different genres. However, does simply producing the music count as being creative?

As touched upon by the author, this could hinge upon the type of learning employed by the AI system. Is the process employing supervised learning (being supervised by a human), or does it draw from unsupervised learning (no human supervision), whereby the definition of both terms can be found in our Living Dictionary. With supervised learning providing the AI with much more autonomy, this is generally seen as the most likely locus for any AI creativity when producing music.

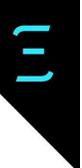

**Music is multi-layered**

The next question would be that, if AI can produce music and be creative in doing so, has it truly produced music? I use 'truly' thanks to the multi-faceted nature of music itself being touched upon by the author. Here, they note how music is not just an assembly of different sounds to a particular rhythm, but rather carries a feeling, a story and is birthed out of a particular social context. Thus, when considering whether such production of a musical piece is to have copyright protection, considerations of the essence of music comes into play as well.

To give an example, the technology housed at Malaga University in Spain by the name of Iamus is able to produce musical scores. However, some may argue that this is but a mechanical production of music, lacking a soul, feeling or emotion to go along with it. To further this, the Rembrandt Project was able to produce, through the power of AI technology such as deep learning (also defined at our Living Dictionary), an additional Rembrandt style piece. While this is a truly amazing feat, what might be argued is that this could never be a truly artistic creation due to the lack of human touch, emotion, context and the like. While such products can be produced by AI, it calls into question the very fabric of the traditions themselves.

**The legal side**

What I can now touch upon is the role that AI can play within the process of producing music and how this affects potentially copyrighting the space. Here, the author distinguishes between AI as an author (AI-generated music) and AI as a tool (augmentative AI). The main problems when trying to apply copyright to the space comes when talking about AI-generated music.

In this regard, it becomes quite difficult to distinguish who the music produced actually belongs to. Given how AI has no legal status, it cannot be attributed to the algorithm, meaning the next of kin becomes the AI designers. Then, with so many different stakeholders involved from designers, to companies, to data scientists, it becomes difficult to say which part of the outcome can be attributed to which person. Questions then come into play about the data being used in the process. For example, should this music output be based on a particular musician (such as creating a new opera piece from a particular singer), there could be legal charges placed against such a use due to the copyright placed over said singer's voice.

The situation becomes much easier when AI is simply being used as a tool, such as to produce library music (utilised for adverts and corporate events). In this context, normal copyright rules apply, such as the UK Designs and Patents Act 1988 labelling the author as the entity who undertakes the necessities to realise their creation. Nevertheless, the questions surrounding the source of the data being used for the AI to aid a composer still arises, especially if such data drawing involved reproducing already copyrighted music.

**Between the lines**

For me, AI being creative and then warranting copyright protection has to go down the line of unsupervised learning. The human guidance involved in supervised learning stifles any form of creativity being associated with the AI, with the algorithm rather being shown what to do and not being creative and coming up with its own path forward.
Having said this, I think that for an AI to ever be creative and derive copyright protection as a result, the need for originality within the unsupervised learning process must be highlighted. Without this, I would argue that AI is simply producing a piece of music, rather than creating a piece of music. Without originality, I don't think AI stands a chance at being labelled as creative. However, should this be achieved, whether what it produces actually becomes classed as music, despite not drawing on any emotion or context, is something to ponder over.

## Fashion piracy and artificial intelligence—does the new creative environment come with new copyright issues?

([Original paper](#) by Heidi Härkönen)

(Research summary by Connor Wright)

**Overview**: With AI becoming more and more independent from the human touch as time rolls on, many questions surrounding whether AI can be a designer, copyright protection and its links with creativity are being asked. Such questions within the fashion context, an industry notoriously difficult to introduce a legal copyright framework, then only get more interesting. In this sense, this paper fittingly asks; can AI actually be original?

**Introduction**

As discussed in my previous research summary, creativity and AI are starting to become more and more intertwined. With the topic bringing new questions surrounding protecting human endeavours through the medium of copyright in a very copyright-tricky field, AI throws an extremely interesting spanner into the works. Through exploring the use of AI in the fashion industry, I'll view its effects on whether AI can actually be original in this current fashion context and merit copyright protection as a result. I'll then conclude that while the road is more promising than previously thought, I think AI's inability to think 'outside of the data box' is its biggest stumbling block to merit such legal protection within the fashion industry.

In order to best contextualize AI within the fashion industry and copyright debate, it's worth highlighting the uses of AI in said industry delineated within the paper itself:

**Being used to predict future trends**
Being able to predict what consumers' tastes are is a key part of effectively designing fashion pieces. What types of silhouettes, colours and materials to use based on consumer preferences can both make effective and heavily influence a fashion designer's creative process. To do this, data mining has been implemented to uncover trends and patterns present among a fashion data set (such as consumer purchases in a country in the last year) to more easily determine what will be popular in the future.

**AI as an assistant to creativity**

AI can help to facilitate a designer's creative process through its supply of helpful tools. For example, being able to analyse a consumer's browser history to figure out which colour of shirt to bring up to better appeal to their particular interests.

**AI as an independent designer**

Probably the most poignant topic within the debate is the possibility of AI being considered a designer in itself. Through the use of general adversarial networks (GANs), AI systems have been able to produce visually similar but different images of the training data to see how customers react to such pieces before producing them. From there, deep learning can also be used to mix multiple styles and come up with some different designs.

**What does AI in the fashion industry mean for creativity?**

Within the paper, a comment I found interesting was how AI designs could be seen as creative through being similar to what a human would've created (such as a similar clothing design). Given this, it's worth briefly exploring a crucial part of creativity: originality.

**Originality**

Originality is a notoriously difficult topic to pin down in the fashion industry. Due to the prevalence of 'inspiration' within the design of fashion pieces, the line between originality and being inspired becomes very blurred indeed. However, such a concept forms a key part of whether something can not only be seen as creative but also be seen as to have AI as an author and thus be protected by copyright law.

I agree with the author as to how AI as a tool for predicting future trends and as an assistant to a designer is not to grant AI being considered as creative. Simply predicting the trends after being fed in the relevant data, while simply being utilised in another tool-like format to present products differently wouldn't help the AI's cause either.

In this sense, the author views AI as lacking the intuition to fully take its claim for originality all the way. The prominent role that data plays in the AI fashion-design process, in my view, stifles the possibility of intuiting a new fashion design process, but rather simply allows the AI to reproduce already established fashion trends and items present in the data. I believe the AI cannot quite think 'outside of the data box' enough to yet be truly considered original, such as the context in which it's producing the piece. This can be further reflected in the current thoughts on potential infringement in the fashion industry.

**Potential infringement in the fashion industry**

What must be said is that the fashion industry needs some form of imitation to function. New styles and twists on classic pieces form a large part of the fashion ecosystem, which means fashion designs already barely scrape the originality criteria even without the introduction of AI due to the similarity between every designer's work. This is to say, it becomes increasingly difficult to decipher between infringement and simply being inspired, proving a difficult problem when trying to regulate any potential copyright infringement.

One clear line agreed upon by all, at least in the European context, is that an author of a creative piece must be a human. One way to then think of it is who would be liable for a copyright infringement in itself? At this point, it would have to be a human involved somewhere in the AI's involvement in the creative process. For example, the AI's use of a data set compiled of original fashion works would mean that any AI creation is going to have to be judged to be sufficiently different from the data set in order to avoid infringement. Whether an AI can produce this difference in the form of originality and merit copyright protection from the data set used is then the key question.

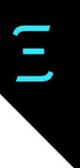

**Between the lines**

While I hold that AI is not creative in the fashion industry and copyright context, it is not solely a passive assistant either. AI is beginning to play a greater role in the fashion design process, and its road to originality and copyright protection isn't as far off as it used to be. However, the key to achieving such a feat, for me, relies on AI's ability to produce a fashion piece different from that of the data set it has been supplied not only to be regarded as creative but to merit the copyright protection such a work would deserve.

# Go Wide: Article Summaries (summarized by Abhishek Gupta)

## 2021 Emerging Trends: How Artificial Intelligence Will Affect Creative Decisions

**(Original article by [Forbes](#))**

**What happened**: Infusion of AI capabilities into software like Luminar, the Adobe Creative Suite, among others has led to a flurry of discussions in terms of what it means for artists and the work that they do. The trend has been that the creative industry is a lot more data-driven compared to before, there are lower barriers to creating more high-quality content, content creation itself has been commoditized, and the emphasis will be on more unique creations given the commoditization of common skills.

**Why it matters**: The trends identified here are important because they indicate the complementarity in the use of AI alongside human creatives. This bucks the trend compared to conversations that center on alarmist portrayals of "robots are going to take our jobs" and present more realistic scenarios of what might actually come to pass.

**Between the lines**: These trends are important to correctly identify so that the skills that are being imparted in the education system align correctly with the actual capabilities and limitations of technology. Humans will continue to play an important role in the creative ecosystem, but the bundle of tasks within their jobs are going to be transformed through the introduction of AI into various tools.

# 4. Environment and AI

**Opening Remarks** by Asim Hussain (Chairperson, Green Software Foundation)

Climate change is a complex topic but at the core of it is a simple truth. The temperature on the surface of the Earth is a function that takes as input the total amount of carbon dioxide in our atmosphere. Of course, there are other variables to that equation, but carbon dioxide is one of them and one we can control.

Climate change has always been happening on Earth, but the rate of change is now too fast for most life to adapt to, especially the way humans live now. For example, it's a lot easier for a tribe to migrate than a city. If we want to slow down the rate of change to a manageable speed, we need to reduce the amount of carbon dioxide in our atmosphere. The unfortunate truth is that not only are we not reducing the amount of carbon dioxide, we are adding carbon dioxide to our atmosphere. The amount we are adding is increasing every year.

Over the last year, a taxonomy of Green software has been emerging in the people who talk about this subject, a taxonomy that's useful in framing the solutions we can employ in the software space to tackle the problem.

**We can also frame Green Machine Learning in the context of this taxonomy.**

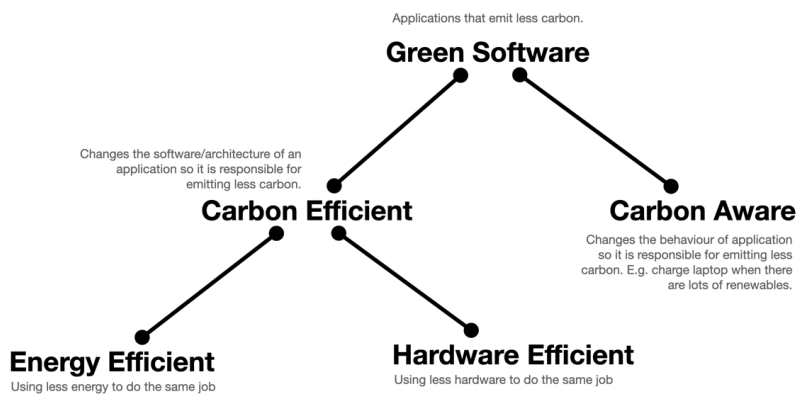

**Green Software**

Carbon is a resource. Spending it releases carbon into our atmosphere, we have a limited amount of it we can spend, so it has value.

We describe Green software as software that is responsible for [emitting less carbon](#) into our atmosphere. We think of the real-world physical emissions that an application is responsible for making and what you can do to reduce those emissions.

**A Green AI is responsible, end to end, for emitting less carbon into our atmosphere.** It takes a lot of skill to build a green application. Building a green application demonstrates a level of excellence in your craft.

**Carbon Efficient Software**

Through changes to its **code or architecture**, a carbon-efficient application uses a resource linked to carbon more efficiently.

The consumer of the software will see no change in the behaviour of a carbon-efficient application. All the features that used to work continue with no difference in the user experience at all. The software is just optimized now with respect to carbon, the same way it might be optimized with respect to performance or memory. **A carbon-efficient AI uses a resource linked to carbon more efficiently.**

**Energy-Efficient Software**

About [40% of all the carbon emissions in the world](#) today are through the creation of electricity. That's because most electricity in the world is created by burning fossil fuels, mainly coal.

Electricity is what we call a [carbon proxy](#). It's a resource linked to carbon that we can measure and optimize. To be energy efficient means you are carbon efficient. So an energy-efficient software is a flavour of carbon-efficient software.

Energy efficiency is not a new area to computing, but it's an area that has been chiefly driven by battery performance. Anything running on a battery has to be energy efficient, or the battery will drain. Thus, energy efficiency has typically never been a question applied to software run on devices with a permanent connection to an AC power supply. **An energy-efficient AI uses less energy to perform the same function.**

**Hardware Efficient Software.**

Another resource linked to carbon is hardware. **Use less hardware as possible to perform the same function**.

Where energy efficiency is a story about the carbon emissions from the creation of electricity, hardware efficiency is a story about [embodied carbon](). All hardware emitted carbon in its creation and will emit carbon in its destruction. Therefore, if our goal is to be carbon-efficient, our goal is also to be hardware efficient.

We can take many directions, from code-level changes to leverage a [GPU instead of a CPU]() to architectural changes that increase the average server utilization. **A hardware efficient AI architecture uses the least amount of hardware possible to do the same job.**

**Carbon Aware Software**

Whereas carbon-efficient involves using a resource linked to carbon more efficiently, being carbon aware is about using a resource linked to carbon more *intelligently*. The resource most often talked about with respect to carbon here is again electricity. An energy-efficient application reduces the total amount of energy it consumes. A carbon aware application behaves more intelligently, it might consume the same amount of energy but intelligently consumes it for positive carbon impacts.

Renewable energy is energy generated from clean, renewable natural resources, like the sun, the wind, the oceans. If we consume more energy than can be provided by renewable power plants, we burn fossil fuels like coal.

The challenge with renewable energy is that we can't control it. If we don't use the energy created when it's windy, we can't store it for use later. It gets thrown away. The more we can respond to the variability of renewable energy, the faster we can transition to a work that is 100% powered by renewable energy. **We need to do more when more renewable energy is available and less when there is less renewable energy available.**

Being carbon aware is about making applications that respond to the availability of the sun, of wind, of waves. Thus, creating applications more aligned with nature. A carbon aware AI behaves intelligently with respect to the availability of renewable energy. Perhaps the training was shifted to a later time when there is greater availability of renewables. Possibly the inference was performed in whatever region of the world is currently 100% powered by the sun or the wind.

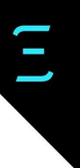

**Summary**

The field of Green Software Engineering is new. The [Green Software Foundation](#) is helping to nurture its growth, a platform where people interested in exploring this place can collaborate.

In the following chapter, we'll talk about *Quantifying the Carbon Emissions of Machine Learning.* Measuring carbon emissions is an essential first step to reducing those same emissions. Machine Learning is typically a high consumer of electricity compared to other cloud workloads, so focussing on the energy efficiency of machine learning models is an important direction to pursue. Hardware also emits carbon in its creation and destruction. We need to use hardware more efficiently, extract the most operations, most value, from each Joule of electricity that we consume. The more we can do with fewer machines, the fewer carbon emissions from creating and destroying those machines.

In the following chapter, we also discuss *Energy and Policy Considerations in Deep Learning for NLP.* If you have made your machine learning model as energy-efficient as possible, the next step is to make sure that electricity consumption drives positive environmental changes. This is where carbon aware machine learning takes the leading role as a strategy. The core challenge facing the world right now is the transition away from fossil fuels into renewable energy. The front lines in that war are happening in our electricity grids. By making machine learning, carbon aware, the industry can help drive the transformational changes we need to enable the energy transition towards 100% renewable-powered grids.

The taxonomy is a valuable framework for discussing ideas and categorizing solutions from more energy-efficient AIs to carbon aware AIs aligned with Earth's natural cycles. There are areas ripe for exploration and experimentation. We've just started to get a shared understanding across the technology space for the challenges we need to solve and the language for how to communicate solutions to each other.

---

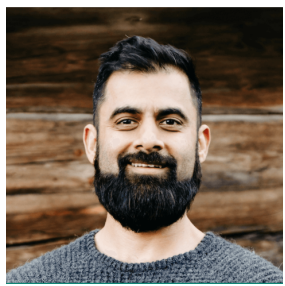

**Asim Hussain**

Asim is a developer, trainer, author and speaker with over 19 years experience working for organisations such as the European Space Agency, Google and now Microsoft, where he is the Green Developer Relations Lead.

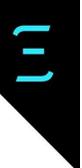

# Go Deep: Research Summaries

**How Tech Companies are Helping Big Oil Profit from Climate Destruction**

([Original paper](#) by Greenpeace)

(Research summary by Shannon Egan)

**Overview**: The tech giants Amazon, Microsoft, and Google have each set ambitious targets for climate action, including the rapid adoption of renewable energy. Yet their contracts with oil and gas producers are absent in the accounting of company CO2 emissions, even though these projects often enable more fossil fuel extraction. The support that these tech companies provide through cloud computing infrastructure and data analytics could cause global increases in emissions and accelerate the pace of climate change.

**Introduction**

Cloud computing is dominated by three American firms: Amazon (AWS), Microsoft Azure, and Google Cloud, who combine for 61% market share of the estimated $130 billion USD industry. These providers are the go-to source for computing needs across industries; the oil and gas sector is no exception.

In this report, Greenpeace unearths Amazon, Microsoft, and Google's major contracts with oil and gas producers. Among these contracts are some of the biggest names in the industry: ExxonMobil, Chevron, and Suncor are highlighted in particular. These partnerships raise the question: Can the tech companies really honour their climate commitments while being entrenched in the fossil fuel business?

Actors in the oil and gas industry are investing heavily in cloud computing infrastructure and artificial intelligence (AI) systems for data analytics, with spending expected to grow from 2.5 billion USD in 2020 to 15.7 billion USD by 2030. Cloud computing's "Big Three" (Amazon, Microsoft, and Google) seem eager to accept this work. While digital innovation could make the oil extraction process more efficient and decrease the emissions per barrel, we should still be wary of the consequences for climate change. The motives of the oil and gas companies are ostensibly to boost overall production, typically implying an increase in global emissions as more fossil fuels are removed from the ground to be burned.

**How can Big Tech help Big Oil?**

It is first important to understand: 1) What parts of the oil and gas business can benefit from technological innovation, and 2) What kind of services are the tech companies providing?

Oil and gas production can be roughly divided into 3 stages: upstream, midstream, and downstream (See Figure 1).

The upstream phase includes exploring for new wells; a highly uncertain process that requires building 3D models of rock formations beneath the earth's surface, usually with very limited data. Machine learning (ML) is used to fill in incomplete datasets by extrapolating complex patterns from more detailed data. Meanwhile, cloud computing supports seismic testing by allowing data to be stored and analyzed in real-time. These contributions can increase the likelihood of identifying new wells, but ML, in particular, requires a lot of existing data. To facilitate these analytics, stakeholders in the oil and gas industry have developed a common database known as the Open Group Open Subsurface Data Universe (OSDU) Forum. Amazon, Microsoft, and Google are all members of OSDU, and likely draw from this database to train their ML algorithms.

The midstream phase also permits interesting applications of technology, particularly through constructing an "Internet of Things" (IoT): a network that processes data from the many physical devices from the well to the customer, coordinating their behavior in real-time. For example, such a network could monitor pipeline pressure and temperature to achieve the optimal flow rate, or quickly identify when a spill has occurred. These measures help improve the efficiency of transportation, but also contribute to increased takeaway capacity: the total amount of fossil fuel product that can be transported from the extraction site towards consumers. Takeaway capacity is often the limiting factor on production; implying that if this is increased, so too will consumption.

**Impact on CO2 emissions**

This leads to an important question: how will the tech sector's involvement in the fossil fuel industry affect global emissions? To get an idea, we focus on one particular contract for which information is available. XTO Energy, a subsidiary of ExxonMobil, will apply Microsoft-developed cloud computing, ML, and IoT technology to their operations in the Permian Basin, Texas. According to a Microsoft press release, their interventions could increase production by as much as 50,000 oil-equivalent barrels per day by 2025. This represents an additional 3.4 million metric tons of CO2-equivalent emissions per year (according to Greenpeace's estimate), about one-third of the annual CO2 emissions of Luxembourg.

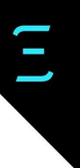

These numbers are especially striking when compared to the ambitious climate commitments that Microsoft has set, which are summarized alongside Amazon and Google in Figure 2. Their goals include powering 100% of their operations on renewable energy by 2025, including their supply chain and energy used to manufacture devices; becoming "carbon negative" by 2030, and even purchasing enough carbon offsets to compensate for all historical emissions by 2050. Including additional emissions generated by oil and gas industry contracts would make these targets much more difficult to achieve. The additional emissions estimated for Microsoft's contract with XTO amount to 21% of the tech firm's total carbon footprint. In contrast, the same margin is only 5% of ExxonMobil's expected production in 2025, for the Permian basin alone.

**Between the lines**

The process of oil and gas production is full of interesting technical problems that can be tackled with cutting-edge data analytics and cloud computing. This, along with the potential earnings, has attracted the biggest firms in the tech world to the industry.

The Greenpeace Oil in the Cloud Report does an excellent job of exposing and determining the nature of partnerships between tech and fossil fuel industry players. However, its call for Amazon, Microsoft, and Google to phase out all of their contracts with "Big Oil" may be misguided. Other companies would inevitably step in, and we cannot impose a global ban on digital innovation in the oil and gas industry. Instead, we need to devise appropriate incentive structures to ensure that this innovation does not lead directly to increased fossil fuel consumption. For this, solutions that work at the level of the entire industry will be the most effective, such as cap-and-trade or carbon taxation policies. However the public should also put pressure on Amazon, Microsoft, and Google to 1) keep track of and disclose how much "additional" fossil fuels were extracted as a result of their technology, and 2) commit to offsetting the resulting emissions for as long as their solution is in use.

## Energy and Policy Considerations in Deep Learning for NLP

(Original paper by Emma Strubell, Ananya Ganesh, and Andrew McCallum)

(Research summary by Abhishek Gupta)

**Overview**: As we inch towards ever-larger AI models, we have entered an era where achieving state-of-the-art results has become a function of access to huge compute and data infrastructure in addition to fundamental research capabilities. This is leading to inequity and impacting the environment due to high energy consumption in the training of these systems. The paper provides recommendations for the NLP community to alter this antipattern by making energy and policy considerations central to the research process.

**Introduction**

We've seen astonishing numbers detailing the size of recent large-scale language models. For example, GPT-3 clocked in at 175 billion parameters, the Switch Transformer at 1.6 trillion parameters, amongst many others. The environmental impact of the training and serving of these models has also been discussed widely, especially after the firing of Dr. Timnit Gebru from Google last year. In this paper, one of the foundational papers analyzing the environmental impact of AI, the researchers take a critical look at the energy consumption of BERT, Transformer, ELMo, and GPT-2 by capturing the hardware that they were trained on, the power consumption of that hardware, the duration of training, and finally, the $CO_2eq$ emitted as a result along with the financial cost for that training.

The researchers found that enormous financial costs make this line of research increasingly inaccessible to those who don't work at well-funded academic and industry research labs. They also found that the environmental impact is quite severe and the trend of relying on large-scale models to achieve state-of-the-art is exacerbating these problems.

**GPU power consumption**

Prior research has shown that computationally-intensive models achieve high scores. Arriving at those results though requires iteration when experimenting with different architectures and hyperparameter values which multiplies this high cost thousands of times over. For some large models, the carbon equivalent rivals that of several lifetimes of a car.

To calculate the power consumption while training large models on GPUs, the researchers use manufacturer-provided system management interfaces which report these values in real-time. Total power consumption is estimated as that consumed by the CPU, GPU, and DRAM of the system multiplied by the Power Usage Effectiveness factor which accounts for the additional energy that is consumed for auxiliary purposes like cooling the system. These calculations are done for Transformer, BERT, ELMo, and GPT-2 based on the values for the hardware and duration of the training as provided in the original papers by the authors of those models.

While there has been prior research capturing values of training such models from an energy and cost perspective, those typically focus on just the final configuration of the model rather than the journey used to arrive at that final configuration which can be quite significant in its impact. Through the experiments conducted by the authors of this paper, they find that TPUs are more energy-efficient than GPUs, especially in the cases where they are more appropriate for the model that is being trained, for example, BERT.

**Iteratively fine-tuning models**

This process of fine-tuning a model through iterative searches of the model architectures and hyperparameter values adds up to massive financial and energy costs as shown in the paper where a single iteration for the model training might cost only ~USD 200, the entire R&D process for arriving at that model which required ~4800 iterations cost ~USD450k which can easily put it out of the reach of those without access to significant resources.

Thus, the researchers propose that when a model is supposed to be further fine-tuned downstream, there should be a reporting of the sensitivity of different hyperparameters to this process to guide future developers. An emphasis on large-scale models furthers inequity by promoting a rich-get-richer cycle whereby only the organizations that have a lot of resources are able to do this kind of research, publish results, and thus gain more funding further entrenching their advantage. Tooling that promotes more efficient architecture searches is limited in its application at the moment because of a lack of easy tutorials and compatibility with the most popular deep learning libraries like Tensorflow and PyTorch. A change in this is also bound to make an impact on the state of carbon accounting in the field of AI.

**Between the lines**

This paper kickstarted a reflection in the field of NLP on thinking about carbon accounting and overreliance on accuracy as a metric for evaluating the value of results in the AI research community. Upcoming efforts such as carbon-efficient workshops at various top-tier NLP conferences have further boosted awareness of these issues in the community. The hope is that there will be sustained momentum around this as we seek to build more eco-socially responsible AI systems. Follow-on research is required, especially to make tooling more compatible with existing deep learning frameworks. Making reporting a standardized process of the research lifecycle will also help with this. Work done at the Montreal AI Ethics Institute titled SECure: A Social and Environmental Certificate for AI systems provides some more recommendations on how we can do better when it comes to building more eco-socially responsible AI systems.

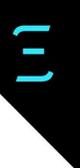

## Is AI Greening Global Supply Chains?

**([Original paper](#) by Peter Dauvergne)**

**(Research summary by Sarah P. Grant)**

**Overview**: Industry leaders are quick to champion AI as a transformative force for environmental sustainability. In this research paper, Peter Dauvergne examines their claims through a critical international political economy (IPE) lens, and finds that AI's environmental sustainability benefits for the global supply chain are overstated.

**Introduction**

*"We can't save the world by playing by the rules, because the rules have to be changed."*

Delivering these words during her TEDxStockholm talk, Greta Thunberg made her position clear: that without tougher laws, the climate crisis can't be tackled effectively.

But in an attempt to ward off tougher regulations, many transnational corporations (TNCs) are trying to promote self-governance, argues Peter Dauvergne in a paper on AI, global supply chains and environmental sustainability.

Here, Dauvergne draws on literature from international agencies, nonprofit watchdogs, journalists, business management consultants, and scholars to examine the issue from an IPE perspective. One of his major questions is whether AI's eco-efficiency gains as promoted by TNCs will contribute to significant sustainability progress–and his answer is a definitive "no."

He finds that AI is often used as a way to project an image of corporate social responsibility (CSR), but that this detracts from the issue of environmental harms posed by AI through impacts such as over-consumption, intensified resource extraction and e-Waste. He concludes that AI is not transforming TNCs into agents of environmental sustainability, and that CSR messaging can create a false sense of security, making it harder to govern industry players.

**AI's core business benefits**

Dauvergne lays the foundation for his analysis by addressing the core business benefits of AI. He references survey data from consulting firm McKinsey to show how supply chain management is one of the major business activities where AI is yielding the most benefits. Manufacturers and retailers in particular stand to profit the most from the optimization of supply chains through machine learning and intelligent automation.

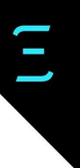

After delving into the business impacts of AI, he then examines how corporations have positioned their cost-cutting and efficiency activities as sustainable practices over the past 20 years. He observes that the value of new technology for solving global problems "has long been a key part of this CSR messaging. This is now the case with the framing of artificial intelligence as a key solution for achieving corporate sustainability."

While Dauvergne's main argument is that corporate metrics and CSR messaging are exaggerating the impacts of AI for environmental sustainability, he does acknowledge that many of the specific claims are true.

He states, for example, that the algorithms can enhance shipping by rerouting in case of bad weather and data-powered apps can cut emissions by reducing idling time for trucks. However, he asserts, these are really just micro-benefits that will only go so far when the main corporate purpose is sustainability in profit-making.

**AI and its environmental costs**

After focusing on how TNCs are highlighting the benefits of AI for environmental sustainability as part of branding, marketing and CSR messaging, Dauvergne then examines four major environmental costs of AI:

**Increased Energy and Mining Demands**

Dauvergne argues that AI will be used to increase efficiency, and history shows that efficiency gains almost always lead to higher resource extraction, production and consumption. Firms from multiple industries promote AI as a greening technology that can be used to protect biodiversity, prevent tropical deforestation, or prevent carbon pollution. In reality, computers, smartphones, robots, data centers, are driving up demand for energy and mining.

**Over-Consumptionist Cultures**

Dauvergne also emphasizes how advertising–critical to the business models of Google and Facebook–is leveraging AI to drive up consumption levels. He illustrates how deep learning techniques could increase the value of the packaged consumer and the global retail sector by turbocharging consumer demand and consumption.

**Impacts on Vulnerable Populations**

Those who benefit from manufacturing outputs and those who are harmed by it are separated by geography, observes Dauvergne. AI is increasing demand for hardware that uses metal tantalum that is extracted from areas where human rights protections are minimal. The hardware also contains rare earth elements that create toxic waste, dumped in areas that are out of sight of populations living in wealthier states.

**Tsunamis of E-waste**

The amount of e-waste is increasing substantially each year–and rose to "a weight equal to stacking up every commercial airliner ever built," writes Dauvergne, referring to previous studies. Smart products are not designed to last, which will only contribute to the "tsunami of e-waste" that is "flooding the developing world with toxic chemicals and hazardous metals."

AI sustainability: The political economy perspective
This research is an initial attempt to probe the environmental consequences of supercharging supply chains with AI. As Dauvergne notes in this paper, at the time that he conducted this research, almost no IPE scholarship had focused on this topic.

He provides several insights for future research and states that a more in-depth consideration is needed to examine how efficiency and productivity benefits can produce environmental harms. Dauvergne makes the case for research that does not take CSR messaging at face value.

**Between the lines**

This is an important article because it demonstrates that AI ethics is about more than simply executing technical fixes to make systems align with human goals. It's also a matter of exposing what different groups in positions of power value, and how by helping them accomplish their goals, AI can in turn produce potentially harmful outcomes.

There is, however, another compelling area that future research could explore that is not covered in this paper. Here, Dauvergne focuses on the role of government as a regulator. In contrast, the economist Mariana Mazzucato has asserted in her book Mission Economy: A Moonshot Guide to Changing Capitalism that wicked problems like the climate crisis require a more flexible role for government, one in which it acts as a co-creator with the private sector–and not just as an institution that "fixes markets."

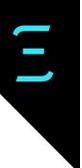

Further research, then, could help address the question of whether AI's potential environmental impacts would be mitigated by a more "mission-oriented" economic approach–one where the private sector's goals are brought into alignment with a broader societal purpose.

## Quantifying the Carbon Emissions of Machine Learning

**(Original paper by Alexandre Lacoste, Alexandra Luccioni, Victor Schmidt, Thomas Dandres)**

**(Research summary by Abhishek Gupta)**

**Overview**: As discussions on the environmental impacts of AI heat up, what are some of the core metrics that we should look at to make this assessment? This paper proposes the location of the training server, the energy grid that the server uses, the training duration, and the make and model of the hardware as key metrics. It also describes the features offered by the ML CO2 calculator tool that they have built to aid practitioners in making assessments using these metrics.

**Introduction**

It goes without saying that the firing of Dr. Timnit Gebru stirred the AI ethics community and highlighted some deep chasms between what is societally good when building AI systems and what serves the financial interests of organizations. In particular, the environmental impacts of large-scale AI systems was a point of concern. This research proposes some hard metrics that we can use to calculate the carbon impact of such systems, using the concept of CO2eq, a comparative metric that makes it easy to analyze carbon emissions from disparate activities. The tool created as a part of this research work is a manifestation of the desire to enable more practitioners to document their carbon impacts. They used publicly available data to provide the base values for these metrics which are populated in the tool for people to generate their own consumption metrics. The researchers found that the location of where training takes place and the hardware on which AI systems are trained have a significant impact on the total emissions and should be key considerations.

**Guiding metrics and results**

At the core of this paper is the idea of using CO2eq, a standardized metric used in the broader climate change community to get a handle on the carbon footprint of various activities in a way that makes comparison easier. Single metrics are not without their flaws, but for a nascent field like carbon accounting in AI systems, it is a great starting point. The first metric to calculate is the energy mix that is being utilized by the data center where the server is located. The researchers use publicly available data, assuming that the data center is plugged into the local grid where it is physically located. They find that there is a great degree of variability with some regions like Quebec, Canada having low values like 20g $CO_2eq$/kWh to really high values of 736.6g $CO_2eq$/kWh in Iowa, USA.

The researchers collected 4 pieces of publicly available data: the energy consumption of the hardware itself (GPU, CPU, etc.), location of the hardware, the region's average $CO_2eq$/kWh emission, and potential offsets purchased by the cloud provider. Keeping these factors in mind, the researchers urge practitioners to choose cloud providers wisely since different levels of renewable energy certificates and carbon offsets purchased by them have an impact on the final output. The power usage effectiveness (PUE) of the infrastructure of different cloud providers also changes the output of the calculation. The PUE is a measure of how much overhead is expended for every cycle of useful computation. In addition, as highlighted before, choosing the right region for training your model also has a significant impact, sometimes to the order of 35x as demonstrated above.

**Potential solutions and caveats**

The AI research journey is not without failed experiments and false starts. We are referring here to different experiments that are run by changing different architectures and hyperparameter values. But, there are efficient methodologies to do so: for example, one can use randomized search for finding optimal values compared to grid search which does so in a deterministic manner and has been shown to be suboptimal. Finally, specialized hardware like GPUs are demonstrably more efficient than CPUs and should be factored in making a decision.

Taking all of these factors into account, the researchers also urge the community to weigh the impacts that such changes might have. In particular, a dramatic shift to low-carbon intensity regions can lead to unutilized capacities elsewhere leading to emissions regardless of usage. In making these calculations, there are a lot of assumptions used since we don't have complete transparency on the actual carbon emissions of data centers and the associated energy mixes for the grids that they draw from. Also, the tool is just focused on the training phase of the AI lifecycle, but repeated inference can also add up to have a sizable impact on the final carbon footprint.

**Between the lines**

The findings in this paper are tremendously useful for anyone who is seeking to address the low-hanging fruits in reducing the carbon footprint of their AI systems. While the underlying data is unfortunately static, it does provide a great first step for practitioners to get familiar with the ideas of carbon accounting for AI systems. The next iteration of this tool, dubbed CodeCarbon, moves closer to what the practitioners' community needs: tools that are well-integrated with the natural workflow. The original formulation was a web-based tool that introduced friction and required the data scientists to manually enter information into the portal. The newer iteration has the advantage of capturing metrics just as is the case with other experiment tracking tools like MLFlow, enabling potentially higher uptake in the community.

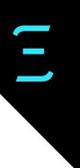

# Go Wide: Article Summaries (summarized by Abhishek Gupta)

## Hundreds of sewage leaks detected thanks to AI

**(Original article by BBC)**

**What happened**: Highlighting an innovative use of AI, the article points to research that has detected spillage of sewage from treatment plants into nearby water bodies. The system was trained on data that captured how water flowed in plants that had normal functioning and one that was erratic. Through this process, the system was able to detect places where leaks were taking place which had passed by undetected before.

**Why it matters**: When we look at the resources that are required to enforce environmental adherence policies, government agencies often find themselves understaffed and having to cut corners to do best with the resources that they have. Automating detection such that leaks can be detected and human resources can be better directed towards addressing problematic cases can help us create stronger adherence to these policies.

**Between the lines**: I am particularly excited to see how such systems generalize beyond the use case in the UK and if there are adaptations required to work in countries where the treatment plants are different from the ones in the UK. There is immense potential for developing nations that face egregious violations of environmental policies to do better monitoring and enforcement of their regulations through the use of AI.

## We finally know how bad for the environment your Netflix habit is

**(Original article by Wired)**

**What happened**: Recent focus on the environmental impact of cloud infrastructure and services delivered to us through that have woken up companies like Netflix that are seeking to get a better understanding of their carbon footprint. Using a tool called DIMPACT, they found that streaming an hour of Netflix emits CO2eq comparable to driving a car for a quarter of a mile. The primary contribution of the tool is that it provides much more accurate estimates of the carbon footprint of these services.

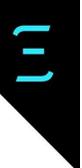

**Why it matters**: This subfield at the moment is rife with misinformation on the actual impact of these services. This research work helps to shed light on what is actually happening which is essential for us to design the right kind of interventions. This doesn't mean that Netflix doesn't have a responsibility to optimize how they deliver their services, but it helps us get a better grasp on what needs to be done.

**Between the lines**: We need to consider the alternatives that people engage in if they are not watching Netflix. Replacing your Netflix habit with walking is different from replacing it with a late-night drive in your SUV. Context and accurate estimates will help us make more informed choices.

## AI and Climate Change: The Promise, the Perils and Pillars for Action

**(Original article by [ClimateAction.tech](ClimateAction.tech))**

**What happened**: Time and again, AI has been positioned as a saviour for our climate change woes through its use in better freight allocation, precision agriculture enabling better use of resources like land and water, better weather forecasting, and optimal heating and cooling of buildings to save energy among other needs. Yet, this often underemphasizes the environmental impact of AI itself and also sometimes overstates the actual impact that the introduction of this technology can have in achieving these goals.

**Why it matters**: A careful analysis of the actual benefits that this technology can bring along with the side effects such as the high energy consumption of training and then running inference from these models should be made more explicit and transparent. As the author points out: having more enabling technical environments that normalize carbon considerations in the use of this technology is a great first step. This can help to integrate with other efforts on climate justice leveraging the expertise and progress from those initiatives.

**Between the lines**: Ultimately, the use of AI is not a replacement for other climate change efforts. It is a tool that can aid us in those efforts but should be done so under the guidance of domain experts. Sharing this expertise through grassroots collaborations are another way that such efforts can be made more potent.

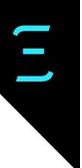

# 5. Geopolitics and AI

**Opening Remarks** by Jeffrey Ding (PhD Candidate, University of Oxford)

The possible effects of AI on geopolitics are too numerous to summarize. Since AI is a general-purpose technology, with possible applications that range from facial recognition to machine translation, it has the potential to affect so many international outcomes. Scholars and commentators regularly link advances in AI to a broad set of geopolitical effects, including but certainly not limited to international economic inequality, the rise and fall of great powers, the resilience of different regime types, and revolutions in military affairs.

The general-purpose nature of AI, however, makes it very difficult to assess the net impact of this technological domain on various geopolitical outcomes. Consider, for instance, the question of whether AI will intensify technological decoupling between the U.S. and China. Some applications of AI, such as the Chinese government's use of facial recognition for surveillance purposes, are fracturing U.S.-China supply chains. Yet other applications, including improved translation capabilities, are strengthening economic interdependence between the two countries. A comprehensive answer would necessitate calculating the net effects on decoupling from the distribution of all still-evolving AI applications.

One thing that we can be certain of is that the relationship between AI and geopolitics is bi-directional. That is, AI is shaping (and could eventually transform) how states project power, but the geopolitical landscape is also shaping the development of AI. As has been the case with past transformative technologies, social forces— people, interest groups, ideas — will shape how AI is deployed in various contexts. These forces will mold this general-purpose technology to serve their own purposes.

This chapter's selections unpack this complicated relationship. One of the articles explores how states should identify "strategic" technologies, such as AI, to gain relative wealth and power, emphasizing that surrounding international economic and political structures shape the degree to which certain technologies are more strategic than others. Another selection examines how automated decision-making technologies infringe the rights of migrant communities. Another outlines discrimination in mortgage applications and market shocks as potential risks from AI-enabled systems. Several others tackle how different countries and communities are developing their own frameworks for AI development, including China, the European Union, Mexico, and indigenous groups.

It is natural to take a system-level view of AI and geopolitics. Undoubtedly, AI is becoming increasingly central to international competition. However, taken together, these chapters remind us that even as geopolitical competition intensifies over AI development and more and more states incorporate AI in grand strategies and high-level plans, each local context is different. Therefore, sound strategic thinking requires reckoning with how these local contexts will change and reshape the trajectory of AI development.

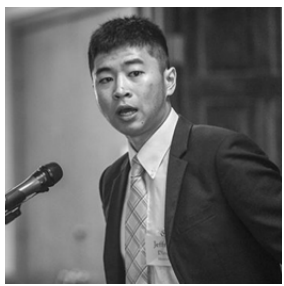

**Jeffrey Ding**

Jeffrey Ding is a PhD Candidate in international relations at the University of Oxford and a pre-doctoral fellow at Stanford's Center for International Security and Cooperation, sponsored by Stanford's Institute for Human-Centered Artificial Intelligence. He is also a research affiliate with the Centre for the Governance of AI at the University of Oxford.

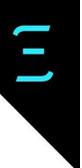

# Go Deep: Research Summaries

**Code Work: Thinking with the System in Mexico**

(Original paper by Hector Beltran)

(Research summary by Alexandrine Royer)

**Overview**: Hackathons are now part of a global phenomenon, where talented youth participate in a gruelling technical innovation marathon aimed at solving the world's most pressing problems. In his ethnographic study of hackathons in Mexico, Hector Beltran illustrates how code work offers youth a set of technical tools for social mobility as well as a way of thinking and working within the country's larger political-economic system.

**Introduction**

Hackathons have emerged as a global phenomenon, where talented youth reunite to showcase their problem-solving and technical skills. The EUvsVirus hackathon held in April of last year saw over 20 000 participants coming up with 2,000 ideas within 62-hrs on how to manage the COVID-19 pandemic. Now sponsored by governments, hackathons have emerged as sites for networking, entrepreneurial training, and to secure top prizes that provide the funds to launch quick solutions to address global or local crises.

Hector Beltran has been following the hackathon movement's evolution and what he terms "hacker-entrepreneurs" in Mexico. Beltran carried out ethnographic fieldwork from 2013 to 2017, travelling back and forth between Mexico City and Silicon Valley. At the time, the country underwent significant political changes as a newly formed leftist party had taken over the long-reining revolutionary party. Against this political backdrop and promises of rapid economic development, hackathons offered youth opportunities to carve out a future without relying on formal institutional support. As stated by the author:

*"For self-identified 'disenchanted' youth in Mexico, skeptical of the promises of social mobility by means of formal education, 'hacking' emerged as a way to make sense of their futures in a precarious state and economy and as a way to let their 'code work' intervene in narratives that had only delivered false hopes."*

**The hacker-entrepreneur**

In the opening paragraphs of his paper, Beltran describes the scene of a long line of young hackers who have gathered to take part in Hack CDMX 2015 in Mexico City. Governments and local city officials are now proud sponsors of such events. The goal of Hack CDMX is to come up in under 48 hours with a technological solution to a pressing problem within the city. Hackathons act as mini innovation laboratories whose time crunch mirrors the rapid rhythm of Silicon Valley. The youth in attendance are presented as exhibiting an ethos of hacking everything, even adding directional arrows to the door's signage.

For these youth, hacking skills go beyond the computer and are part of their outlook on society; hacking is a mindset. As encapsulated by Beltran, "these young people exhibit a sensibility for modifying, tweaking, and finding ways to exploit vulnerabilities in systems and structures, from the text on the sign to the practices of corrupt police officers."

Beltran's guiding question is to understand hackathons' continued popularity even though many projects born out of these events do not receive the funding nor support guaranteed by city officials. Narrating his experience with young computer programmers, the author introduces the term 'hacker-entrepreneur', a reflection of how youth come to realize the promises of social mobility through technology. These young hacker-entrepreneurs use code work and coding logic as a way of reconfiguring their relationships with the state and the economic institutions that benefit from their hacking. For Beltran, their code work "refers to both the ways research participants use the logics underlying coding and design principles as they develop software systems and how these logics become 'good to think with', about the institutions and systems that function as elements in state-driven infrastructures that spatialize unequal opportunities."

**Finding a place in the neoliberal economy**

During the early to mid-2010s, Mexico was poised to emerge as the next "tiger" economy due to its rapid industrial growth. The country was producing graduating classes of engineers and technologists in numbers that rivalled those of the US. Developmentalist narratives of an emerging "Aztec tiger" armed with a skilled workforce form the broader economic context behind Mexico's hackathons. Hackathons were part of the state's ambition "to keep these recent graduates busy, as potential generators of companies that would create jobs for them and their colleagues, and as the type of infrastructure that could help Mexico emerge on the global innovation stage."

Attendees are well aware that hackathons are enmeshed within governmental politics, commenting that such events were a way for officials to "further their own political agendas" through photoshoots with "poster politicians." These forward-thinking and "socially conscious" youth with entrepreneurial spirits are celebrated through photo-ops and congratulatory handshakes with city officials for living up to the ideal of self-reliance in the neoliberal capitalist economy. While Mexican politicians promoted these model entrepreneurial hackers, thousands of Mexican youth looking to gain meaningful employment found themselves disconnected from institutional support and disenchanted by governments' unfulfilled promises.

**The hacker ethos**

The Mexican hacking community is not a homogeneous group, with individuals coming from a cast of varying characters, backgrounds, and motivations for diving into code. One side of the spectrum is UNAM-educated computer science graduate Hugo, who travels two hours on public transportation to attend such events. On the other is El Pato (the duck), the iLab director and start-up entrepreneur who boasts a private school background. Within Hackathon events, individuals from contrasting socioeconomic backgrounds come to occupy the same space, share their technical abilities, and are part of a common global "hacker culture".

Hackers use the same set of tools, coding and code work to navigate institutions and critique the systems with which they interact. Within the hacker community exists a tension in following the "real" hacker ethos, the anti-corporate and covert figure, and depending on financing from tech companies. Blurring the lines of the hacker identity is the emergence of hacker schools, with programs specifically designed to connect trainees to companies by ensuring that each "batch" finds their right institutional match.

**The social-technical fabric of code work**

Anthropologist Nick Seaver, based on his own research with software developers, encourages us to view algorithms as "complex socio-technical objects". Following Seaver, Beltran demonstrates how code work is a "socio-technical fabric" that is woven within domains outside of the coding space and into the wider political and economic arena. For these veteran and budding hackers, "the code work allows them to think and act alongside the 'the system'." Hackers think of their own economic disposition through coding terms, with independent contractor Hugo referring to his work arrangement as "loose coupling". Loose coupling is defined as "a robust way to write code where data structures can use other components' interconnected systems without knowing the full details of their implementation."

Hugo's use of the term "loose coupling" signals an awareness of his autonomy and replaceability within the tech ecosystem. Loose coupling reflects the "aspirations of hackers to benefit from their involvement with tech companies versus the practice of being exploited by them." Hackers, cognizant of the labour market's volatile conditions, seek to exercise a degree of control over their labour through the way they enact and envision codework.

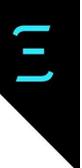

**Hacking — a toolbox and a mindset**

Bringing us back to the hackathon, Beltran observes how the team behind the winning project, a cycle-route sharing system, expressed their certainty that the app would likely never be implemented. Despite these unfulfilled promises, many of the same individuals kept returning to hackathons, choosing to participate for "the code work, not necessarily the 'results'." Understanding the software behind new technologies allows youth to recapture them for their own means and carve out a future within the constraints of Mexico's neoliberal economy and political climate. As summarized by the author:

*"Between the code worlds and social-political worlds, the code work gives hacker-entrepreneurs the tool kit for social critique; it gives them the tools to think with the system, whether that system is the latest software infrastructure, socioeconomic program or political reform."*

**Between the lines**

Beltran's in-depth analysis of hackathons in Mexico provides a nuanced outlook towards the possibilities and potentials offered by a new technological infrastructure that counterbalances simplistic understandings of technology as guaranteeing economic development and social mobility. For the youth taking part in hackathons, coding became a heuristic. Hacking provided an opportunity for youth lacking institutional support for social and economic advancement, and the tools for a social critique of those very institutions. Over macroeconomic narratives of technological change, such micro-cases offer culturally situated and contextual presentations of code work's impact in shaping Mexican youth's relations to their environment and identities as techno-citizens.

## 10 takeaways from our meetup on AI Ethics in the APAC Region

**(Event recap written by Connor Wright and Shannon Egan)**

**Room 1**

Being a region usually estranged from the AI discussion in the West, the true power and utility of considering such a perspective shone through. Ranging from trust in governments, the multiculturality of the region and the impact of COVID-19, AI is truly considered in a refreshing light. So, here are the top 5 takeaways from Room 1's discussion.

**The multiculturality of the APAC region has to be considered**

The APAC region contains around 70% of the global population/economy, which encases a myriad of different cultures and perspectives, especially in terms of views on what matters within the AI debate. We have seen how Europe has paid the price for having a unilateral view of AI (such as in the UK government's use of a grading algorithm). Hence, the room questioned whether there should be, or could actually be, a policy to bring the APAC region under one umbrella.

**Trust in government**

A prominent part of a multicultural approach is the faith that different cultures have in their different form governments to solve different country crises, even when not democracies. For example, South Koreans were observed to hold a higher trust in government, as seen in allowing the government to use their credit card histories to track potential COVID spreading.

The room then noted how this bears a surprising resemblance to the 'collectivist' attitude towards handling the pandemic. The government does not boast a large pool of confidence from the population, but the population still felt it necessary to trust in its decisions.

**It's easy to build AIs, but what's there to stop them?**

One benefit of a more centralised governance model is the ability and confidence this brings when having to affront AI-orientated problems. For example, the government in Beijing has confidence to shutdown AI companies if they have to, whereas other countries in the APAC region, perhaps, do not. One worry is then a potential power-dynamic being generated between countries like China and less centralised members of the APAC region.

In this case, the controlling of funding for AI-orientated companies was proposed as a way to help empower less centralised countries. However, defining what the benefits of this are in different contexts is extremely hard.

**AI is not quite established in some countries in the APAC region**

The room observed how some APAC members (such as Australia) have more time in learning how to best introduce AI into the country. At this point in time, it may be that Australia doesn't have enough data for AI to be instantiated appropriately in the Australian context. However, as necessity is the mother of all invention, should more sophisticated data storage or ways of sourcing such data be required, there certainly will be efforts to put them into place. For, what would be the worst case scenario is countries like Australia being stuck in a tech war between them and the more advanced countries of the region.

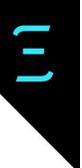

**COVID and its effects in the APAC**

One war that for sure each individual country of the region is going through is that of combatting the pandemic. One observation brought up by the room is how during war times, the liberties taken away were done so on a more temporary basis. However, it appears that the digital implementations being enacted now are here to stay. In this sense, it may not be the end of the world that people have mountains of COVID-related health data now as it decreases in value over time, but the question then remains whether it expires quickly enough.

**Some concluding thoughts**

What the room's discussion has clearly demonstrated is just how multi-faceted and rich the APAC experience of the AI landscape truly is. Different methodologies to approaching AI, the pandemic and even governance in itself proves to bring the refreshing notion of awareness to the fore. AI is going to influence every corner of the globe, which in turn means every corner has an opportunity to have their say, with the APAC region proving a particularly fruitful perspective to listen to.

**Room 2**

Room 2's discussion flowed through a series of questions, all connected by the central theme: how will AI alter our relationship to government, to industry, and to one another?  Specifically, we asked: How are companies adapting their approach as public consciousness of ethical issues in AI increases?  Is it justified to have a national database used to surveil the population?  Can cultural sensitivity be built into our AI systems to address regionally-specific bias?  All of these questions were tackled with emphasis on the implications for human rights and social justice.

**The near future of autonomous vehicles**

The advent of autonomous vehicles is both exciting and worrying.  On one hand, they have the potential to transform our daily experience and our economy in the next few years.  On the other, there are likely to be bugs in the system that cause harm to their drivers and others on the road.  Despite this danger, the EU's recent proposal for harmonised rules on artificial intelligence contains very few mentions of self-driving cars. Failing to regulate the airline industry early resulted in many lives lost, so we should move quickly to establish autonomous vehicle safety standards. The good news is – flying is now one of the safest means of transportation, thanks in part to strong regulation.

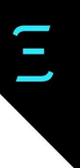

**The business case for AI Ethics**

As more people understand the role that AI plays in their lives, especially with respect to their personal data, tech companies have come under increased scrutiny. With the status quo being a lack of transparency or informed consent, many firms have tried to gain competitive advantage by positioning themselves firmly in the opposite direction. The terms "Responsible AI" and "Explainable AI" now surface frequently on company websites. Does this represent progress, or simply the emergence of a new marketing tactic?

**Surveillance and facial recognition technology**

Facial recognition technology (FRT) has become increasingly controversial due to its potential applications in national surveillance. This has pushed firms like Amazon and Microsoft to ban sales of FRT to police, while IBM has pulled out of the business entirely. But they cannot prevent clients from using technology that has already been purchased, and many firms still exchange data with governments around the globe. Public discomfort with FRT varies greatly by culture. In Singapore, GovTech's Lamppost as a Platform (LaaP) project has installed cameras equipped with FRT in many public spaces. The project has received little backlash from Singaporeans, which may be attributed to an exceptionally high trust in government.

**Addressing algorithmic bias across cultures**

When we discuss combatting algorithmic bias, we must ask: whose bias? AI algorithms are often deployed without correcting for biases present in their training data. But even if they are, the priorities are heavily influenced by the cultural and professional context of the developer.

The APAC region comprises an extremely diverse set of cultures, ethnicities, and religions. As result of geographic and cultural distance, as well as the complexity of interactions between groups, developers at powerful firms in America are unlikely to address the particular forms of discrimination present within APAC countries, such as caste-class convergence in India.

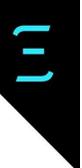

**Decolonizing AI**

Colonization has had a marked impact in shaping the APAC Region, in many cases forming inequitable power structures which still exist today. AI has the potential to reinforce this legacy, by concentrating power among those with access to large datasets for AI training, and to resources required to run computationally expensive algorithms. Prosperity generated by the AI economy should also benefit indigenous cultures across the region, even those with reduced access to technology.

**Some concluding thoughts**

The specific issues discussed all point to an important conclusion: developers of AI must identify where their systems could cause harm in the real world, and take steps to mitigate the negative impacts. The political and cultural context in which the technology is used, as opposed to where it is developed, is the most important factor in these considerations. Furthermore, the diversity of cultures and political systems in the APAC region mean that the solutions cannot be one-size-fits-all. Just as ethics vary by culture, so too should the ethics of AI.

## The Logic of Strategic Assets: From Oil to AI

([Original paper](#) by Jeffrey Ding and Allan Dafoe)

(Research summary by Connor Wright)

**Overview**: Does AI qualify as a strategic good? What does a strategic good even look like? The paper aims to provide a framework for answering both of these questions. One thing's for sure; AI is not as strategic as you may think.

**Introduction**

Is AI a strategic good for countries? What is strategic nowadays? The theory proposed serves to aid policymakers and those on the highest level to identify strategic goods and accurately interpret the situation. What a strategic good involves will now be discussed, both in terms of the importance of externalities and whether AI qualifies.

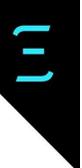

**Key Insights**

**What is a strategic good?**

The crux of the paper centres on the problem of accurately identifying a strategic good. The paper suggests that such goods "require attention from the highest levels of the state to secure national welfare against interstate competition". While this may be wide-reaching, the authors offer the following formula:

*"Strategic level of asset = Importance x Externality x Nationalization"*

The importance of the asset is based on both military and economic terms. For example, oil that fuels a country's naval fleet vs cotton being used to manufacture high-end fashion.

The externality part is about positive externalities. Here, the more positive externalities produced, the more strategic the product. Private actors are discouraged from investing in the good as they cannot receive all the positive externalities exclusively. For example, wind turbines offer positive externalities in clean energy, but private actors can't exclusively own this.

Nationalisation then focuses on how localised the externalities are. The good becomes less strategic if the externalities derived from it can spread to other countries.

**Strategic goods in terms of externalities**

The externalities brought by strategic goods can be classed in three ways: cumulative-strategic logics, infrastructure-strategic logics and dependency-strategic logics:
- Cumulative-strategic logics term how strategic goods are to possess high barriers to entry. This leads to low market investment and the need for government consent for the product to be purchased (such as aircraft engines). On the other hand, Uranium isn't a cumulative-strategic logic as a country's purchasing of uranium doesn't put up barriers to entry for others.
- Infrastructure-strategic logics note how strategic goods in the form of fundamental technologies tend to upgrade society. The diffuse positive externalities produced echo throughout the community and the military, such as the steam train in the Industrial Revolution.
- Dependency-strategic logics focus on whether extra market forces and few substitutes determine the supply of a good or not. For example, the good becomes more strategic if a nation can cut supplies of a specific good to other countries (such as lithium).

As a result, a strategic good is based on the good itself and the country's strategy with it. For example, the US's use of oil in 1941 allowed them to be the supplier of 80% of Japan's oil. Hence, when the US decided to cut the oil supply to Japan as part of the war effort, it had devastating effects on the Japanese military.

It's important to note how the good's positive externalities must be both important and strategic, as seen in this case. For example, oil was able to produce positive externalities in the form of modernising travel. However, standard-issue military rifles can be necessary for a country's military, but not strategic. They are easy to manufacture (cannot produce a dependency-strategic logic), do not have high barriers to entry, and do not change society too much. Hence, the more logics employed at the same time, the more strategic the good is.

**What this theory means for strategic goods**

A strategic asset is then where "there is an externality that is both important and rivalrous [(strategic)].". Strategic goods are no longer based on military significance, where a good would be strategic if it could be used in the war effort. Under this framework, such goods would not require a high level of attention, so they would not be classed as strategic. Instead, the important and rivalrous externalities derived from technology that can reduce $CO_2$ emissions solely in the country that uses it can be tagged as strategic.

**The strategic aspect of the development of AI**

AI then becomes an interesting case in determining whether it is a strategic asset or not. Here, there is a low rate of cumulative-strategic logics. There are no high barrier entries to AI while also possessing high infrastructural logics through its potential to modernise society. From there, a potential emerging dependency-logic between the US and China could begin to surface, with time only telling whether the US's computing power can be restricted to China. If so, a dependency-logic can be taken advantage of, and if not, China can continue to surge in the AI power rankings.

**Between the lines**

AI can certainly be classed as a strategic good in my book, but I thought it would be classified more strongly according to the formula at hand. At times, the lower barrier to entry to gain a foothold in the AI arena is often overlooked. This sobering realisation can contribute to what I believe in strongly: seeing AI for what it is.

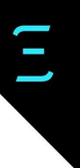

# Artificial Intelligence and Inequality in the Middle East: The Political Economy of Inclusion

([Original paper](#) by Nagla Rizk)

(Research summary by Connor Wright)

**Overview**: The Middle East and North Africa region could prove an exciting incubator for the positive use of AI, especially given how different countries have prioritized its development. However, the tendency towards economic growth over social service improvement seems to have deprived a generally youthful population of expressing their digital literacy skills. Regional peculiarities complicate this further, requiring a conducive environment to develop AI within the region becoming ever more apparent.

**Introduction**

Could the Middle East and North Africa (MENA) region be a promising region for AI to be used for good? Within the MENA region, AI can exacerbate current social and economic divides by concentrating power in the hands of the few but also help smaller businesses gain access to the market. Nevertheless, as a region that has not benefited from trickle-down economic policies, the multifaceted and multilayered inequality, stretching from inequality of opportunity associated with religion and class to economic inequality, has the potential to get only deeper. The improvement or worsening of this situation is tied heavily to the current data situation in the region, with the current attitude towards education and regional peculiarities complicating matters even further.

**Regional particularities**

The first thing to note within this scenario is the particularities of the MENA region itself. Boasting an oasis of different cultures, dialects, and traditions, it also contains a wide-ranging scale of economic prowess and sources of income. Countries such as the UAE can dedicate more resources to the development of AI systems (even having a designated government minister to AI). In contrast, others, such as Mauritania, just don't have the necessary means to do so.

Nonetheless, what unites the region through a common thread is the youthful population present within each country and the relentless pursuit of tech development and economic growth being the norm. This has meant that aspects such as economic development, inclusionary practices, political participation, and other social service development have not reached the top of any political agenda. One significant cost being education.

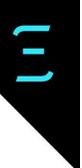

**The need for education**

An overarching acknowledgment in AI is the need for specific education surrounding the area to better prepare the population for adapting to the current times. However, despite many countries within the region's working population being included in the bracket of medium-low skilled workers (whose jobs are most at risk of automation), projects to up-skill such workers have not been fully pursued. This is certainly not the case because of a lack of digital literacy within the countries involved. Most of the populations of said countries have internet access and are familiar with its usage.

However, such skills cannot be utilized by said populations, with mass unemployment running rampant through the region. One potential solution can be to resort to entrepreneurship to take advantage of these digital skills, but a severe lack of data accessibility amongst other factors (such as non-affordable cloud data structures) proves severely debilitating for efforts to do so.

**Data inaccessibility**

With the states within the region being primarily in control of the national statistics centers that harbor most of the country's data, the lack of available open data then inhibits young entrepreneurs in these regions from being able to further their own projects. Given how most data collection is then centralized, extremely accurate local data is harder to overcome through a lack of local data repositories and initiatives. This then brings to the surface how if there is now locally centered data collection, how representative and accurate will the data be?

**Data quality**

With a centralized role being played by the state in these regions, there runs the risk of minority and more micro-level experiences being overlooked. For example, data being collected by the state about hard-to-reach rural communities (who do not enjoy a high level of internet access) is unlikely to be carried out to such great lengths to compensate for the lack of access. In this sense, it is likely that the data that is being collected on these areas through the areas that are slightly more connected are then taken as representative of the whole of these regions. Certain inaccuracies about the experience of such communities can then lead to some inaccurate assumptions about what those experiences consist of. These assumptions would then be translated into the AI design itself.

The major problem is how a country-wide AI policy or usage cannot represent all thanks to the presence of these assumptions, being labeled as "data blindness" in the paper. In this sense, the data is 'blind' through not fully representing the community it is being utilized to describe, contributing to a "data myopia." In this sense, the paper notes how national, aggregate-level methodologies for data collection cannot accurately reflect reality when it only reflects the existence of a particular segment of the population.

**Between the lines**

It is important to note, still, that the MENA region cannot be treated as a singular entity. Various countries such as Tunisia, Jordan, and Egypt are all at different legislative stages of the AI and data process, with the uneven economic development mentioned beforehand playing a part in this. In this way, only some members of the MENA region can take advantage of implementing AI positively. For me, lacking in education or digital literacy is not the most significant problem in allowing all in the region to take advantage of AI, but the lack of opportunity to express such skills in the work environment. More investment is not to be seen as the panacea but rather the construction of an environment conducive to innovation and development that will genuinely allow the region to take off in the AI space.

# Go Wide: Article Summaries (summarized by Abhishek Gupta)

## China Poised To Dominate The Artificial Intelligence (AI) Market

**(Original article by [Forbes](#))**

**What happened**: ByteDance — the company behind TikTok, is said to be valued at more than $180bn which firmly puts it at the top of the unicorns in the world. Yet, its renown in the western world is fairly limited. The rise of the Chinese ecosystem of technology including strong AI companies like Baidu, Alibaba, Tencent, and of course ByteDance makes it a strong contender for AI supremacy alongside the US. That said, we should be careful in framing the development of AI as an adversarial undertaking.

**Why it matters**: 3 key factors driving this are a massive domestic market with an appetite for automation, a clear national strategy backed by resources and willpower to realize that strategy, and strong import & export controls that limit domestic competition allowing these companies to flourish. While blindly emulating these factors is a bad strategy, adopting decisive action at the national level will be key for other countries that are looking to harness the potential of AI for their own economies. The recent NSCAI report pushes in that direction for the US (MAIEI contributed to that report.)

**Between the lines**: There are several factors like immigration policies and openness to exchanging with academic and industrial labs that will become essential cogs in the success of any national strategy to level up their AI capabilities.

## EU to launch first AI regulations

**(Original article by [Unite.AI](#))**

**What happened**: In a leaked draft document, we get a glimpse of some proposed regulations for AI in the EU. This would be the first such regulations that place clear limits on the use of AI in a wide range of high-risk contexts. Such red-lined zones include the uses of AI in credit scoring, criminal justice, and the provision of social services. It also prohibits the use of AI to manipulate decisions, behaviours, and opinions of people which would put commercial and political manipulation at risk.

**Why it matters**: Just as was the case with the GDPR, this has the potential to rewire the industry in terms of the ethical use of AI. It will push the industry towards being more careful and deliberate in their use of AI, and perhaps even steer them out of wanting to develop and test systems in these domains in the first place. The financial fines are in line with those under the GDPR.

**Between the lines**: As of now, the jurisdiction remains unclear, much more so than was the case with GDPR because data in an AI context might be used to train the AI system elsewhere and then distributed for consumption, making the questions of scope murky.

## How China turned a prize-winning iPhone hack against the Uyghurs

**(Original article by [MIT Tech Review](MIT Tech Review))**

**What happened**: An annual hackers conference centred on zero-day exploits gave a window to the rest of the world on how cybersecurity vulnerabilities are discovered and subsequently exploited for various motives. In this case, as the Chinese government caught wind of the work being done by their citizens in external fora for monetary rewards, they created an internal equivalent of a popular competition called Pwn2Own as the Tianfu Cup that encouraged researchers and hackers to uncover vulnerabilities and get rewarded for them while keeping knowledge internal to China. At one such edition of the competition, a researcher named Qixun Zhao found a vulnerability that enabled breaking into new iPhones through an exploit for the Safari web browser on the device.

**Why it matters**: This hack while the fix was being generated by Apple allowed malicious actors to further inflict harm on Uighurs in China who are already subject to a lot of surveillance and oppression by the Chinese government. While Pwn2Own worked in close partnership with companies whose software vulnerabilities were discovered, allowing them an opportunity to address those issues. This new formulation that takes this work behind the scenes creates a black market for zero-day exploits that can significantly harm the overall state of cybersecurity in the world.

**Between the lines**: Human rights abuses facilitated by vulnerabilities in consumer technologies will unfortunately continue to become a more heavily utilized avenue until we achieve better global cooperation on managing these risks and sharing information across geographical boundaries. Technology doesn't stop at the borders in a globalized world, especially when you have manufacturers like Apple whose devices are used throughout the world.

## US-China tech war: Beijing's secret chipmaking champions

(Original article by [Nikkei Asia](#))

**What happened**: A range of Chinese chip manufacturers have embarked on a gargantuan effort to trace their entire supply chain to figure out the provenance of their bill of materials. This is in response to a call to develop self-sufficiency fueled by the ongoing tensions between the US and China, especially as companies get put on the US Entity List that prohibits US companies from supplying to Chinese manufacturers. This has had ripple effects whereby companies from other countries are on tenterhooks in supplying these companies. In addition, given the immense complexity of the chip manufacturing supply chain coupled with extreme concentration of manufacturing equipment and other raw materials in different parts of the world, for the foreseeable future, a clean decoupling is going to be really hard. This is not just from the Chinese perspective, but also from a US perspective, where China still accounts for a large fraction of the supply of raw materials for chip manufacturing.

**Why it matters**: From the Chinese perspective, calls for self-sufficiency have been made by the government for many years, the current trade tensions only accelerate that trend. It also provides the efforts of the government with some cover to boost local initiatives and bring home some of the knowledge and expertise required to successfully produce chips, arguably the hottest commodity given the rising pace of digitalization. From a US perspective, the decoupling is seen as something of strategic value where they want to have "neck-choking" capabilities to achieve other strategic aims using choke-points in the supply chain to negotiate.

**Between the lines**: A part of this development trend is that we will get more distributed knowledge and potentially increased competition in the very concentrated nature of the supply chain today. On the other hand, perhaps there will also be a limited sharing of knowledge if countries begin to realize that they need to shore up their domestic competence in case the world becomes more siloed and a country loses its technological edge because of a lack of access to chips which have become a flashpoint in the technological ecosystem.

## Chinese Users Do Care About Online Privacy

**(Original article by [Protocol](#))**

**What happened**: The trope that gets tossed around quite a bit is that privacy is a second-class citizen to convenience in China. Yet, a number of cases, as highlighted in this article, show that Chinese users, companies, and even the government do care about privacy. In a momentous case, someone sued a national park for requiring facial recognition to enter the premises and won. Now, inspired by the push that the EU has done on privacy regulation, there is a possibility that China gains some privacy laws that end up being stronger than elsewhere in the world.

**Why it matters**: While the bill is still in draft mode, and it could end up being watered down, this is a positive sign that policy and regulation changes in different parts of the world can have an impact in other places. The tech governance and development ecosystem has the ability to co-evolve in a way that ultimately brings benefits to the users. It requires careful shepherding to get there though.

**Between the lines**: Having articles that highlight balanced viewpoints is important. It is easy to fall into hype and tropes that skew meaningful public discussion. When it comes to different tech ecosystems around the world, speaking with local scholars and folks on the ground will help us gain a better understanding of what actually matters to those people rather than imposing our assumptions on how that system operates.

## Understanding Contextual Facial Expressions Across the Globe

**(Original article by [Google AI Blog](#))**

**What happened**: Facial expressions are said to vary across different parts of the world. The sentiments we try to express might manifest differently on our faces, potentially depending on what culture and part of the world we grew up in. But, before this research, there was scant evidence, often with contradicting results because of the difficulty in amassing and analyzing a large dataset. Google collated a dataset with ~6 million videos across 144 countries and checked them for 16 facial expressions.

**Why it matters**: They found that there was 70% consistency in facial expressions used to convey different sentiments across the world. They also found that the social contexts within which the emotions were expressed had the strongest correlation; it held across different regions, especially so for regions that are closer together showing the pattern of spread of human culture. To check for biases that might arise in such an analysis, the researchers checked across demographic groups and also analyzed the results across video-topic and text-topic analyses to check for consistency in the findings.

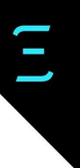

**Between the lines**: While there are many problems with relying on automated tools to make inferences about expressed emotions and what they might mean, reading through the findings of this research paper and applying them carefully and in a context-appropriate fashion can yield some interesting applications. It also showcases how machine learning can be used as a tool to strengthen results from other fields when data collection and analysis might be hard and those studies were previously limited to small sizes.

## Across China, AI 'city brains' are changing how the government runs

**(Original article by [South China Morning Post](#))**

**What happened**: The idea of smart city has been around for a long time but now with the ability to access large-scale storage for data, easy deployment of sensors, and of course AI to analyze the reams of data, cities are experimenting with the idea of a "city brain". The article gives examples of a few cities in China that have attempted to use this approach. In particular, Hangzhou has a city brain put in place by Alibaba that has led to some benefits like more efficient traffic routing. In other cities, the use of this approach has been deployed with the intention of improving the efficacy of bureaucratic processes like power consumption management.

**Why it matters**: These approaches are not without their costs. In particular, there are significant financial costs to deploy the sensor infrastructure, data storage, and compute capabilities which cost approximately 1 billion yuan in Hangzhou. In addition, there are significant costs in terms of privacy and surveillance that are not welcome by both everyday citizens and some city officials as well. This also has the potential to reshape the urban landscape as people start to alter their behaviour to preserve their privacy and freedom.

**Between the lines**: The success of such programs needs to be squared against the costs that they impose on the people inhabiting the cities. Unfortunately, some of these changes are macro and slow-moving and will only become evident in the long-term during which there might be a regime change in the city administration. Being more proactive and deliberate will be important in how these technologies are deployed. In particular, working with the residents of the city is crucial to get this right, in a way that brings them benefits where they feel the city is currently lacking.

## TikTok changed the shape of some people's faces without asking

**(Original article by [MIT Technology Review](#))**

**What happened**: In May this year, TikTok users reported that the in-app camera altered their images. On further investigation, they weren't able to find a way to turn that off. A user, for example, said that her jawline was altered, changing her appearance to be more "feminine." The author of the article found that the issue was isolated to Android phones, and when probing further with the company, they took down this change and said it was a temporary glitch.

**Why it matters**: Altering the appearance of someone without their consent and without giving them an option to opt-out is a violation of the trust and the implicit contract that users have with the app and its maker. Especially in the case of a large maker who has worldwide reach, this seems to be an imposition of patterns practiced in one part of the world (notably in China, where other apps apply beauty filters automatically) to the rest of the world.

**Between the lines**: Censorship can take many forms, and one might construe this to be a form of norms imposition originating in a particular culture. Transparency in the experiment from TikTok would have increased the trust that users have in them compared to the blowback they are now facing. Users found themselves stuck without options, and more importantly, without an explanation for what went down.

## Chinese AI lab challenges Google, OpenAI with a model of 1.75 trillion parameters

**(Original article by [PingWest](#))**

**What happened**: The Beijing Academy of AI (BAAI) has released a very large-scale multi-modal model called Wudao that is about 10% larger than the current largest model called SwitchTransformer from Google and 10x larger than the very famous GPT-3 from OpenAI. BAAI announced stellar results on many different benchmark tasks including examples of how it can beat DALL-E from OpenAI in generating images from a text description.

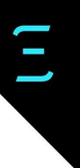

**Why it matters**: Multi-modal models that are able to operate on different kinds of inputs like images and text are becoming more and more important in AI and Wudao presents a giant leap forward. For example, when we look at some of the disinformation and hate speech on social media platforms today, they take the form of memes which are multi-modal in that they combine images and text. Automated content moderation for this kind of content has been challenging and such models could present us with viable options to address this challenge.

**Between the lines**: Unfortunately, there is still an adversarial dynamic between China and the US in terms of AI advances where developments are always touted as a race against each other. Increased collaboration will yield even better results, especially on the front of making sure that these systems are ethical, safe, and inclusive. What is also important to note here is the vastly different funding model that BAAI has compared to the privately-funded labs like OpenAI and DeepMind. It will definitely lead to different priorities in terms of what is researched and hence having greater collaboration between such entities can have a normalizing effect on the overall state of the ecosystem.

# 6. Outside the boxes

**Opening Remarks** by Cristian Gherhes (Research Fellow, Oxford Brookes University)

Artificial intelligence has been with us for a while and is only becoming more present in our lives. While some believe that AI will usher in a prosperous future for all, others point to numerous ethical concerns and potentially devastating consequences such as socio-economic inequality, algorithmic bias, and job losses. However, we assume that the development, use, and impact of AI is or will be the same across the world, but this is not the case. The rise of AI has spurred a proliferation of domestic AI strategies, with approaches varying significantly. While some countries have detailed national AI strategies, others have no formal AI policy or strategy. Some have chosen to focus on specific sectors while others have set out to tackle socio-economic and environmental challenges.

As a social scientist and economic geographer, I am interested in how social phenomena and their consequences vary across places. This includes how the development, application, and outcomes of AI vary across the world. In writing the opening remarks for this chapter, I want to remind us that context matters when it comes to AI. While the capabilities of AI may be universal, its application and use differ considerably across nations, resulting in a mosaic of approaches to AI ethics.

While there has been a proliferation of AI guidelines and principles developed by myriad organisations worldwide, there are questions about their actual impact. In "Explaining the Principles to Practices Gap in AI", Daniel Schiff and colleagues point to the significant gap in the implementation of AI principles in practice, as many of these remain rather vague. They warn that, without specificity on measuring the impacts of AI, these documents risk becoming promises without action, and propose the implementation of impact assessment frameworks to meaningfully go beyond mere speculation and "ethics-washing".

We have seen AI permeate almost all aspects of human life, and this includes intimate relationships and dating, yet this comes with important ethical implications that need to be addressed. In "Dating Through the Filters", Karim Nader explores the ethical considerations of using collaborative filtering algorithms on dating apps, flagging the problems with using biased data to make recommendations to users of dating apps. As these can result in racial algorithmic bias, Karim concludes that dating apps should perhaps remain outside the scope of algorithmic control. Are we prepared to accept the potential pitfalls of online dating? What is socially acceptable and ethical may look differently across cultures and countries.

Importantly, some places are more advanced in their AI journeys than other. But how did they get there? In my own article, my colleagues and I tell "The Story of AI in Montreal", from AI winter to the AI hype that we see today, showing how the work, ambition, and vision of a few AI researchers became reality for the many and started a global AI revolution. We highlight the importance of human agency in building the thriving AI ecosystem that we see in Montreal, and Canada more broadly. The article demonstrates the importance of collaboration between government, industry, and academia as well as the public and civil society in co-developing AI ecosystems. Critically, we highlight that Montreal's AI ecosystem reflects and upholds the values of the Canadian society, advocating and striving for the ethical and socially responsible development and use of AI.

But are these values present everywhere? And are they followed by all actors of our society? The short answer is not really. In "The Sociology of Race and Digital Society", Tressie McMillan Cottom examines how digital technology is reshaping ethnicity, race, and racism. The author highlights the problems with platform capitalism which, rather than unlocking economic opportunity for all, often fosters exclusion and preys on the vulnerable—[we need not look further than the "gig economy"](). Tressie proposes a series of solutions that can help tackle these issues, suggesting that these need to be embedded at all stages of the machine learning development lifecycle. Indeed, we need to strive to do better so that AI benefits us all.

Further, can we really trust AI to do what is best for us? Mark Ryan questions the concept of trustworthy AI and argues that AI is not something that has the capacity to be trusted. The author of the research summary offers a more positive perspective, suggesting that AI cannot be trusted "yet". Indeed, the answer is not so simple, and here, too, context and culture play a role. [A recent MIT-BCG survey tells us that 86% of users in China trust decisions made by AI, only 39% of Americans and 45% of Europeans, respectively, do so]()—still not convinced that geography matters?

In "One Map to Rule Them All? Google Maps as Digital Technical Object", Scott McQuire argues that it does, just in a different way. Scott focuses on the techno-geographical milieu created by Google Maps which has reshaped the way we experience space, time, and social life through the datafication of our physical world. The article zeroes in on the ethical implications of commercially-driven innovations by tech giants, highlighting data control asymmetries and divergence from public interest as key issues. But while Google Maps seems ubiquitous, it should not have free rein in shaping our social realities.

It was about time we brought the practices of tech giants under scrutiny. On that note, in "Slow AI and The Culture of Speed", John Tomlinson reminds us of the importance of slowing down in a world that has become obsessed with efficiency and productivity. With the digitalisation of society, everything has become available "now", anywhere, anytime. But this can have negative implications for designing AI, with our obsession with speed having adverse consequences in many instances. In the US, for example, [Amazon workers are fired by artificial intelligence algorithms without any human input](#), which are used, presumably, for the sake of efficiency. The author of the research summary argues that sometimes ethical AI means slow AI and reminds us that ultimately it is us who need to decide on the values that we want expressed and enacted through AI—a statement I could not agree more with.

So, what can we do about all these ethical challenges? In "Algorithmic Impact Assessments – What Impact Do They Have?", the authors advocate for the use of Algorithmic Impact Assessments (AIAs) to identify the potential benefits and harms of algorithmic systems. Besides ensuring a diversity of voices in shaping AIAs, they argue that AIAs need to represent harms as accurately as possible and require adequate accountability mechanisms to ensure that organisations "walk their talk". But how likely is it that AIAs will become established practice everywhere? The answer is again: "it depends"—it depends on where companies are operating and the rules that are in place to enforce these practices.

In "AI Ethics in the Public, Private, and NGO Sectors", the authors explore public, private, and non-governmental approaches to AI ethics and find significant differences across the three sectors. While NGOs show more ethical breadth, the public sector focuses more on economic growth and unemployment, whereas the private sector prioritises client-facing issues and technical fixes. The authors suggest that we may need to challenge our notions of what is desirable and possible in AI ethics practice and policy. An extension here is also examining differences across countries to determine whether context plays a role in these differences.

The development of commercial AI has been closely monitored as of late, especially with the rise of tech giants and AI-enabled solutions across industries. Financial services have particularly come under intense scrutiny time and again over the past few decades. In "Survey of EU Ethical Guidelines for Commercial AI", Jimmy Huang and colleagues evaluate the efficacy of the current EU ethical guidelines for commercial AI in financial services. They identify important gaps and highlight key risks such as the use of inappropriate metadata, discriminatory practices, and opaqueness that can potentially harm users. The ethical guidelines shows that the EU is forward-looking and ready to ensure the ethical development and use of AI solutions. Yet this is not the case everywhere, and [where such initiatives are absent, customers still bear the negative consequences](#).

Finally, while [Fei-Fei Li, former chief scientist at Google AI, claims that the benefits of AI "have no borders"](#), institutions, rules, regulations, societies, and cultures are confined by them. Geographical borders, and the people and institutions within them, directly shape how AI is developed and deployed within nations worldwide. AI is therefore not a globally uniform technology, but a global technology with national inflections, whose outcomes can vary from country to country and from culture to culture. While we celebrate our differences, it is important to ask ourselves what are the common values that we want to embed in our AI systems? While the specifics may vary across nations, we need to strive for agreement and achieve consensus. In this context (pun intended) a key question remains: can we have AI ethics without borders in a world where we seem to be increasingly defined by them?

---

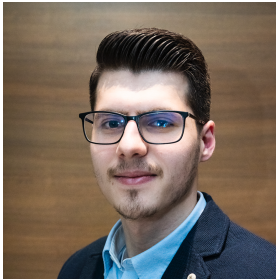

**Cristian Gherhes**

Cristian Gherhes is Research Fellow and co-director of Oxford Regions, Innovation & Enterprise Lab (ORIEL) at Oxford Brookes Business School, Oxford Brookes University. His research centres on entrepreneurship, economic development, and innovation. As a social scientist, he has a particular interest in the development and diffusion of Artificial Intelligence and implications for organisations, policy makers, and society. Cristian also leads on research projects focusing on AI adoption and diffusion in professional services firms, AI readiness, geographies of AI, and Industry 4.0.

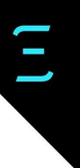

# Go Deep: Research Summaries

**Mapping the Ethicality of Algorithmic Pricing**

**(Original paper by Peter Seele, Claus Dierksmeier, Reto Hofstetter, Mario D. Schultz)**

**(Research summary by Shannon Egan)**

**Overview**: Pricing algorithms can predict an individual's willingness to buy and adjust the price in real-time to maximize overall revenue. Both dynamic pricing (based on market factors like supply and demand), and personalized pricing (based on individual behaviour) pose significant ethical challenges, especially around consumer privacy.

**Introduction**

Is it moral for Uber's surge pricing algorithm to charge exorbitant prices during terror attacks? Using automated processes to decide prices in real-time (i.e. algorithmic pricing) is now commonplace, but we lack frameworks with which to assess the ethics of this practice. In this paper, Seele et al. seek to fill this gap.

The authors performed an open literature search of Social Science and Business Ethics literature on algorithmic pricing to identify key ethical issues. These ideas were filtered into an ethics assessment – categorizing the outcomes of algorithmic pricing practices by the level of society which they impact. "Micro" for the individuals, "meso" for intermediate entities like consumer groups, or industries, and "macro" for the aggregated population. The outcomes were further sorted as morally "Good", "Bad", or "Ambivalent" from an ethics standpoint, forming a 3×3 table that can be used to generalize the ethics assessment.

For all levels of the economy, the authors identify good, bad, and ambivalent outcomes that are likely common to most implementations of algorithmic pricing. Personalized pricing presents additional ethical challenges, however, as it requires more invasive data collection on consumer behaviour.

**What is algorithmic pricing?**

The idea of dynamic pricing has been popular since the 1980s, but the practice is increasingly powerful and profitable in the age of the internet economy. While it is easy to define algorithmic pricing in technical terms, the authors identified the need for a definition that is useful in a business ethics context. The result is as follows:

*"Algorithmic pricing is a pricing mechanism, based on data analytics, which allows firms to automatically generate dynamic and customer-specific prices in real-time. Algorithmic pricing can go along with different forms of price discrimination (in both a technical and moral sense) between individuals and/or groups. As such, it may be perceived as unethical by consumers and the public, which in turn can adversely affect the firm."*

While there are a variety of approaches, all pricing algorithms typically have the same mandate: predict the consumer's willingness to buy at a given price, either on aggregate (dynamic pricing) or at the individual level (personalized pricing) in order to maximize profit. The use of cutting-edge algorithms like neural networks and reinforcement learning, as well as increased tracking capabilities via browser cookies, enable companies to do this in an increasingly sophisticated way.

**Ethical concerns**

Although algorithmic pricing primarily alters the individual consumer's experience with a merchant (micro), the ripple effects of this practice extend upwards to the organization level (meso), and further to society and the entire economic system (macro).

Evidently, firms benefit from the increased sales that algorithmic pricing facilitates, but could this practice also contribute to the common good? The answer is yes, with some caveats. Algorithmic pricing doubles as a real-time inventory management mechanism. Good inventory management can lead to reduced waste in the product supply chain, thereby decreasing both costs to the firm and the carbon footprint of the production process. The firm will enjoy increased profits, which can also be considered a moral "good" if prosperity gains are shared with the rest of society; either through innovation, increased product quality, or wealth distribution.

The major ethical dilemma of algorithmic pricing comes from the collection of fine-grained consumer behaviour data, as well as the lack of transparency around that data collection. The driving force for algorithmic pricing models, especially personalized pricing, is tracking cookies. This data, which includes browsing activity such as clicks, and past page visits, as well as personal information entered on the site, can be used to finely segment consumers according to tastes, income, health etc. in order to display the most advantageous price for the merchant.

Many consumers are unaware that this information is even being collected, and merchants do not have to ask explicit consent to use tracking cookies for pricing purposes. The onus is left to the consumer to protect themselves if they do not want to be targeted for personalized marketing. This data collection creates a massive informational advantage for companies, which may offset the price advantage that a consumer can gain due to the ease of searching online. It also limits a consumer's ability to plan future purchases, as prices constantly fluctuate. These ethically "bad" features of algorithmic pricing may limit the economic freedom of the consumer, even if the moral "goods" tend to enable more choices.

Other outcomes of algorithmic pricing fall into a grey area. Surge pricing is effective at managing demand to ensure that services/goods can be delivered to all who need them, but this method occasionally creates a trap of choosing between forgoing an essential service or paying an exorbitant price.

Ultimately, any application of algorithmic pricing requires a case-by-case treatment to ensure that the good outcomes are enhanced, and the bad are appropriately mitigated.

**Between the lines**

Algorithmic pricing is already transforming our economy, and we need to adapt our understanding of economic systems to one where price reflects not only the value of the good but the consumer's perception of that value.

The problem of ensuring ethical algorithmic pricing has intersections with many existing domains: privacy, data sovereignty, competition law, micro and macroeconomics. In some cases, existing laws and principles from these related fields can be extended to address algorithmic pricing. However brand new incentives and regulations specific to algorithmic pricing are also needed. For example, policymakers should investigate limiting what time frame the information can be stored for, as well as granting consumers "the right to be forgotten" so that their ever more detailed consumer profile can occasionally be erased.

# Disaster City Digital Twin: A Vision for Integrating Artificial and Human Intelligence for Disaster Management

**(Original paper by Chao Fan, Cheng Zhang, Alex Yahja, Ali Mostafavi)**

**(Research summary by Connor Wright)**

**Overview**: With the potential for a climate crisis looming ever larger every day, the search for solutions is essential. One such solution could then be found in the construction of a disaster city digital twin, allowing governments to simulate the potential effects of different disasters on physical cities through running simulations on their digital replicas. Doing so can help make more tangible the effects of a climate crisis, whether it be through flooding or unusual weather. AI, as a result, has a key part to play in making this a reality.

**Introduction**

Ever wondered what your digital twin city would look like? Well, the clue is in the title. A digital twin is a digital representation of a city, which the authors then utilise to explain how to integrate disaster city-modelling and AI in order to better predict potential disasters — for example, the effects of a climate crisis on the city. In order to do this, the improvement of information and communication technology is essential, with AI playing a key role. In order to best capture the paper, I'll elaborate on 5 sections: the first is the importance of simulation. After that, I launch into the 4 aspects of the simulation process elaborated by the authors. I then offer some concluding words on how such modelling can play a key role in not only forecasting how to keep cities safe from a climate crisis, but also how to best keep our future cities safe and prepared.

**The importance of simulation**

How could such simulation actually help utilise AI to fight a climate disaster? Well, here are my 3 reasons I've taken from the text:
- AI will help improve the effectiveness and efficiency of disaster management, mainly seen through its ability to scan large amounts of data and report back in real-time.
- Disasters are super dynamic and so are constantly evolving and changing. This real-time data collection thus, becomes even more important.

- Thanks to this ability, AI will help decision-making, resource allocation and coordination through running simulations as part of the disaster city digital twin (DCDT), running off the data it has collected.

One aspect that the authors mention, which I believe lines up really well with what we do at MAIEI, is how "there is a dire need for interdisciplinary convergence towards a unified vision" (p.g. 2). Here, the DCDT model serves as a common output for multidisciplinary research to converge and aim towards. In order to confront a potential disaster effectively, the use of resources and services from all different parts of society will be required, hence, making collaboration between these different disciplines, paramount.

The authors identify 4 different aspects in this DCDT simulation process:

**Sensing the data**

Multiple source data sensing will consist of collecting data from multiple different sources to have a clearer picture of what's going on in the digital city when a disaster takes place. Having a variety of sources then presents a higher probability of information being actualised in real-time, whether it comes from local fire departments or residents' social media accounts.

To do this, the remote sensing used takes advantage of UAV and satellite technology to paint a picture of how the city was before, during and after a disaster. Thanks to AI technology's scanning ability, objects having changed locations can then be flagged and sorted into relevant regions where such change is taking place. For example, the AI could identify how a building now no longer exists in the new set of images coming in, using its geolocation to then label its relevant region. This can potentially be represented through a human instructing the AI to compare the two images and then produce a similarity score, with a low score being down to the building collapse.

**Collecting the data**

In order for this to be possible, the data needs to be collected. Such data comes in 3 different departments:
- The aforementioned remote sensing with the UAV and satellites providing images on the current state of the disaster city in the simulation.
- Social sensing then focuses on social media accounts of local residents, gathering data on societal impacts due to infrastructure disruption and any personal emergencies.

- Crowdsourcing then makes sure that data is coming into the AI from different locations. This allows the AI to analyse the most up-to-date data and produce the most accurate analysis, which the CIEM in Norway is taking full advantage of.

It almost goes without saying that this collection raises some serious privacy concerns. However, such discussion deserves a whole other research summary, so I'll move on to the analysis of the data for now.

**Analysing the data**

Due to the focus on crowdsourcing, data will be coming in thick and fast to the AI, as well as in all types of different formats, such as social media posts, images or even audio. So, part of the analysis process is to homogenise the data in order for the AI to better cope. Such analysis could consist of the following:
- A dynamic network analysis examining the type of coordination being done between different actors.
- A meta-network analysis extracting what each actor is doing within the simulation. For example, the previously mentioned dividing up of images into regions through using natural language processing to scan street names.
- The above-mentioned image-tracking can then better identify the different infrastructures being lost or moved.
- The already touched upon the use of social analysis social media being used to understand human behaviour during a disaster by providing real-time reactions.

**Decision-making based on the data**

Having now collected the data, how are decisions to be made as a result? Here, a Serious Game Learning Environment is developed based on the digital city, proving to be the " use of scientific game-engine simulations of real-world phenomena to solve a problem.". Here, what can be seen is how different areas are affected infrastructurally (such as flooding due to global warming), which can then allow researchers and developers to cater to the different needs of those regions. Resultantly, the development of common areas of interest in disaster prediction can be established, better allowing for multi-actor cooperation. For example, the simulation of the disaster city allows actors to see how their decisions can affect not only the situation and the people in danger but also other actors themselves.

**Between the lines**

I'm sure that we've all found predicting the future pretty hard at times given the current pandemic, even thinking ahead to the next day! When trying to forecast for years in advance, this difficulty then multiples further. However, what DCDT aims to do is give us a better chance to do so. Already being heavily conducted in China, DCDT could hold a key role in being able to accurately predict how we are to go about keeping our future cities safe from a potential climate crisis and prepared should one occur. Whether it be assessing how to guard against tsunamis or preserve wildlife, AI's role in this process is not to be underestimated, both in humanistic and environmental terms.

# AI Ethics in the Public, Private, and NGO Sectors: A Review of a Global Document Collection

**(Original paper by Daniel Schiff, Jason Borenstein, Justin Biddle, and Kelly Laas)**

**(Research summary by the authors)**

**Overview**: While recent studies suggest that a global consensus about AI ethics is emerging, this paper finds meaningful differences across the public, private, and NGO sectors. Our team evaluated 112 AI ethics documents from 25 countries, scoring the presence or absence of 25 ethical concepts, the documents' level of engagement with law and regulation, and how participatory the authorship process was. Overall, NGO documents reflect more ethical breadth, public sector documents prioritize economic growth and unemployment, and private sector documents emphasize client-facing issues and technical fixes.

**Introduction**

Is there global agreement on which ethical issues surrounding AI are most important and how these issues should be understood?

It's an important question. How developers, policymakers, publics, and other key actors who are shaping the future of AI think about ethics determines which issues leading decision-makers emphasize—and which issues they don't. Agreement across different stakeholders could facilitate responsible innovation and better domestic and international coordination. Disagreements could lead to conflict over the best course for AI research, development, and policy. Yet agreement may be superficial or even misplaced, while conflict can be important and productive. Understanding consensus and dissensus is thus critical and some of the best research to date suggests a global consensus is emerging around 5-10 AI ethics issues deemed most important, such as bias, transparency, and privacy.

We put the question to the test, looking specifically at differences across public, private, and non-governmental approaches to AI ethics. To do so, pairs of two coders from a team of four researchers scored the importance of 25 unique ethical concepts across 112 AI ethics codes, principles, frameworks, and policy strategies, resulting in a new open-access dataset, the AI Ethics Global Document Collection.

We find that private-sector documents reflect less ethical breadth, are less participatory in their creation, and are less engaged with law and regulation. Public sector documents are disproportionally focused on economic growth and employment concerns. NGO sector documents reflect the most ethical breadth overall, addressing issues like cultural sensitivity and human-human interaction that are often neglected. Finally, there are other topics, like the psychological impacts of AI, that are rarely covered in AI ethics documents from any sector, a finding that may warrant further attention.

These differences may result from how organizations frame their own responsibilities, how they interpret and trade off different ethical concepts or goals, and whether they rely on individuals internal to their organization or more diverse voices. Yet regardless of the cause, understanding differences in ethical priorities and interpretations is important to shaping the future of AI ethics practice and policy positively.

**A Summer for AI Ethics Principles**

Those interested in AI ethics know that the last five years have seen a frenzy of action, as companies, governments, civil society organizations, and academics offered up their own vision of AI ethics. A wide variety of documents—AI ethics codes, guidelines, principles, frameworks, and even policy strategies— seek to define the problems and (sometimes) propose solutions to address AI's social and ethical implications.

Despite optimism that this newfound attention holds promise for the future of responsible and ethical innovation and policymaking, some have cautioned that AI ethics principles are currently "toothless." According to that view, AI ethics may be trapped in ethics-washing and window-dressing, perhaps serving as a strategy by the private sector to avoid costly regulation. As others have also argued, we suggest that what is needed is to translate the principles to practices and to ask deeper questions about the socio-technical possibilities for ethical AI beyond common ideas of algorithmic fairness and transparency.

In our related research, we considered how to measure the efficacy of AI ethics documents, what kinds of impacts they might have, and what motivations characterize their creation. We argued that motivations should not be evaluated in binary terms but are rather a complex combination of efforts to strategically inform internal and external changes, genuinely drive responsible outcomes or further competitive goals, or even signal one's own leadership. Most importantly, we believe the discourse around AI ethics is worthy of study, including because it drives critique. Ethics principles are playing a fundamental role, for example, in how EU and US regulation is developing, and in how companies are devising impact assessments.

It is important to consider then, do members of the public, governments, different cultures and regions of the world, large and small companies, and so on, agree about what AI ethics means and what is important to prioritize?

**Measuring AI Ethics in the Public, Private, and NGO Sectors**

Our study sought to tackle part of this question, namely: Is there agreement or disagreement across various AI ethics documents produced by organizations in the public (e.g., governments), private (i.e., companies), and NGO (e.g., civil society, academic) sectors?

To help answer this question, pairs of members of our team of four (a policy scholar, two philosophers, and an information scientist) manually coded over 4000 pages of AI ethics documents published between 2016 and the middle of 2019. Our final sample of 112 English-language documents (including official and high-quality unofficial translations) comes from 25 countries. While prior research in this field has sought to reduce the complexity of AI ethics, for example, by arguing the discourse can be understood according to 5-10 ethical concepts, our work embraced the complexity by analyzing 25 distinct ethical topics ranging from concepts such as accountability and inequality to privacy and workforce diversity. Two coders scored each document and topic as either absent (0) or as a minor (1) or major (2) topic, allowing us to quantify the presence, relative priority, and overall ethical breadth and depth of ethics topics across documents and organizational sectors.

**Consensus, Omission, and Disagreement in AI Ethics**

This fine-grained approach confirms that some topics are shared priorities across sectors, including transparency, fairness, privacy, and safety. Notably, there is also a consensus in which topics are neglected or omitted, such as concerns about existential risk and cultural sensitivity. These omissions are worthy of attention and highlight the fact that paying attention to consensus alone can mask important dimensions of AI ethics.

We found that certain topics are ranked and prioritized differently across sectors. For example, public sector documents are disproportionately focused on issues related to economic growth and unemployment, while the private sector pays little attention to these issues. This may indicate that governments are disproportionally focused on innovation and economic issues, potentially to the exclusion of other important topics. Those concerned about how the financial incentives of the private sector influence AI ethics may thus also find it important to think more about the public sector, especially in light of forthcoming regulation.

Our results indicate that the NGO sector, as defined, has the broadest and most inclusive treatment of AI ethics. While private and public organizations pay little attention to how AI can lead to psychological impacts or influence human-human relationships, NGOs do raise these issues. NGOs also emphasize accountability, suggestive of their role as an external watchdog, attempting to influence the other sectors. Another key finding is that private-sector documents reflect the least ethical breadth overall, with greater attention to areas where there are ostensible technical 'fixes' to AI ethics issues, like privacy and explainability.

**Participation and Engagement with Law**

We also coded documents for the degree to which there was diverse and public participation in the creation of documents, for example, through public hearings, workshops, and multiple authors. Public and NGO documents are far more participatory in terms of author representation. Approximately three-quarters of public sector documents are produced through an 'open' or 'semi-open' process according to our coding, whereas less than a quarter of private-sector documents are open or semi-open. Similarly, we coded documents for their level of engagement with issues of law and regulation, which may suggest whether organizations have considered how to implement formal regulation in practice, versus favoring self-regulatory strategies. Private sector documents are again less engaged with formal regulation and law.

**Learning from Sectoral Differences**

**What can we learn from these differences?**

To begin an answer, our paper draws on the notion of "ethical frames," defined as a set of grounding assumptions about ethical problems and solutions. One explanation for our findings is that organizations may think that they are only responsible for a subset of issues—such as private companies wishing to avoid harms to customers, or governments aiming to maximize the economic growth of a country. Organizations may also be pulled between pro-social and competitive goals and may make assumptions about who is worthy to define the problems and solutions surrounding AI ethics in the first place.

In our view, these beliefs, assumptions, and the ethical frames they constitute are worthy of examination. For example, the roles and responsibilities of organizations can be questioned, expanded, and informed by the collective need to shape AI ethically. It may be important to seek more voices and challenge our notions of what is desirable and indeed possible in AI ethics practice and policy. Further, whether ethics and innovation are necessarily in tension with one another is not a settled matter. In sum, we suggest the trajectory of AI ethics need not be locked in stone—more possibilities can be revealed by challenging the terms of the debate itself.

**Between the lines**

There is much still to understand. Our analysis looks only at AI ethics documents, not the broader body of work done by organizations, nor at differences across regions, for example. Further, there are many ways to aggregate, interpret, and define AI ethics concepts, and these subtleties can have important consequences, as our study shows. Going forward, much more needs to be understood about the kinds of impacts AI ethics documents and the principles articulated in them are having. Yet despite and perhaps even because of contention and criticism surrounding AI ethics principles, the associated discourse matters and will continue to matter as the AI community sets a course for action.

## One Map to Rule Them All? Google Maps as Digital Technical Object

**([Original paper](#) by Scott McQuire)**

**(Research summary by Alexandrine Royer)**

**Overview**: Few among us can now navigate unknown spaces without relying on the assistance of digital maps. Scott McQuire reveals how Google Maps operates as a digital technical object that works to reconfigure our understanding of time, space and contemporary social life.

**Introduction**

In the early 2000s, we all remember entering our address into Google Earth and watching in wonder as the planet would tilt and turn and eventually zero into our neighbourhood. Access to satellite imagery via the Internet allowed users to discover remote parts of the globe that were previously little known, shrinking the distance between our locals and faraway places. Digital maps have reconfigured how we navigate and explore urban areas to the extent that Google Maps defines our understanding of space more than we do it. For Scott McQuire, Google Maps has become a key "socio-technical 'artefact'" that works to "reconfigure the nexus between technology and spatial experience in the 21rst century".

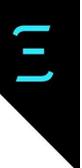

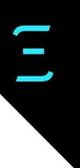

One of the early lessons of geography is that maps are by no means neutral objects. They carry assumptions about the world order in their distribution of borders, territory and spaces. As noted by McQuire, "a map is less the representation of a pre-existent world but constitutive of world-view." While mapping was once the task of geographers, cartographers, and other academic elites, the rise of digital maps has led to what Crampton terms 'populist cartography'. Digital platforms have enabled a reinvention of mapping aimed at the general consumers. Online geomedia, such as Google Maps, carries for McQuire, "distinctive lessons for how we might understand the implication of media technology in the remaking of contemporary social life."

**Mapping success**

McQuire points to four factors that contributed to Google Maps' position at the top of the digital mapping market. For one, Google already had a running geospatial data visualization tool through Google Earth, with the software purchased from Keyhole Technologies. Satellite imagery was no longer restricted to military personnel and space agencies. Its easy online access contributed to Google Maps' popularity while feeding the data needed to improve its maps. The second factor was Google's adoption of a 'participatory strategy' through opening Maps Application Programming Interface (API) to developers back in 2005. The release of Maps API meant that it could be integrated into external websites, boosting third-party development and securing Maps' as the largest maps provider in the world. Another factor was the addition of Street View to Maps in 2007. Through Street View, photo images of passing streets becoming "an exclusive and proprietary data source that is now fundamental to Google's mapping capabilities." The fourth factor was the launch of the Google maps app in 2008, initially featured as the default map on iPhone and later on Androids.

The trajectory of Maps development shows how the platform is a "technology in motion." It also reveals Google's modus operandi of "launch fast and work out any problems later." The release of Street View came with a set of concerns over Google's ownership of data and privacy. Google Maps further presented a change in the temporality of the map. Physical maps and directories can become out-of-date as soon as they are published. With digital maps, changes in urban environments are reflected in real-time. For McQuire, "map platforms are symptomatic of changing social relations in which speed has become paramount. To put it another way, the temporalization of the map as a networked platform is symptomatic of the digital technical object".

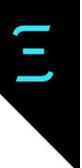

**Changing the digital ecosystem**

Google Maps' ubiquity as the online mapping tool was not always a foreseeable guarantee. It had to confront rivals such as MapQuest, Amazon, Yahoo and Microsoft in the 'map wars.' Aside from the factors previously listed, Google Maps' success can be attributed to its proprietary internal software platform, Atlas. Atlas is the 'ground truth' on which additional data streams can be integrated, such as aerial, satellite imagery and crowdsourcing features (e.g. suggest an edit). Google's purchase of the verification software reCAPTCHA in 2009 also serves to interpret images from Street View and update the information back into Google Maps.

Google Maps impressive evolution as a mapping device has made the platform an engrained part of the digital ecosystem. It is the platform through which other software operations are built upon, creating an additional source of revenue for the multinational. Maps' success also rests on the synergy between advertisers and internet users, who can 'freely' access maps and in return maintain their accuracy. Users' data can further signal areas with significant commercial activity, adding to the new 'areas of interest' feature. As noted by McQuire, the "capacity to generate new features through combinations of extracted and processed data is now a strategic frontier in digital mapping."

Google Maps' ambition is to offer granular-level information on urban spaces that other commercial products will come to depend on. The high-resolution images captured by Street View and information fed by users can facilitate ride-hail drivers in finding correct entrances to buildings. Google's detailed mapping capacities further its advantageous position towards Smart City developments and the use of driverless cars. Control over these projects is what Bergen suggests to be the new 'maps wars.' "Wars that are not only spatial but also temporal, based on the operationalization of 'real-time' feedback systems capable of connecting multiple data streams to specific places." The never-ceasing extraction of data conducted by Google maps on spatial locals, for McQuire, reflects how the physical world is transformed as data.

"In the twenty-first century, it is the stuff of human life itself- from genetics to bodily appearances, mobility, gestures, speech and behaviour- that is being progressively rendered as a productive resource that can only be harvested continuously but also subject to modulation over time".

**From open platforms to proprietary infrastructure**

To continue operating as a high-speed data processing platform, Google Maps requires expensive infrastructure ranging from server farms and proprietary software. As previously mentioned, part of the 'big data' that runs Google Maps is publicly sourced. While the platform is designed to incentivize users to tinker with results, users receive no corporate stake for their labour. The company refuses to relinquish corporate control of what the OSM has termed a 'common resource'. Members of the public cannot re-use citizen-sourced information without the company's permission.

Beyond the data control asymmetries created and enforced by Google Maps, the platform has become a "powerful force in shaping social realities". Referring to the Ta no Mapa (It's on the map) Google Maps' project, residents of Brazilian favelas were asked to provide points of interest within their communities. Researchers noted a disjuncture between residents' perspectives and Google's commercial interests. As indicated by McQuire, "Google maps has the potential to impose a similar commercial-commodity orientation over online digital mapping." Social changes fostered by digital mappings are happening without public oversight over platforms' decisions and operations.

Governmental efforts to combat disinformation have revealed how regulation of 'big data' services is highly challenging, especially given their global reach. While these platforms rely on user-created content, peer-based communication and other forms of horizontal collaboration, companies continue to operate as oligarchies without public consent nor compensation. For McQuire, "the decentralized technical architecture of the Internet and its related digital ecology… have also become the architecture for mass data capture and repackaging user-created content as proprietary services".

**Living with technical objects**

Citing Stiegler, McQuire mentions how the rapid expansion of profit-motivated technological invention causes frictions with other domains in society, such as politics and socio-cultural norms. As a form of 'technological leadership,' Digitization has created new relations of time and space. As noted by McQuire, "technological 'leadership' now extends to how the recursive data-fication enabled by digital communication networks converts the world at large into a 'techno-geographical milieu'." With innovations such as Google maps, technical objects lay down their terms of operation first, and social and legal instruments bend to it. The real-time effects of digital networks are reflective of what Siegler termed "hyper-capitalism", being defined as a "system in which traditionally different sectors – production, logistics, entertainment, marketing and finance – become increasingly integrated and even 'synchronized' across the globe."

**Between the lines**

McQuire's exploration of the trajectory of Google Maps provides important insights into how digital platforms more broadly are changing our temporal and spatial experience of the non-digital world. The profit-seeking ambitions of Google mean that the world will be filtered according to these hyper-capitalist interests, with users actively contributing, albeit somewhat unknowingly, to this perception. Although Google Maps may seem like an all-knowing digital technology, many parts of the world remain unmapped.

The datafication of the physical world is skewed to the interests of companies in the Global North. Like those residing in urban slums, specific communities remain in the shadows of global platforms and unlikely to benefit from positive innovations, such as food delivery services, enabled by these technologies. Increased documentation and surveillance of spaces through Street View may also run counter to these communities' safety and privacy interests. In all, Google Maps merits greater governmental and public scrutiny when it comes to its terms of operation.

## Making Kin with the Machines

([Original paper](#) by Jason Edward Lewis, Noelani Arista, Archer Pechawis, Suzanne Kite)

(Research summary by Muriam Fancy)

Overview: Indigenous epistemologies are able to develop the ethical frameworks and principles to understand how to build and see our relationship with AI and machines. The essay discusses how we can understand creating kinship with AI through Indigenous epistemologies while valuing respect and reciprocity. The authors draw upon Cree, Lakota, and Hawaiian cultural knowledge to create and acknowledge the responsibility to include computational creation in the circle of relationships.

**Introduction**

Joi Ito's "Resisting Reduction" manifesto describes a core principle to Indigenous knowledge, and that is that everything is connected. The author's purpose in this essay is to build upon the proposal of "extended intelligence" and propose a "circle of relationships' which includes human and non-human relationships, and in this case with artificial intelligence. Through utilizing Indigenous epistemologies, the understanding of how to create these relationships and understand interconnectedness is much more dynamic and acknowledges how these relationships are not generalizable, but rather are situated in territories, protocols, and genealogies and therefore imparting the relationality as one of the core principles for this relationship.

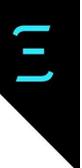

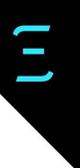

**Indigenous epistemologies**

As the authors note, the epistemologies discussed in the essay are not monolithic, rather the epistemologies referenced in this essay are specific to Cree, Lakota, and Hawaiian Indigenous communities.

As mentioned previously, a core epistemological principle is relationality. The difficulty of applying relationality to cyberspace and computational software on how it exists on Indigenous peoples territory. As the authors note, "…how do we as Indigenous people reconcile the fully embodied experience of being on the land with the generally disembodied experience of virtual spaces? How do we come to understand this new territory, knit it into our existing understanding of our lives lived in real space, and claim it as our own?". It is through Cree, Lakota, and Hawaiian knowledge that this can be further explored to creating kinship and understanding computational creations in the circle of relations.

**Hāloa: the long breath**

Kānaka maoli (Hawaiian people) ontologies can conceptualize these relationships through seeing AI as ĀIna, which is a play on the word of āina meaning Hawaiian land. AI thus should be seen as a relation that supports and "nourishes".

The story of Hāloa serves as an understanding that what is understood as cognition is not static, but rather continues to evolve and change throughout generations. As the authors describe cognition as understanding and acquiring knowledge through thought, experience and senses, and in Hawai'i that also includes the 'ike knowledge of previous generations. And through reflecting on the history of colonial practices, instead of extractive behaviour, principles of balance and abundance must be valid. The history of colonization and western values continue to value the benefit of one over the other, rather the authors say that there must be multiplicity and understanding of how connectedness is found through multiple relations.

**wahkohtawin: kinship within and beyond the immediate family, the state of being related to others**

Nēhiyawēwin (the Plains Cree language) divides all things into two categories: animate and inanimate, one is not better than the other, but they are "different states of being". The author for this section explores how AI would thus be viewed, and how it would fit into the circle of relations. To which the author found when consulting friends, depended on factors of "humanness", "naturalness", and in what capacity could the AI be included. The author explores what it would mean to include AI into the author's cultural practices.

However, a significant concern that should be highlighted, is the way in which AI can harm communities. The way in which AI is built is often exclusionary, harmful, and without thought of accountability, all of which were principles that supported genocide against Indigenous peoples.

The author proposes that ways to mitigate this threat include creating programming languages that are in nēhiyaw nisitohtamowin (for Cree people), as well as other cultural frameworks for other Indigenous communities that would also like to address this issue. AI that has nēhiyaw nisitohtamowin in its language would be a mechanism that would catalyze its inclusion in the community cultural processes.

**wakȟáŋ: that which cannot be understood**

"In order to create relations with any non-human entity, not just entities which are human-like, the first step are to acknowledge, understand, and know that non-humans are beings in the first place." Lakota cosmologies provide an understanding of how to create ethics for humans and our relations to everything. Lakota ethical ontologies are understood through protocols, which speak to prioritizing non-human interiorities (mind, soul or consciousness which is also intentionality, subjectivity, reactivity, feelings, and expression).

Resisting reduction: an Indigenous path forward
Singularitarian, as Ito notes in their "Resisting Reduction" essay is a significant concern. The driver behind singularitarianism shares a similarity to the biases of those who support colonization and slavery. Building and deploying AI for power, exclusion, and extraction is what perpetuates harm.

A path forward includes understanding what kinship and the inclusion of AI in the circle of relations can look like through a multiplicity of Indigenous epistemologies.

**Between the lines**

I strongly urge everyone to read the essay by the authors in its entirety. This research summary discusses the high-level points brought and discussed in the essay but does not do it justice for how it is explained and grounded in the essay by the authors. If I have made many mistakes in this summary, I do sincerely apologize, and please note that I am more than happy to change it so that it better reflects the authors and the essay.

As the authors discussed, AI and other machines are built using western ideologies of extractivism, singularity, and without accountability. One of the many reasons why I thoroughly enjoyed this essay is because it further proves how the inclusion of Indigenous epistemology (which is not static but evolves through generations) can better speak to how to build a better future with AI.

## In AI We Trust: Ethics, Artificial Intelligence, and Reliability

([Original paper](#) by Mark Ryan)

(Research summary by Dr. Andrea Pedeferri)

**Overview**: The European Commission's High-level Expert Group on AI (HLEG) has developed guidelines for a trustworthy AI, assuming that AI is something that has the capacity to be trusted. But should we make that assumption? Apparently no, according to this paper, where the author argues that AI is not the type of thing that has the capacity to be trustworthy or untrustworthy: the category of 'trust' simply does not apply to AI, so we should stop talking about 'trustworthy AI' altogether.

**Introduction**

Trust is an essential feature of our social life. Trust is an attitude that we possess and use when we engage in interpersonal relations, when we believe and have confidence that an agent will do what we ask them to do. Trust is not just a blind bet in the trustee to do something: they have to possess some degree of trustworthiness. However, it's still a bit of a risky business since trustees could break our trust thus "betraying" us. This is why we usually put care in choosing trustworthy agents, to minimize the risk of being betrayed. This is important at the personal level as well as different social levels. It's critical for social, economical, political agencies to be trustworthy. Accordingly, trust has also implications about possible regulations that we can impose or demand to "control" the trustworthiness of those multi-agent structures.

As you (supposedly) trust your well-paid financial advisor in providing you with the best (that is, the best for your interest) financial strategies, should you also trust an AI that does exactly the same job? In a recent deliberation, the European Commission's High-level Expert Group on AI answered: yes. In his, also recent, article "In AI We Trust: Ethics, Artificial Intelligence and Reliability", Mark Ryan gives us the opposite answer: no.

**Why AI can never be trusted**

Ryan comes to the conclusion that "AI cannot be something that has the capacity to be trusted according to the most prevalent definitions of trust because it does not possess emotive states or can be held responsible for their actions". Our common tendency to anthropomorphize AI by attributing it to human-like features, like mental states, is therefore an incorrect way to characterize AI that does not allow us to treat AI as a trustworthy agent.

In order to understand how Ryan came to this quite drastic conclusion, we need to look at the assumptions he draws from. First, he focuses on what is commonly referred to as "Narrow AI", that is, "a system that is designed and used for specific or limited tasks", which is different from "General AI" which is a "system with generalized human cognitive abilities". Second, he relies on a synthesis of the "most prevalent definitions" of trust. AI simply can't, says Ryan, fit in these definitions. As a result, it cannot be taken as a trustworthy agent at all. Let's briefly see the definition used in the paper to better understand Ryan's conclusion and its implications.

**What is trust?**

Ryan proposes a definition of trust that encompasses three main accounts usually associated with trust: the rational account, the affective account and the normative account.

Previously I described trust as a sort of bet on an expected future behavior of an agent. If we read trust in these terms, we could think of trust as the result of a (rational) choice made by the trustor by comparing pros and cons. So, according to this rational account, trust is the result of a calculation by the trustor, and the prediction of whether the trustee will uphold the trust placed in her has nothing to do with any motivation she could possess. Calculations are what machines are usually very good at, so we could say that according to this rational account AI is trustworthy and we can thus trust it. However, Ryan is very skeptical that this account represents what we usually mean by trust; he thinks that this describes just a form of reliance we can have toward AI. This is because rational trust has a total "lack of concern about the

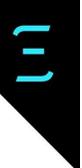

trustee's motivation for action". The presence of those motivations is essential for having trust instead of just reliance. To explain the difference between trust and reliance Ryan provides few examples such as the 'sexist employer' presented originally in a paper by Potter:

"There is a sexist employer who treats his female staff well because he fears legal sanctions if he does not. Because he has not done anything inappropriate to his current female employees, they may consider him reliable, but not trustworthy. 'The female employees might know that their employer treats them well only because he fears social sanctioning. In that case, he could not betray them [because they did not place any trust in him to begin with], although he could disappoint them. However, the rational account of trust would state that the female employees can trust the sexist boss because this type of trust only focuses on the trustee's past behaviour to predict whether they should be trusted. "

Ryan argues that AI deserves similar treatment: we can rely on it to do the job, but it lacks all the features moral agents need in order to be considered trustworthy. So, what are those features?

According to Ryan, the full definition of trust (A trusts B) is:
1. A has confidence in B to do X.
2. A believes B is competent to do X.
3. A is vulnerable to the actions of B.
4. If B does not do X then A may feel betrayed.
5. A thinks that B will do X, motivated by one of the following reasons:
    a. Their motivation does not matter (rational trust)
    b. B's actions are based on goodwill towards A (affective trust)
    c. B has a normative commitment to the relationship with A (normative trust)

The affective and Normative accounts differ from the rational account because "they state that betrayal can be distinguished from mere disappointment by the allocation of the intent of the trustee". So, in order to have "real" trust, the trustee has to possess motivation(s) for action. The rational account can well do without any motivation. Why can't we talk about motivations when it comes to AI? The idea behind the rational account is that reliability is only based on predictions that rely on past performance. However, there are many situations where our decisions about trust cannot be taken by looking at reliability alone.

For example, let's suppose we want to establish a peace treaty with an enemy that fought against us till this moment. By the rational account, they should not be trusted because clearly unreliable. However, that would rule out any possible assignment of trust and therefore any chance for peace between us. Of course, it is important to know, collect and analyze past data when it comes to informing our choices about trust. But, as Ryan points out, "Trust is separate from risk analysis that is solely based on predictions based on past behaviour […] While reliability and past experience may be used to develop, confer, or reject trust placed in the trustee, it is not the sole or defining characteristic of trust. Though we may trust people that we rely on, it is not presupposed that we do".

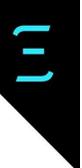

This is because in trust we form expectations that entail the presence of emotive states and motivational states together with psychological attitudes. This is described by the affective and normative accounts of trust. The core principles of those two accounts are motivational states that according to the author are uniquely human. Or, better: "AI may be programmed to have motivational states, but it does not have the capacity to consciously feel emotional dispositions, such as satisfaction or suffering, resulting from caring, which is an essential component of affective trust." This makes AI incapable of complying with the three-component of affective trust, that is,
- the trustee is favourably moved by the trust placed in them;
- the trustee has the trustor's interests at heart;
- and the trustee is motivated out of a sense of goodwill to the trustor.

Moral responsibility is at the center of normative and affective accounts, leaving no hope for AI to be regarded as trustworthy. In Ryan's view, AI is just as a normal artifact in being not a recipient of any moral responsibility which, on the other hand, falls on their developers and users. In fact, according to Ryan even if "AI has a greater level of autonomy than other artefacts, [this] does not constitute an obfuscation of responsibility on those designing, deploying, and using them".  This does not change even if we think at AI as part of complex multi-agent systems. As when we think at the level of complexity of a corporation we associate the concept of trust to the corporation itself and not, for example, to a single employer, so AI continues not to be a "trustworthable" agent even when it's understood as part of complex, multi-agent systems.

**Between the lines**

As a human artifact, AI is still in its "infancy", continuing to develop. One of the greatest challenges of AI development is how to embed some sort of consciousness in it. Assuming that it will eventually be possible to build a "conscious AI" that would not necessarily make it a moral agent. However, that could reposition AI with respect to the three accounts of trust used by Ryan. In this respect, Ryan's conclusion could be used not as a definitive claim about AI and trust but as a stimulus to reach the level of affective and normative trust that AI seems to lack. Accordingly, we can give a more positive reading of the relation between AI and trust, by claiming that the answer to the question of whether we can trust an AI is a more flexible and open "not yet".

## From AI Winter to AI Hype: The Story of AI in Montreal

**(**[Original paper](#) **by Cristian Gherhes, Tim Vorley, Paul Vallance, Chay Brooks)**

**(Research summary by Dr. Cristian Gherhes)**

**Overview**: This paper explores the emergence of AI as a new industry in Montreal, Canada. It highlights the key roles that different actors (i.e., individuals/organisations/institutions) played individually and collectively over three decades in creating the thriving AI ecosystem that put Montreal on the world AI map.

**Introduction**

How do new industries come to life? And why do they grow in some places and not others? One view is that this has a lot to do with serendipity and the type of industrial activity already present in a region. But what about new, radical technologies like AI? How do such technologies become established and develop into new industries? Focusing on Montreal, this paper shows that the emergence of AI in the region cannot be attributed to historical accident but is the result of sustained work by a number of key actors who contributed to building an AI technological innovation system over three decades.

The paper refers to these actors as "system builders" and they include trailblazers (i.e., pioneering actors with a fundamental role in the development of AI in Montreal, like AI scientists, venture capital firms (VCs) and AI start-ups), anchors (i.e., universities, public research labs, and multinational enterprises who contribute to knowledge creation and talent development), and the state (federal, provincial, and local government actors). Each of these actors performed different roles, contributing in specific ways to the development and growth of the AI ecosystem. The paper highlights their role across two phases of system building: a long phase of scientific discovery and exploration, followed by a more strategic phase of intense system building. We show that AI did not just happen to Montreal; it took the efforts of multiple actors to build a thriving AI ecosystem.

**Before the hype: the scientific exploration phase (early 1990s–2016)**

It took more than two decades for the regional AI ecosystem to develop and put Montreal on the world AI map. This is because AI has not always been a hot topic, and this was especially the case in the 1980s and 1990s, a period also known as the second "AI winter", when public interest in AI waned and funding dried up.

But all of this did not matter for a small group of researchers led by Prof Yoshua Bengio who carried on with their research on artificial neural networks when the world was ready to move on. Here is the catch though: they would not have been able to do so without the critical government funding that kept flowing to support their work. Key actors here are the state and the Canadian Institute for Advanced Research (CIFAR) in particular—a publicly funded organisation whose ethos of scientific curiosity and exploration promotes long-term fundamental research. Very importantly, the research grants it awards are not tied to commercial objectives, which enables researchers to tackle big scientific questions by working together over long periods of time. Luckily for Montreal, and for Canada, AI was one of the programmes funded by CIFAR that kept research in neural networks alive during the AI winter.

While impossible to predict at that point that AI will grow into the industry that it is today, public funding was critical to its development. It supported Prof Bengio's work at the Montreal Institute for Learning Algorithms (MILA), which he founded in 1993, creating a small network of AI researchers. MILA ultimately played a key role in advancing the field of AI. Trailblazing through this period of uncertainty, researchers at the institute made major scientific contributions, including breakthroughs such as deep learning, curriculum learning, and generative adversarial networks (GANs) which underpin many of the AI innovations that we see, and use, today. They also contributed to building a strong academic pillar and to training the next generation of AI researchers—a critical resource for the ecosystem that will later fuel its growth.

However, this is not all. A hidden trailblazer—a VC firm—entered the nascent ecosystem in the late 2000s. On a mission to build an entrepreneurial ecosystem in Montreal, which back then had a weak entrepreneurial scene, the VC helped build the infrastructure that now supports start-up activity in AI and beyond. This involved launching accelerator programmes, building international links and bridges between industry, academia, and government, and starting much-needed seed funds. This early activity, while not directly linked to AI, promoted the development of an entrepreneurial culture and paved the way for growth.

**Riding the AI hype: the strategic development phase (2016 onwards)**

The scientific breakthroughs revived global interest in AI and propelled Montreal's nascent ecosystem onto a more intensive system-building phase. Around 2016, AI activity in Montreal saw a boom, and it was an AI start-up that started it and put Montreal in the global spotlight. Founded in 2016, Element AI quickly became the fastest-growing AI start-up in the world, raising what was then considered record funding for an AI company. The fact that Prof Yoshua Bengio, who by this point became known as one of the godfathers of deep learning, was a co-founder, boosted the company's, and the ecosystem's, credibility. Its rapid success catalysed the growth of the AI ecosystem which became a magnet for international talent, VCs, and multinationals—all attracted by the concentration of AI expertise.

What followed was a wave of start-up activity and new actors entering the ecosystem. Very prominent among these are foreign tech giants like Microsoft, Google, and Facebook, who were among the first to open research labs in Montreal and to anchor themselves in the ecosystem by working with star academics, building links with universities and start-ups, and providing research funding—others soon followed suit. This gave credibility to the ecosystem and signalled its potential to investors, talent, and entrepreneurs both within Canada and internationally.

The AI hype that followed helped attract critical resources, particularly money and talent, and the actors that paved the way for growth in the early AI days became part and parcel of the growing AI ecosystem. The renowned AI researchers attracted more AI researchers and funding, which attracted more entrepreneurs and companies into AI, which attracted more VCs, and the impact compounded, turning Montreal into an AI hotspot. The Canadian Government also stepped up its role as an enabler through key strategic AI investments alongside the Government of Québec. These prominently include the CIFAR-led Pan-Canadian Artificial Intelligence Strategy and the Innovation Superclusters Initiative which saw CAD$230m invested in the Montreal-based SCALE.AI supercluster. Besides these, a range of tax incentives at both federal and provincial levels, investments in AI accelerators, and a friendly immigration policy have made it attractive for AI start-ups, multinationals, and talent to establish in the region.

The ambition is to make Montreal the Silicon Valley of AI. And it looks like Montreal has thought of everything. Just when concerns relating to the development and use of AI in the economy and society started to make the headlines, Montreal was already advocating for ethical and socially responsible AI through the Montreal Declaration for a Responsible Development of Artificial Intelligence. Other initiatives quickly followed, including the Montreal AI Ethics Institute and the publicly funded International Observatory on the Societal Impacts of Artificial

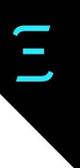

Intelligence and Digital Technologies (OIISIAN). But the journey does not stop here. The key question is: is there more to this hype?

The evolution of the AI ecosystem over the three decades is summarised in the figure below.

**Where next for Montreal?**

The AI hype got Montreal dreaming of AI unicorns. It is the collective vision of those involved in AI in Montreal that the city becomes the home of the next tech giants. However, doing AI research is one thing, commercialising it is another. There are concerns that the scientific breakthroughs have created inflated expectations and that benefitting economically from AI is easier said than done. While no doubt Montreal has earned its status as a centre of excellence in AI research, questions remain over its ability to mature and generate the next generation of AI companies.

Just last year, something that many feared, happened. In November 2020, Element AI, Montreal's AI poster child, announced its acquisition by American software company ServiceNow. This is far from the successful exit that everyone envisaged at the start, given that the company was sold for less than the total amount of capital raised—a significant loss for the investors, including the taxpayers. Beyond that, the exit raises important questions for the future of the AI ecosystem, which lost a key anchor firm. Will Montreal lose credibility? Will investors be more cautious? Will talent stick around? Will another AI winter come?

Some factors are beyond anyone's control and only time will tell, but Montreal has built an incredible technological innovation system around AI, and those who helped build it plan to stay and to continue to build. With the highest concentration of academic researchers in deep learning in the world, a growing fundraising scene, more than $2bn in foreign direct investment in 2019, and new AI companies starting up or joining the ecosystem every year, there are reasons to be optimistic.

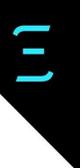

## Dating Through the Filters

(Original paper by Karim Nader)

(Research summary by Karim Nader)

**Overview**: This essay explores ethical considerations that might arise from the use of collaborative filtering algorithms on dating apps. Collaborative filtering algorithms learn from behavior patterns of users generally to predict preferences and build recommendations for a target user. But since users on dating apps show deep racial bias in their own preferences, collaborative filtering can exacerbate biased sexual and romantic behavior. Maybe something as intimate as sexual and romantic preferences should not be the subject of algorithmic control.

**Introduction**

Dating apps have allowed people from extremely different backgrounds to connect and are often credited with the rise of interacial marriage in the United States. However, people of color still experience substantial harassment from other users that can include racial generalizations or even fetishization. This bias can extend from the users to the algorithm that filters and recommends potential romantic and sexual partners. Dating apps algorithms are built to predict the intimate preferences of a target user and recommend profiles to them accordingly, but biased data leads to biased recommendations.

This research establishes that the data that is fed to the algorithm on dating apps reflects deep racial bias and that dating apps can perpetuate this bias in its own recommendations. Further, since recommendations are extremely effective at altering user behavior, dating apps are influencing the intimate behaviors of their users. A look into the philosophy of desires further complicates the issue: intimate biases are often seen merely as personal preferences. But since users have little control over algorithmic filtering, dating apps can come between users and their romantic and sexual autonomy.

**Collaborative filtering**

Collaborative filtering works by predicting the behavior of one target user by comparing it to the behavior of other users around them. For example, if a majority of users who buy chips also buy salsa, the algorithm will learn to recommend salsa to anyone who buys chips. This way, filtering algorithms can build recommendations that reflect general patterns of behavior. And it turns out that they are highly effective at doing it! However, collaborative filtering has a tendency to homogenize the behavior of users on a platform without necessarily increasing utility. Moreover, studies on YouTube's recommender system show that, through algorithmic recommendation, reasonable searches can quickly lead a user to videos that promote conspiracy theories. Algorithmic filtering can thus normalize problematic patterns of behavior through gradual technological nudges and pressures. Is the same true of dating apps? To show that, we'd have to establish that dating app users themselves are feeding the algorithm biased data through their activity.

**Race and online dating**

Christian Rudder, founder of OkCupid, explains that match scores (OkCupid's compatibility score which is calculated by an algorithm) are the best way to predict a user's race. In other words, the match scores of users of different races will show patterns that are distinct enough that we can identify the race of the profile simply by seeing which profiles the algorithm believes is a good match to them. Again, algorithms learn from user data so what kind of data is leading to this kind of racial algorithmic bias on dating apps? Well, it turns out that dating app users show distinct patterns of preference when it comes to race. Several empirical studies confirm those trends: users on online dating platforms seem to segregate themselves based on race and so, prefer people of their own race. Most users exclude people of color from consideration, except those of their own race, and generally show a preference for white men and women. People of color are more likely to include the profiles of white users for consideration, but white people are not as likely to include the profiles of people of color. Since correlations lead to recommendations, users on dating apps will be recommended to other users of their own race and will receive more recommendations for white users.

**Shaping sexual and romantic preferences**

Now, we've established that the algorithm behind dating apps can exacerbate some kind of racial bias. The problem is that it is not clear if this is a problem that needs to be addressed. Surely the Spotify algorithm favors some artists over others, but when it comes to personal taste like music, bias is simply a preference. Sexual and romantic biases might similarly be simple preferences. However, sexual and romantic biases reflect larger patterns of discrimination and exclusion that are grounded in a history of racism and fetishization. And so, there might be some justification for us to raise a moral objection to the use of collaborative filtering on dating apps. After all, recommendations can and do change the behavior and preferences of users. Studies show that if two people are told they are a good match, they will act as if they are regardless of whether or not they are truly compatible with each other. Regardless, the issue might be that users have absolutely no control over the filtering that determines who they see on dating apps. Explicitly stated preferences are sometimes overridden by algorithmic predictions. Using collaborative data in the context of dating apps seems to undermine extremely personal sexual and romantic desires that should not be 'predicted' by an algorithm.

**Between the lines**

Most of the research on dating platforms has focused on dating websites that allow users to browse through a collection of profiles with little to no algorithmic intervention. However, dating platforms have evolved substantially and algorithmic suggestions play a powerful role in the experience of dating app users. This research brings attention to the reach of algorithmic bias on platforms that researchers often overlook.

While people of color anecdotally report lower success rates and occasional harassment and fetishization, those concerns are not taken seriously because personal romantic preferences are seen to be outside of the realm of moral evaluation. Philosophers and moral experts need to pay closer attention to biases that evade ethical scrutiny in this way.

While this research is an important step towards bringing race, romance and attraction into discussions of algorithmic bias, it is merely a conceptual, philosophical and ethical analysis of the question and more empirical work needs to go into understanding the algorithms behind dating apps and the experience of users on those platforms.

## Explaining the Principles to Practices Gap in AI

**([Original paper](#) by Daniel Schiff, Bogdana Rakova, Aladdin Ayesh, Anat Fanti, Michael Lennon)**

**(Research summary by Abhishek Gupta)**

**Overview**: As many principles permeate the development of AI to guide it into ethical, safe, and inclusive outcomes, we face a challenge. There is a significant gap in their implementation in practice. This paper outlines some potential causes for this challenge in corporations: misalignment of incentives, the complexity of AI's impacts, disciplinary divide, organizational distribution of responsibilities, governance of knowledge, and challenges with identifying best practices. It concludes with a set of recommendations on how we can address these challenges.

**Introduction**

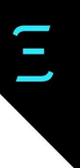

Have you found yourself inundated with ethical guidelines being published at a rapid clip? It is not uncommon to feel overwhelmed with many, often conflicting, sets of guidelines in AI ethics. The OECD AI repository alone contains more than 100 documents! We might also experience a gap in the actual implementation of these guidelines leaving much to be desired after several rounds of discussions. The authors attempt to structure these gaps into some common themes. They emphasize the use of impact assessments and structured interventions through a framework that is broad, operationalizable, flexible, iterative, guided, and participatory.

**What are the gaps?**

The paper starts by highlighting some initiatives from corporations outlining their AI ethics commitments. What they find is that these are often vague and high-level; in particular, without practical guidance for implementation and empirical evidence on their effectiveness, the claims of being ethical are no more than promises without action.

Starting with the incentives gap, the authors highlight how an organization should be viewed not as a monolith but as a collection of entities that have different incentives that may or may not be aligned with the responsible use of AI. They also warn people that companies might engage in the domain of AI ethics to ameliorate their position with their customers and to build trust, a tactic known as ethics shopping, ethics washing, or ethics shirking. Such an approach minimizes accountability on their part while maximizing virtue signaling. Thus, aligning the organization's purpose, mission, and vision with the responsible use of AI can help alleviate this challenge, utilizing them as "value levers."

AI's impacts are notoriously hard to delineate and assess, especially when they have second- or third-order effects. We need to approach this from an intersectionality perspective to better understand the interdependence of these systems on the environment surrounding them. This is important because the harms from AI systems don't arise in a straightforward way from a single product.

Thinking about these intersectional concerns requires working with stakeholders across disciplines but they come from different technical and ethical training backgrounds that make convergence and shared understanding difficult. Discussions also tend to focus sometimes on futuristic scenarios that may or may not come to pass and unrealistic generalizations make the conversation untenable and impractical. Within the context of an organization, when such discussions take place, there is a risk that the ethicists and other stakeholders participating in these conversations don't have enough decision-making power to affect change. There is often a diffusion of responsibility laterally and vertically in an organization that can make concrete action hard.

Finally, there is now a proliferation of technical tools to address bias, privacy, and other ethics issues. Yet, a lot of them come without specific and actionable guidance on how to put them into practice. They sometimes also lack guidance on how to customize and troubleshoot them for different scenarios further limiting their applicability.

**What an impact assessment framework can do**

The authors propose an impact assessment framework characterized by the following properties: broad, operationalizable, flexible, iterative, guided, and participatory with brief explanations of each of these tenets. This framing also includes the notion of measuring impacts and not just speculating about them. In particular, contrasted with other impact assessment frameworks, they emphasize the need to shy away from anticipating impacts that are assumed to be important and being more deliberate in one's choices. As a way of normalizing this practice more, they advocate for including these ideas in the curricula in addition to the heavy emphasis that current courses have on privacy and bias and their technical solutions. The paper concludes with an example about applying this framework to forestation and highlights how carbon sequestration impacts should also consider the socio-ecological needs, for example, those of indigenous communities.

**Between the lines**

It's great to see frameworks that are centred on practical interventions more than abstract ideas. The gap between principles and practices today is stark and such an ontology helps an organization better understand where they can make improvements. We need more such work and the next iteration of such a research endeavour is to apply the ideas presented in this paper in practice and see if they hold up to empirical scrutiny.

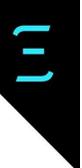

# Go Wide: Article Summaries (summarized by Abhishek Gupta)

## New Algorithms Could Reduce Racial Disparities in Health Care

**(Original article in [Wired](#))**

**What happened**: An AI system used in analyzing knee-radiology images when trained by using the patient feedback as the ground truth label compared to the doctor's determination was found to lead to better outcomes for patients. It was able to uncover blindspots that the doctors had in reading those images because of the historical bias in how the assessment is done and which factors are used in making that assessment.

**Why it matters**: Healthcare automation is seen mostly as a source of bias today — not as a tool that can help improve patient outcomes. So, flipping the script in terms of how the systems are trained in the first place leading to better outcomes for Black people who were being under-diagnosed and not treated well, is a great step forward.

**Between the lines**: As we start to see more scrutiny on the way AI systems are deployed in the real world and we unearth more failure scenarios, I suspect that innovative ways of addressing biases and thinking outside the boxes will lead to more successful outcomes than just relying on purely technical measures.

## Why Is Big Tech Policing Speech? Because the Government Isn't

**(Original article in the [NY Times](#))**

**What happened**: A reflection on the incidents from the beginning of 2021 when #stormthecapital led to real-world harm that originated through organized efforts on the Twitter-alternative Parler. The subsequent actions by Google and Apple to ban the app from the app store and further by Amazon in denying their cloud services to Parler stopped the social media site in its tracks. It was a coup by the tech giants in stopping harm and violence from spreading further during a tumultuous time in US history.

**Why it matters**: While the action makes sense in the short run, such "de-platforming" has raised deeper questions about who should be the arbiter of such decisions. When Trump was removed from Twitter, Merkel from Germany found it to be potentially setting the wrong precedent because such actions should stem from laws (as would be the case in Germany) whereas in this case it was shaped by the political climate (though both Facebook and Twitter deny that it had anything to do with that).

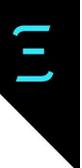

**Between the lines**: There is discomfort in having the government regulate speech online, as is the case with large corporations doing so. There is also discomfort when nothing is done at all making this a particularly hard challenge to solve. Balancing the public good against freedom of speech needs to be done in a way that services democracy, not just to stand in isolation.

## The Little-Known Data Broker Industry Is Spending Big Bucks Lobbying Congress

**(Original article by [The Markup](#))**

**What happened**: The Markup revealed the amount of money spent lobbying by data brokers to rival the money spent lobbying by large tech corporations. Perhaps unsurprisingly, a lot of those lobbying dollars went towards bills that looked at privacy and AI. Many of these data brokers are not names that people are generally aware of but they play an outsized role in how the entire data economy operates.

**Why it matters**: Higher scrutiny on how such companies operate and the impact that their lobbying efforts have is important for people to know so that our activism and other efforts are well-aligned. Given that these companies are only required to self-declare as data brokers in California and Vermont, there is a dire need to bring to light what they're doing.

**Between the lines**: As we start to see more privacy laws enacted in different parts of the world, we need supporting mechanisms to hold companies like data brokers accountable that form the fabric of the entire data economy in the first place. Public awareness and education are going to be key drivers towards creating an ecosystem that is tuned for public welfare rather than private profits.

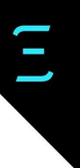

## How AI Can Help Companies Set Prices More Ethically

**(Original article by Harvard Business Review)**

**What happened**: Providing a few concrete examples of how companies can play a role in accessibility to products and services during uncertain times, the article talks about the role that AI is playing in hyper-personalized pricing which can lead to discrimination. It provides a crucial 3-step framework talking about what you are selling and how you are selling it to enable companies to operate in a way that favours socially beneficial outcomes.

**Why it matters**: During the recent events in Texas with electricity outages and with COVID-19 (think toilet paper and webcams) through 2020 and 2021, it has become clear that dynamic pricing can have horrific downstream impacts when goals are specified only towards profitability which can nudge the system towards preying on people's vulnerabilities during difficult situations.

**Between the lines**: We often talk here about how AI is causing harm but perhaps there are ways for it to act as a diagnostic tool in surfacing where inadvertent harms are being inflicted. Acting on those insights to improve the organization's contribution to society can be a great step in the right direction.

## How Facebook's Ad System Lets Companies Talk Out of Both Sides of Their Mouths

**(Original article by The Markup)**

**What happened**: The Markup discovered that companies are using different messages based on political leanings to target people on Facebook. While targeted advertising is not new, what was interesting to observe here was the use of radically different phrasing and messages based on whether someone leaned conservative or liberal.

**Why it matters**: The degree of granularity offered by Facebook to target people is quite significant and companies like Comcast and Exxon Mobil that don't have the best public images can use these kinds of advertising tools to ameliorate their image. They do so by phrasing text and creating images that are likely to appeal to specific audiences. While this was still done to a certain extent before, the granularity of targeting and variations in messaging is much more acute now.

**Between the lines**: Continued pressure from organizations like The Markup uncovering such uses along with tools like the Citizen Browser and the NYU Ad Observatory will play an important role in bringing to light the long road that we still have ahead. Funding more open source tooling and studies will be another essential arrow in our quiver.

## Blood, Poop, and Violence: YouTube Has a Creepy Minecraft Problem

**(Original article by [Wired](#))**

**What happened**: Investigation by Wired unveiled that there are disturbing thumbnails on highly viewed videos on YouTube under innocuous topics like Minecraft and Among Us, which are primarily played by children. While the video content didn't contain as much inappropriate content as some previous scandals like Elsagate, the thumbnails are still prominently present in easily accessible places frequented by children.

**Why it matters**: YouTube has consistently struggled with moderating content and its inability to do so effectively in the case of children is particularly egregious. With the pandemic hitting busy parents hard, a lot of them have been relying on letting their children watch YouTube as a way of catching a break. Problematic content like this that can show up easily on children's screens is a grave danger.

**Between the lines**: YouTube Kids is supposed to be the clean, child-friendly version of the website but in the past with incidents like Elsagate, we've seen that it isn't immune to adversarial manipulation and requires ample work before it becomes a place where children can go unescorted.

## Error-riddled data sets are warping our sense of how good AI really is

**(Original article by [MIT Tech Review](#))**

**What happened**: Researchers from MIT discovered large errors in standard datasets that are used to benchmark the performance of AI systems. Datasets like ImageNet and Quickdraw are estimated to have ~6% and ~10% errors in the labels. While large datasets are already known to have sexist and racist labels, wrong labels in neutral categories exacerbate the problem.

**Why it matters**: Performance evaluation and model selection are done on the basis of metrics evaluated against such benchmarks and if they have incorrect labels (in the context of supervised learning), we get an inflated sense of the capability of the system. What they found out was that simpler models performed much better than complex models when the erroneous labels were corrected, strengthening the case for simplicity in our modeling approaches.

**Between the lines**: Data hygiene, which is the practice of ensuring that the datasets we use are clean and correct, is an important facet of good machine learning practices. It will also have ethical implications especially when used in high-impact scenarios and must be prioritized in any AI systems development.

## Shadow Bans, Dopamine Hits, and Viral Videos, All in the Life of TikTok Creators

**(Original article by [The Markup](#))**

**What happened**: TikTok has been accused in the past of taking a heavy-handed approach to content moderation. As noted in the article, they are known to employ a mixture of human and machine approach (common to a lot of platforms) but whereas platforms like Facebook embody a "keep" approach, TikTok has been known to be swift with content takedowns. Yet, there are other, more effective ways to deploy content moderation, notably through shadow bans whereby content is silently downgraded from appearing in consumer feeds. Such bans are often predicated on characteristics of the content creators like their looks and skin colour which leads to discrimination.

**Why it matters**: The opacity of how the TikTok algorithm operates (as is also the case for other platforms) has led to a lot of speculation and unwanted behaviours from content creators who are consistently stressed about how to maintain earnings which are predicated on the number of views they get on the platform.

**Between the lines**: As more self-help groups emerge addressing the woes of content creators, such as those on Reddit, we might start to level the playing field between the corporations and people using these platforms. Until then, informal fora and tips & tricks might be the best bet for content creators to fight against the unfairness of the system.

## Want to Get Along With Robots? Pretend They're Animals

**(Original article by [Wired](Wired))**

**What happened**: Kate Darling, a robot ethicist from MIT, talks about how we can reframe the conversation on human-machine interaction through the lens of the relationship between humans and animals. Drawing from historical texts when animals were tried for wrongdoings in courts, Darling makes a comparison that comparing robots to animals helps to give us a more nuanced perspective: animals have acted alongside humans as co-workers, companions, soldiers, etc. where they have brought complementary skills to the fore.

**Why it matters**: This reframing is important because it helps us better understand not only how we interact with robots but also how we think about issues like accountability when it comes to technological failures harming humans. It also helps move away from frequent dystopian narratives into something that is more realistic in the role that robots play, and will play, in our society.

**Between the lines**: Discussions like these help to move the field forward especially in an environment today where we have frothy discussions about robot rights and other conversations that try to grapple with problems that aren't quite relevant or might not be relevant in the near future. We have real AI systems that are deployed in the world today interacting with humans, manipulating them to different ends. Finding new ways to have these discussions and inviting people from various backgrounds will help to elevate the level of the conversation.

## Hard Choices: AI in healthcare

**(Original article by [Yale Medicine](Yale Medicine))**

**What happened**: In a discussion with researchers from the renowned Yale Interdisciplinary Center for Bioethics, this article highlights how issues of doctor's agency and problems of amplification of bias are at the heart of using AI in healthcare. Readers of the AI Ethics Brief are already familiar with bias concerns in AI systems, but the problem of reduction in the agency of doctors as they start relying more on machines is grave. This can happen both intentionally (when doctors might try to avoid litigation in a place like the US where the system has been demonstrated to be right many times in the past) or unintentionally (a sort of learned behaviour, akin to automation bias).

**Why it matters**: This is critical to examine because of direct implications on human life. Doctors hold a significant role in society where we trust our lives in their hands in our most vulnerable moments. If they are in turn placing that trust into the digital hands of machines, we run the risk of being subject to inexplicable decisions from machines that are built by corporations who might have optimized the machines to achieve goals that may be orthogonal to patient welfare.

**Between the lines**: Retraining of doctors and education in medical school on the problems such automated systems bring and how doctors can avoid those pitfalls will be essential. To that end, involving AI researchers and developers in bringing that knowledge to medical schools and learning lessons from doctors so that these AI developers can incorporate that in their own system design will be essential if we are to make meaningful progress on building more ethical, safe, and inclusive AI systems for healthcare.

## Google Promised Its Contact Tracing App Was Completely Private—But It Wasn't

**(Original article by The Markup)**

**What happened**: The Exposure Notification Framework which was rolled out jointly by Google and Apple last year as a way of enabling digital contact tracing has been found to have some flaws in its privacy claims. While both companies had said that no private data left the device unless the user elected to do so or identified themselves as being COVID-positive, researchers have otherwise. Bluetooth identifiers and other sensitive information is logged temporarily in system logs on the phone which are accessible to pre-installed apps from the phone manufacturer and could theoretically be exfiltrated to their servers as a part of usage and crash report analytics performed by the device.

**Why it matters**: Given the large number of people who have downloaded the app based on this Framework from around the world, the potential for leaking sensitive information is large. Device identifiers and other information can be correlated to track people, though both the researchers and Google point out that there is no existing evidence that this has been the case.

**Between the lines**: The change as mentioned by the researchers is something simple to implement and won't change the core functioning of the Framework and the apps based on it.

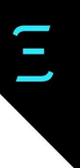

But, Google has repeatedly denied the severity of the issue and only after The Markup reached out to them did they take the concerns from the security researchers seriously. Another incident where Google has demonstrated that they respond seriously only when there is a threat of a PR fiasco.

## To Be Tracked or Not? Apple Is Now Giving Us the Choice.

**(Original article by NY Times)**

**What happened**: The latest iOS update now comes with a mandatory pop-up that apps need to enable that will ask users for their consent to be tracked by the app for advertising and other purposes. While individual settings allowed for that in some measures in different apps, this update makes it a universal requirement. Additionally, the change in UX also makes it more obvious and adds more agency for the users.

**Why it matters**: While developers have found ways to operate on the fringes of what is permissible in the hopes of continuing to suck up data about users to target them with ads and other revenue-generating activities, this new update forces them to face the consequences. It will also make privacy concerns a bit more centred in the UX and hopefully elevate the discussions further as GDPR had done in 2018.

**Between the lines**: This isn't a panacea. There are other ways to track users like fingerprinting of user activities where a number of different behaviours are lumped together to create a unique identifier for a user that doesn't rely on the unique device identifier to track the user. This update will also urge developers to potentially search for other means of tracking. It is an inherently adversarial game and privacy researchers and advocates always need to be on the lookout on how to combat breaches and subversion attempts by companies trying to make an extra buck using our data.

## Facebook Oversight Board Upholds Social Network's Ban of Trump

**(Original article by NY Times)**

**What happened**: The Oversight Board shared its decision on the Trump ban advising Facebook that the indefinite suspension was inappropriate since it isn't an action detailed in their policies. In a 6-month period, Facebook now needs to decide how they act on the recommendations of the Board that suggests they either provide a time-limited ban or issue a permanent ban that is in line with the standard punishment protocol on the platform.

**Why it matters**: This "judgement" will set a precedent for how high-stakes cases will be handled by large organizations. Specifically, Facebook's response will be under greater scrutiny than other decisions made by them given the highly charged political implications of how they act and what precedents are created. It will also be watched closely for the amount of influence that the Oversight Board has on the decisions that are actually made by Facebook. It has been criticized in the past for not having enough power to compel the platform to act in alignment with its recommendations though in the past 4 out of the 5 decisions, Facebook went along with those recommendations.

**Between the lines**: The Oversight Board acts as a first step towards creating greater transparency and accountability in the operations of content moderation. However, there is a lot more to be done and I believe that we need to find ways to work together with the platforms to implement practices that will help us achieve our final goals of having a healthier information ecosystem.

### The four most common fallacies about AI

**(Original article by [VentureBeat](#))**

**What happened**: As we discuss more about the ethical implications of AI, we must also examine how we perceive the capabilities and thus the limitations of these systems. In this article, the author covers some work from Melanie Mitchell scrutinizing the various forms in which we interpret intelligence and how we project those ideas onto machines. Specifically, it examines biases that we have in anthropomorphizing the capabilities of an AI system, how we might be generalizing too soon from narrow AI capabilities, our disconnect between the role of the brain and the rest of the body in realizing intelligence, and inaccurate communication of scientific results.

**Why it matters**: A more accurate understanding of the actual capabilities of AI systems will be essential if we are to make meaningful regulations, policies, and other measures to address some of the ethical challenges in the use of AI systems. Specifically, if we misunderstand (under- or overestimate) the capabilities of AI systems, we might be trying to solve for the wrong problems and set forth erroneous precedents.

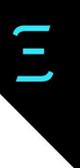

**Between the lines**: In my experience with the domain of AI ethics, as we've had more people pour into the domain, the lack of a shared, well-grounded, and scientifically oriented understanding of the true capabilities and limitations of current and near-term AI systems has led to a lot of people making recommendations that are ineffectual in the goals that they are trying to achieve, both because they are looking at the wrong problems to solve (because they might not be problems at all) or because they are looking at problems that may never materialize which they falsely believe to actually already exist today.

## Nevada Lawmakers Introduce Privacy Legislation After Markup Investigation into Vaccine Websites

**(Original article by [The Markup](#))**

**What happened**: Cookies are used to track users, and data associated with website visits, potentially collated with other sources can lead to rich profiles about any individual. An investigation by The Markup revealed that the COVID-vaccination website run in Nevada had more trackers in it than 46 of the lowest state websites combined! But, the creators of the website changed this, removing outdated cookies, after being shown the results from the investigation. Nevada is also putting forth privacy legislation that has to follow privacy guidelines when they are in public service or have been contracted to provide public service.

**Why it matters**: While cookies can enable functionality on websites, the more invasive kinds of cookies track and share information that can lead to privacy violations galore. More so, given complicated prompts to choose settings, users are often perplexed and end up leaving them enabled even when they don't want to. Such privacy legislations that target at least those websites and services that are run for the general public can help alleviate part of the user burden in this regard.

**Between the lines**: Tools built by The Markup like Blacklight that was used to run the analysis on these COVID-vaccination websites are essential in our fight to protect our liberties and rights. They require an investment to develop but once created they can become powerful instruments in unearthing things like privacy violations in a systematic and scalable manner. We need to encourage more development of such tools, especially when they are in the open-source domain and accessible to everyone.

## Language models like GPT-3 could herald a new type of search engine

**(Original article by [MIT Tech Review](#))**

**What happened**: Google has come out with a new proposal to alter the search experience. They propose to use large language models to replace the current paradigm of the find-index-rank approach to presenting results for our search queries. Typically, we only receive a list of potential matches for things that we search online. The final decision of what is relevant to us is decided by us. A helpful example in the article is comparing this approach to asking your doctor a question and them responding to you with a series of documents that you have to read to find what you are looking for.

**Why it matters**: The obvious concern with utilizing large language models is the degree to which they can change the power dynamics, which already skew away from people being able to tell apart false sources correctly. Answering queries in a natural language fashion through a black box without any explanation for how they arrived at the result is problematic because it can lean towards either algorithmic aversion or automation bias.

**Between the lines**: Such an approach has tremendous potential to alter our relationship to information gathering. In particular, this has the potential to make knowledge discovery more accessible, especially in cases where English (since that is the lingua franca of the web) might not be native to users and the information that they are hunting for might be in English and they are unable to craft the right query to surface that information.

## Twitter's Photo Crop Algorithm Favors White Faces and Women

**(Original article by [Wired](#))**

**What happened**: A few months ago, Twitter users had noticed that when there were multiple people in an image and the image was cropped for display in the timeline, the Twitter cropping system disfavored women and Black people. An analysis on close to 10000 image pairs by the Twitter ethics team has unveiled that indeed there was bias in the saliency algorithm used by the system and they have discontinued its use.

**Why it matters**: While there was anecdotal evidence and small-scale analysis done by researchers in the wild, a more systematic analysis undertaken by Twitter showcasing the same concerns is validation for how community-generated insights can be used to drive AI ethics research. What is also interesting is that an explanation for what was happening has been provided in the paper that has been arXiv.

**Between the lines**: In recent months, as questions have been raised about the efficacy of corporate AI ethics teams, such a study adds a glimmer of hope that useful things can emerge from such endeavours. More so, as a part of our work at the Montreal AI Ethics Institute, we are exploring how a more cooperative approach between civil society and corporations can actually yield better outcomes than the current adversarial state of affairs between the two communities (which has legitimate grounds given the historical interactions between the two groups).

### The Future Of Work Now: Ethical AI At Salesforce

**(Original article by [Forbes](#))**

**What happened**: The article offers insights into the Ethical AI practice at Salesforce with some practical lessons on how they have scaled internally and how they have provided support to external customers. It features Kathy Baxter, the Ethical AI Architect at Salesforce articulating their strategy which follows engaging, advising, and adopting as key pillars to actualize ethical AI practice within the organization.

**Why it matters**: Such details on how companies are actually implementing AI ethics in practice are essential to build trust with customers and the public, more so in a time where moves by tech companies are being heavily scrutinized. As a practitioner, what caught my attention was their use of a model card generator as a way to make model cards more practical and real compared to the largely theoretical construct before.

**Between the lines**: I foresee more companies becoming transparent about their AI ethics methodologies! There is a lot to be learned from each other's work, especially in the nascent stages of this field. The recent case study published by the World Economic Forum talks about Microsoft's Responsible AI practice with a lot of details that can serve as a blueprint for other organizations that are seeking to get started.

## Google's New Dermatology App Wasn't Designed for People With Darker Skin

**(Original article by Vice)**

**What happened**: In an app that is used to assist doctors to do dermatological analysis for skin diseases, it was found that the app had severe limitations in the outcomes from the app for those with darker skin tones. In particular, in a paper that was published in Nature Medicine some time ago, the results from the app were popularized as performing well on people of different ethnicities, much more so than previous solutions that attempted to find a computational solution to detecting skin diseases.

**Why it matters**: Something that readers should pay attention to is that the results for various minority groups were based on self-identified ethnicities rather than Fitzpatrick skin types, in particular Type V and Type VI which were severely underrepresented or absent in the dataset used to train the system. For something where skin type can have significant impacts on the outcomes, relying on self-identified ethnicities doesn't serve as any meaningful proxy for the skin type and can severely overestimate the efficacy of the solution on the non-majority demographics.

**Between the lines**: A problem that continues to pervade in dermatology research is the lack of sufficient or comparable datasets for darker skin tones compared to lighter skin tones. And this only gets amplified in computational approaches as well. In particular, without deep investments in building up more representative datasets first, any future research in applying these methods will continue to suffer more similar failures and ink will be spilled (here and elsewhere!) pointing out the same errors again and again.

## The New York Times Uses the Very Dark Patterns it Derides

**(Original article by Nir and Far)**

**What happened**: Nir Eyal, author of Hooked and Indistractable, highlights how the NYT uses a dark pattern for its unsubscribe workflow. Dark patterns are something that their journalists have chided other companies for yet at this point it is quite well known that NYT makes it incredibly difficult for subscribers to get out. Eyal positions this as the roach motel model and through screenshots demonstrates the ugliness embodied by the NYT subscriptions team in how they handle their customers.

**Why it matters**: Eyal provides recommendations like usability testing and the "regret test" which can help companies get a sense for whether what they are doing follows ethical practices or not. A regret test basically seeks an answer to the question if the user would take an action knowing everything that the designer knows about the product or service. This is a great smoke test to get the basics right in ethical product building.

**Between the lines**: The NYT is a long-time offender in the use of this particular dark pattern. Yet, they don't change their ways but offer a cautionary tale to other companies who engage in such practices that over time, all does come to light and can hurt their long-run sustainability. As Eyal points out in the article, it is perhaps just a function of a large organization with misaligned incentives where someone in charge of subscriptions and revenue decided that such a dark pattern was ok to meet their quotas, disregarding the negative ethical implications that the use of such a pattern has.

## How a largely untested AI algorithm crept into hundreds of hospitals

**(Original article by [Fast Company](Fast Company))**

**What happened**: Epic, one of the largest health data companies in the US, deployed a non-peer-reviewed system called the Deterioration Index across many hospitals amidst the rush unleashed because of the pandemic. In particular, this system is used to aid doctors in triaging a patient to allocate intensive care beds. The typical workflow in the case of a medical system is to subject the system through rigorous peer-review before allowing it to be used in live settings.

**Why it matters**: The biggest flaw emerging from the system, in light of all that we know about the huge issue of bias in medicine, is that it is proprietary. While the doctors are given some guidance on the importance of different factors that go into arriving at the final recommendation from the system, they are not allowed under the hood. This has tremendous potential to amplify pre-existing biases along the lines of race and gender.

**Between the lines**: On the one hand, it is not surprising that a system was rolled out hastily without the regular review process given the enormous strains that medical institutions have faced in the wake of the pandemic. But, as has been articulated in many pandemic playbooks before, this should not be used as an excuse for releasing and using untested technology, especially when it can have significant impacts on the lives of people. A similar argument has been made in the case of facial recognition technology as well as many states rolled that out in a hurry to monitor and enforce social distancing rules among other use cases.

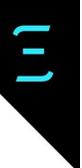

**Dark Patterns that Mislead Consumers Are All Over the Internet**

**(Original article by [The Markup](#))**

**What happened**: Building on the NYT article, the problem of dark patterns continues to plague us and is seemingly all around us. In this piece by The Markup, the article talks about ABCMouse utilizing automatic subscriptions after a free trial to trick customers into giving them money. This is a very common pattern (and the article has a quiz that you can take to test your skills at spotting these) that just doesn't seem to go away. Other companies are also named in the article including Amazon that makes it hard to cancel Prime subscriptions.

**Why it matters**: Dark patterns essentially nudge users into taking actions that they wouldn't take otherwise. This is a huge problem, especially when you have users who are not that tech-savvy who can fall into these traps. There have been attempts in the past to regulate these dark patterns like the DETOUR Act but there needs to be a systematic effort to root these out. A website linked in the article documents many other cases where this takes place.

**Between the lines**: Ethical design practices should be something that is ingrained at a very early stage in the education of designers. More so, this should be reinforced at organizations by way of correctly setting up incentives in the design practice so that the chance of stumbling into, or intentionally practicing, dark patterns becomes minimized.

**How censorship became the new crisis for social networks**

(Original article by [Platformer](#))

**What happened**: While previously the biggest concerns facing social media platforms were that they were leaving up problematic content, now the pendulum has swung the other way - they are taking down too much content, sometimes under pressure from the local governments in the places where they are operating. Trust between users and corporations has deteriorated to the point that even seemingly business-motivated moves like how to rank stories vs. native posts on Instagram are seen as political moves as it disproportionately affects activists who use this as a way of drawing attention to their causes.

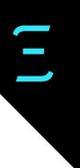

**Why it matters**: This tussle between social media platforms, users, and governments will continue to oscillate if we don't get clearer guidance on how to operate in a way that respects the rights of people while keeping up a healthy information ecosystem. The article makes some interesting arguments on the role that the platforms played in the rise of authoritarianism and how that subsequently was used by the very same people to raise even more concerns.

**Between the lines**: In work that the Montreal AI Ethics Institute did on online content moderation, we lay out a lot of guidance expanding on the work done by the Santa Clara Principles as a way to tackle some of these thorny issues. Most importantly, our emphasis on taking a data-driven approach to assess the impacts of each of these decisions and sharing those transparently with a larger audience has been at the heart of those recommendations.

## 10 steps to educate your company on AI fairness

**(Original article by [World Economic Forum](#))**

**What happened**: With the large gap between principles and practice, the WEF convened a group of experts to propose a set of practical interventions to make Responsible AI a reality. The recommendations notably include: assigning the role of Responsible AI education in the organization to a Chief AI Ethics Officer. Our team covered this role a few weeks ago here. Clear communication of the AI ethics approach is another recommendation that resonated well with us, something that we covered at the start of 2021. Finally, the inclusion of a "learn-by-doing" approach was a welcome change from other recommendations made in numerous documents and guidelines because we are still not sure which, if any, of these approaches are going to work well and this requires experimentation and documentation of the results.

**Why it matters**: This is a great shift from a high-level organization to change the pace and direction of the current discourse in AI ethics towards something that is a bit more solution-oriented rather than just identification of problems which has been the dominant focus of the conversations over the last couple of years.

**Between the lines**: While the recommendations do give some actionable advice, for the most part, they are still quite abstract and require a lot more empirical evidence and consultation with people who have actually implemented some of these ideas and others in their organizations. Reading them as they are presented at the moment still leaves much to be desired, partially undercutting the emphasis that they want to make on them being practical interventions.

### These creepy fake humans herald a new age in AI

**(Original article by [MIT Tech Review](#))**

**What happened**: As regulations tighten around data-use around the world, understandably, there are some organizations that are choosing to use synthetic data as an alternative to "messy, real-world data". The article documents the work of some of the companies that provide synthetic data to other organizations based on tailored needs. While there are benefits of having such curated data, the article makes the case that it is not all roses because there are some shortcomings that hinder their efficacy and also compromise their promises in terms of mitigating biases and protecting privacy.

**Why it matters**: Discussions around the use of synthetic data have always been explored when one finds themselves short of data to train specific ML systems that require large amounts of data to function well. But, recent conversations including this one add a socio-technical lens which is much needed if (and we suspect that this will be the case) synthetic data becomes more commonplace in the development of modern AI systems.

**Between the lines**: Something that needs a bit more analysis and perhaps something that is worthy of a short research endeavour is how some guidelines and standards can be established in the use of synthetic data to build AI systems on synthetic data that meet our needs of mitigating bias and protecting privacy amongst addressing other ethical concerns.

### Can Schools Police What Students Say on Social Media?

**(Original article by [The Markup](#))**

**What happened**: A student faced repercussions from her school based on some remarks that she made on social media igniting an intense debate on what the boundaries of free speech should be and what role schools play in that. While there are many different viewpoints that have been offered, at the moment, as per reporters, the Supreme Court is siding with the student in support that the school overstepped its boundaries. The local ACLU chapter is fighting the case on behalf of the student while the legal team from the school has been silent in its comments.

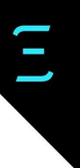

**Why it matters**: It is well accepted that schools have authority on their premises to discipline students when they cross boundaries that can cause harm to staff and other students on campus. But, increasingly as the definition of campus extends into the digital world, the boundaries blur quite significantly. What students do online might remain outside the purview of the school authorities, but there is ample risk of "outside" online activity pouring back into activities on campus and this is what is causing the dilemma in determining appropriate boundaries in this case and others similar to it in the past.

**Between the lines**: This raises classical questions about the boundaries of free speech and how they may infringe on others' rights. In particular, it is even more important in an age where the internet allows any voice to be heard and amplified quickly escalating arguments. The boundaries for what forms free speech and how it may or may not be policed have been studied for many decades and emerging literature in applying these ideas in the internet age will play a significant role in the future of online discourse.

### 'Care bots' are on the rise and replacing human caregivers

**(Original article by [The Guardian](#))**

**What happened**: When we imagine carebots, we think of friendly-faced robots zipping around a healthcare facility, but the truth is much more mundane: they are embedded in the software that is responsible for allocating care hours, monitoring health of patients, and directing staff in these facilities towards different patients based on needs that are assessed in an opaque fashion. The authors in the article argue that while these carebots are already here, the future need not be dystopic; we can still shape how they integrate with our healthcare facilities by being more deliberate about ensuring autonomy, dignity, and warmth in tending to humans at their most vulnerable.

**Why it matters**: Attention for algorithms deployed in a healthcare context tend to focus a lot on issues of bias, privacy, etc. that fall under the large umbrella of Responsible AI. We need to consider how they impact people's perception of care and how comfortable and cared for patients feel in interacting with these systems. We also need to think about the impact that they have on the workflows of healthcare workers.

**Between the lines**: As companies rushed to bridge the care worker shortage in the wake of the pandemic, issues of privacy and bias were swept aside in the interest of expediency. We can't let our guard down as we bring in these systems that are yet to prove their usefulness in a manner that is consistent with the values that we care for in a healthcare setting. We have to be vigilant and most of all deliberate in our integration of these systems in the existing healthcare system. Most of all, we need to involve domain experts, including the healthcare workers who will bear the brunt of the decisions made by these systems in addition to the patients themselves.

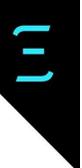

# 7. Community-Nominated Spotlights

[**Editor's Note:** The community-nominated highlights in this section are generous contributions, recommendations, and pointers from the global AI ethics community, with the goal of shedding light on work being done by people from around the world representing a diverse mix of backgrounds and research interests. We're always looking for more nominations, and you can nominate someone for our next report by emailing support@montrealethics.ai]

### Spotlight #1: Sun Sun Lim (Professor, Singapore University of Technology and Design)

Arising from my multiple complementary roles as a researcher, educator and advocate, I believe my work helps fill several gaps in AI ethics.

As an academic researcher, my innate curiosity about how people engage with technology has driven my investigations of technology domestication within the home, the workplace and our diverse lived environments. These settings where cultures, values and lifestyles bump up against technological systems are where we can really witness the strengths and limitations of AI, as well as emerging ethical issues. Surveillance of children by parents, technological intensification of teens' peer pressure, anxieties about displacement from workplace digitalisation, fears of automated systems taking over public services – these are all deep-seated concerns that people have shared with me in the course of my research as an ethnographer. I have sought to capture and represent these views through my research so as to give voice to the visceral human concerns that people have about AI, to thereby make it more human-centric.

As an educator therefore, positively shaping the next generation technologist is a principal goal of mine. We need to focus our energies upstream to ensure that our engineers, designers and technologists of the future have a firm grasp of various facets of society that extend well beyond their technical knowledge. To this end, I have actively leveraged my position as Head of Humanities, Arts and Social Sciences in the Singapore University of Technology and Design to better expose our engineering, design and architecture students to disciplines such as history, literature, philosophy, anthropology, sociology, psychology and communication. I believe that a cohesive and well-rounded interdisciplinary STEAM education will nurture technologists who can make AI more empathetic, ethical and sustainable.

As an advocate, I seek to highlight where we can do more to ensure that policies, provisions and public education can compensate for technological shortfalls and make AI work better for society. I have drawn on my involvement in various committees and in parliament to raise issues such as governance of the use of big data, priorities in digital literacy education, improvement in data sharing, regulatory overreach in eradicating online falsehoods, support for gender diversity in the technology sector, and digital rights for children. Most recently, I have collaborated with colleagues in academia and social services to push for universal digital access. In a digitalising world, digital access is not a want but a need and must be provided universally to ensure a level playing field for everyone.

Since I was a kid, I've been fascinated by how shiny new gadgets and interfaces lure people, while simultaneously thrilling and frustrating them. From when the telephone company sent someone to set up our land line phone, to the first time I heard the hiss and screech of a dial-up modem, to observing toddlers play with interactive screens in malls, I've always been intrigued by the human-technology relationship. I like technology and I can imagine many marvellous possibilities with it! But as a social scientist, I recognise the risks that come with technology if individual competencies, corporate practices, regulatory regimes and social policies are outpaced by technological transformations. Hence, my research has sought to uncover how people adopt and adapt to technology, while reshaping it in the process. What people do or don't do with technology reveals a lot about our hopes and expectations for ethical AI.

In the wake of the pandemic making us so reliant on technology, along with the unfortunate occurrence of some harrowing natural disasters, I believe there will be more vociferous questioning around the long-term environmental impact of AI on climate change. Separately, the thirst for information about the pandemic and our growing turn towards digital connectivity raise issues about the veracity of information, the robustness of our information landscape, and the role of information gatekeepers. If these are increasingly being managed by machine learning algorithms, how can we ensure that they are acting in society's best interests? Finally, the prolonged pandemic lockdowns have made more young people go online for longer, to do more, and at much younger ages. These trends demand urgent attention to [children's digital rights](#) because their online activity is systematically tracked and the data gathered is being used for a heady variety of commercial purposes. We must act to ensure that children's personal data is processed fairly, lawfully, accurately and securely, for specific purposes and with the free, explicit, informed and unambiguous consent of children and their parents.

I'm always open to collaborations on all of the areas I conduct research on. Currently, I would most like to broaden the global conversation around [universal digital access](#). How can we distil the best practices from efforts in different countries to develop a model for universal digital access that is practicable, replicable and sustainable? How can we create people, private and public sector partnerships that can offer stable and reliable digital access and comprehensive digital literacy education to support the digitally disconnected or underserved? I'm excited to learn more from the diverse and talented community of people connected to the Montreal AI Ethics Institute.

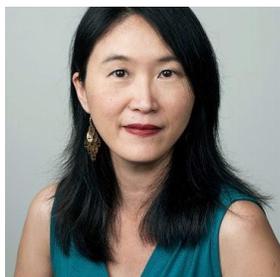

**Sun Sun Lim**

Sun Sun Lim is Professor of Communication and Technology and Head of Humanities, Arts and Social Sciences at the Singapore University of Technology and Design. She has extensively researched and published on the social impact of technology, focusing on technology domestication by families, workplace digital disruptions and public perceptions of smart city technologies. She authored *Transcendent Parenting: Raising Children in the Digital Age* (Oxford University Press, 2020) and co-edited *The Oxford Handbook of Mobile Communication and Society* (Oxford University Press, 2020). She serves on the editorial boards of eleven journals and several public bodies including the Social Science Research Council, Singapore Environment Council and Media Literacy Council. From 2018-2020, she was Nominated Member of the 13th Parliament of Singapore. She was named to the inaugural Singapore 100 Women in Tech list in 2020 that recognises women who have contributed significantly to the technology sector. She frequently offers her expert commentary in international outlets including *Nature*, *Scientific American*, *Channel NewsAsia* and *The Straits Times*. She has won eight awards for excellent teaching. See www.sunsunlim.com

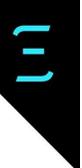

## Spotlight #2: Claudia May Del Pozo (Director of the Eon Resilience Lab at C Minds)

**Creating bridges to include the Global South's perspective in global conversations**

Nowadays, it seems a lot of the talk around AI ethics, policy and regulation falls into a conversation about seeking global consensus. With the lack of physical borders in the digital world and the current state of globalization, this makes sense, but it is hardly the time to be expecting concrete results, or at least inclusive ones. While many countries in the northern hemisphere may have experience dealing with AI systems and, to a certain extent, their impact, countries in the Global South are starting to uncover these technologies' potential, integrating them into existing systems and observing their impact, intended or otherwise. Of course curiosity naturally leads the Global South community to look to the North in terms of AI innovation, particularly in regulatory and ethics principles matters, which then leads us to the question: could that approach work over here?

More often than not, certain aspects get in the way, be they legal, infrastructural, cultural, to name a few, which means that leaders in the Global South need to conjugate foreign frameworks to their realities, creating a plethora of rich and different approaches to similar AI ethics-related challenges. As the world's economies digitalize further, we will have a better understanding of the challenges AI systems pose in different contexts and only then can we truly start conversations about a potential global consensus, one where each country has had iterative experiences with AI systems as a core part of their society and reality. Having this truly inclusive consensus, however, requires that the South amplify its voice and the North to make more room at the table. That point at the intersection of the North and South is exactly where C Minds stands.

Through a variety of initiatives, the Mexican women-led think-and-do-tank seeks to include Mexico and, more generally, Latin America in international conversations, contributing findings from our state-of-the-art research and frontier pilot projects that offer additional insight from a different perspective or generate a completely new dimension for conversations, regionally and globally. Our work at C Minds, which includes the participation of countless regional and international partners, placed Mexico among the first 10 countries to work towards a National AI Strategy, offering other countries a Global South, specifically Latin American, perspective on the matter. Via the Eon Resilience Lab, Mexico is now carrying out the world's very first public policy prototype for transparent and explainable AI, the third AI governance policy prototype at a global level, bringing brand new learnings about the feasibility of such projects, the reality of applied AI ethics, as well as the perspective of Mexican companies, government and other stakeholders on the matter. Together with a regional organization, the lab is also seeking to put forth new priorities on the global agenda, for instance, that of the urgent need for measures to be put in place to protect children's privacy in light of the rise of online education adopted as a response to the social distancing required by the pandemic. While such conversations are only budding on the international stage, Latin America is ready to contribute a report with key findings for and from the region.

On a personal note, I've always been intrigued by the potential of technology to improve our quality of life. The opportunity to be able to work collaboratively on the latest AI-ethics challenges, contributing specific insight to the global arena while also generating a considerable impact in the countries we operate in, is absolutely thrilling. Disproving Mexico as a jet-lagged leapfrogging economy is one of the things my colleagues and I do best. We do this both via the projects we carry out and by using our platform to amplify the visibility of other Latin American projects. As a resident of Mexico, I am able to observe firsthand the impact of C Minds' different projects around me and having a job with a true and tangible impact is any millennial's dream.

The good news is that a lot of what I mentioned above is already starting to happen. On the one hand, global events and projects are increasingly making space for Latin America and the Global South at the table. On the other hand, many initiatives are being implemented in the region to promote a more ethical use of AI, exploring the applicability of best practices from different international AI hubs to the local context and producing our own. For Latin America's AI ecosystem, I am hopeful that 2021 will be the year that the industry strengthens their commitment to ethical and responsible AI with concrete action and I hope it will mark a turning point for governments who have yet to adjust their normative frameworks to the new realities brought about by AI. Slowly but surely, I hope to see these moves by the AI ecosystem generate ripple effects that lead to more awareness of AI systems and their potential impacts among the general public, a key agent if we want to move forward responsibly.

As fellow AI ethics enthusiasts and professionals, we would love to invite you to follow C Minds via our social media channels and help spread our work across the world with a simple share or retweet. And if you find your values and narrative to be very much aligned with C Minds' and your organization would like to explore potential collaborations with us in Latin America, please reach out to us, we'd love to hear from you: claudia@cminds.co

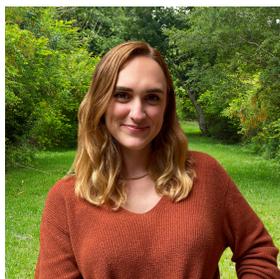

**Claudia May Del Pozo**

Claudia May Del Pozo is the Director of the Eon Resilience Lab of C Minds, a Mexican action tank she collaborated in the design of. C Minds was the first organization of its kind in Latin America, seeking to accelerate social and economic development via the strategic and ethical use of new technologies such as Artificial Intelligence. The Eon Resilience Lab, its latest area, focuses on futureproofing people, helping them understand and make the most of the digitalization of our world by working in collaboration with the industry, Big Tech, governments, and multilateral organizations.

Claudia is currently working together with Facebook, the IDB and with the support of Mexican data regulators to carry out the world's very first public policy prototype for transparent and explainable AI systems. She is also helping Mexico prepare for the future of work and harness AI for post-COVID economic recovery, in a variety of initiatives including local government, national government, the industry, academia and multilateral organizations. In the past, she co-authored a report on AI for Social Good in Latin America, launching the IDB's AI for Social Good Observatory. She also coordinated and co-authored Mexico's National AI Agenda from the IA2030Mx Coalition, placing Mexico among the 10 first countries to have an AI strategy, and has provided training on the responsible use of AI to a wide variety of stakeholders, including policy-makers and the Mexican military.

In the past, Claudia worked at IBM Germany and graduated from Warwick Business School in the United Kingdom. She has co-authored various reports and policy recommendations to promote digitalization and the ethical use of AI in Latin America, a book on Open Banking and is a TEDx speaker.

# Closing Remarks

Wow! We got to tread through a lot of different areas in incredible depth in the past ~200 pages!

We tried to create a more evergreen version of the report with this edition through the publication of topical analyses and sharing our insights on some core areas that we felt are going to be essential to the development of the field. I particularly enjoyed the "Critical Race Quantum Computer" as it provided a very interesting take on how we can think about our perceptions of each other, how we might codify those in machines, and most importantly, it emphasized for me how nuance is the most critical component when think about how to examine the relationship between technology and us, and our relationships with each other as mediated by technology.

What were your biggest takeaways from the report? We are always delighted to hear from you and incorporate your suggestions in making this report a more valuable resource for the entire community. Please don't hesitate in reaching out to us at [support@montrealethics.ai](support@montrealethics.ai) to chat and share your ideas.

In the meantime, hope you stay safe and healthy! Until the next time when we meet in these pages, I'd like to leave you with these lines from Maya Angelou:

*"Do the best you can until you know better. Then when you know better, do better."*

---

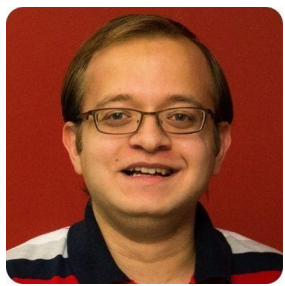

**Abhishek Gupta ([@atg_abhishek](@atg_abhishek))**
Founder, Director, & Principal Researcher,
Montreal AI Ethics Institute

Abhishek Gupta is the founder, director, and principal researcher at the Montreal AI Ethics Institute. He is also a machine learning engineer at Microsoft, where he serves on the CSE Responsible AI Board. He also serves as the Chair of the Standards Working Group at the Green Software Foundation.

# Support Our Work

The Montreal AI Ethics Institute is committed to democratizing AI Ethics literacy. But we can't do it alone.

Every dollar you donate helps us pay for our staff and tech stack, which make everything we do possible.

With your support, we'll be able to:

- Run more events and create more content
- Use software that respects our readers' data privacy
- Build the most engaged AI Ethics community in the world

Please make a donation today at **montrealethics.ai/donate**.

We also encourage you to sign up for our weekly newsletter *The AI Ethics Brief* at **brief.montrealethics.ai** to keep up with our latest work, including summaries of the latest research & reporting, as well as our upcoming events.

If you want to revisit previous editions of the report to catch up, head over to **montrealethics.ai/state**.

Please also reach out to **Masa Sweidan** masa@montrealethics.ai for providing your organizational support for upcoming quarterly editions of the *State of AI Ethics Report.*

**Note:** All donations made to the Montreal AI Ethics Institute (MAIEI) are subject to our **Contributions Policy**.